\pdfoutput=1
\documentclass[final,3p]{elsarticle}
\usepackage{graphicx}
\usepackage{lipsum}

\usepackage[utf8]{inputenc}
\usepackage[T1]{fontenc}
\usepackage{textcomp}
\DeclareUnicodeCharacter{2212}{-}
\usepackage[english]{babel}
\usepackage{amssymb}

\usepackage[version=4]{mhchem}
\usepackage{siunitx}
\usepackage[svgnames]{xcolor}
\definecolor{blues2}{HTML}{0068B4}

\newcommand*\rev[1]{\textcolor{black}{#1}}

\journal{}

\makeatletter
\def\ps@pprintTitle{%
 \let\@oddhead\@empty
 \let\@evenhead\@empty
 \def\@oddfoot{\centerline{\thepage}}%
 \let\@evenfoot\@oddfoot}
\makeatother

\begin{document}
\mhchemoptions{text-greek=default, math-greek=default}
\begin{frontmatter}


\title{Metal--Organic Frameworks in Germany: from Synthesis to Function}



\author[TUDresden]{Jack D. Evans}
\author[TUDresden]{Bikash Garai}

\author[CAU]{Helge Reinsch}

\author[TUMunich]{Weijin Li}
\author[TUMunich]{Stefano Dissegna}

\author[TUDresden]{Volodymyr Bon}
\author[TUDresden]{Irena~Senkovska}

\author[TUMunich]{Roland A. Fischer}
\author[TUDresden]{Stefan Kaskel\corref{mycorrespondingauthor}}
\cortext[mycorrespondingauthor]{stefan.kaskel@tu-dresden.de}
\author[Dusseldorf]{Christoph Janiak}
\author[CAU]{Norbert Stock}
\author[Augsburg]{Dirk Volkmer}

\address[TUDresden]{Department of Inorganic Chemistry, Technische Universität Dresden, Bergstraße 66, 01062 Dresden, Germany}
\address[CAU]{Institut für Anorganische Chemie der CAU Kiel, Max-Eyth-Straße 2, 24118 Kiel, Germany}
\address[TUMunich]{Fakultät für Chemie--Anorganische und Metallorganische Chemie, Technische Universität München, 85748, Garching, Germany}
\address[Dusseldorf]{Institut für Anorganische Chemie und Strukturchemie, Heinrich-Heine-Universität Düsseldorf, Universitätsstraße 1, 40225 Düsseldorf, Germany}
\address[Augsburg]{Chair of Solid State and Materials Chemistry, Institute of Physics, University of Augsburg, Universitätsstraße 1, 86159 Augsburg, Germany}

\begin{abstract}
Metal--organic frameworks (MOFs) are constructed from a combination of inorganic and organic units to produce materials which display high porosity, among other unique and exciting properties.
MOFs have shown promise in many wide ranging applications, such as catalysis and gas separations.
In this review, we highlight MOF research conducted by Germany-based research groups.
Specifically, we feature approaches for the synthesis of new MOFs, high-throughput MOF production, advanced characterization methods and examples of advanced functions and properties.
\end{abstract}


\end{frontmatter}


\section{Introduction}

Metal--organic frameworks (MOFs) are coordination polymers composed of mono- or multinuclear complexes (clusters) and organic multifunctional ligands (linkers) forming extended, regular network structures generating defined pore spaces in a crystalline lattice with a high degree of functionality \cite{10.1126/science.1230444}.

\subsection{Motivation}
Since the first MOFs appeared in the 1990s, this research field has experienced tremendous growth and attention from a wide interdisciplinary community going far beyond coordination chemistry, including solid state, organic, physical, colloid, and, more recently, electrochemistry.
Today, MOF research groups are in engineering, environmental science, earth science and many other fields and the community is still growing, as reflected by the increasing number of participants at the international MOF conference \cite{MOF2018}.
Industrial efforts in major industrial enterprises, small and medium-sized enterprises and startups are spread all over the world. MOFs are record materials in terms of porosity achieving specific surface areas up to \SI{7000}{\square\metre\per\gram} and methane storage capacity of \SI{\approx240}{\cubic\centi\metre\per\cubic\centi\metre} \cite{10.1021/ja3055639,10.1038/nmat5050}.
Moreover, rational functionalization at the molecular level allows for the integration of a wide range functionalities such as optical, magnetic, electronic, chiral, catalytic, or sensor functions stemming either from the ligand or cluster moieties. Early MOF discoveries, such as {MOF-2} \cite{10.1021/ja981669x},  {HKUST-1} \cite{10.1126/science.283.5405.1148} or \ce{CuSiF6}(4,4'-bipyridine)$_2$ \cite{10.1002/1521-3773(20000616)39:12<2081::AID-ANIE2081>3.0.CO;2-A} had relatively low specific surface area, while important milestones, such as {MOF-177} \cite{10.1038/nature02311}, {MIL-101} \cite{10.1126/science.1116275}, and {NU-110} \cite{10.1021/ja3055639} may be considered as important benchmark materials.

In our present review we describe the contribution of Germany-based research teams towards progress in metal--organic framework research in various areas from fundamental research towards applications.

\subsection{History}

\begin{figure}[tbhp]
\begin{center}
\includegraphics[]{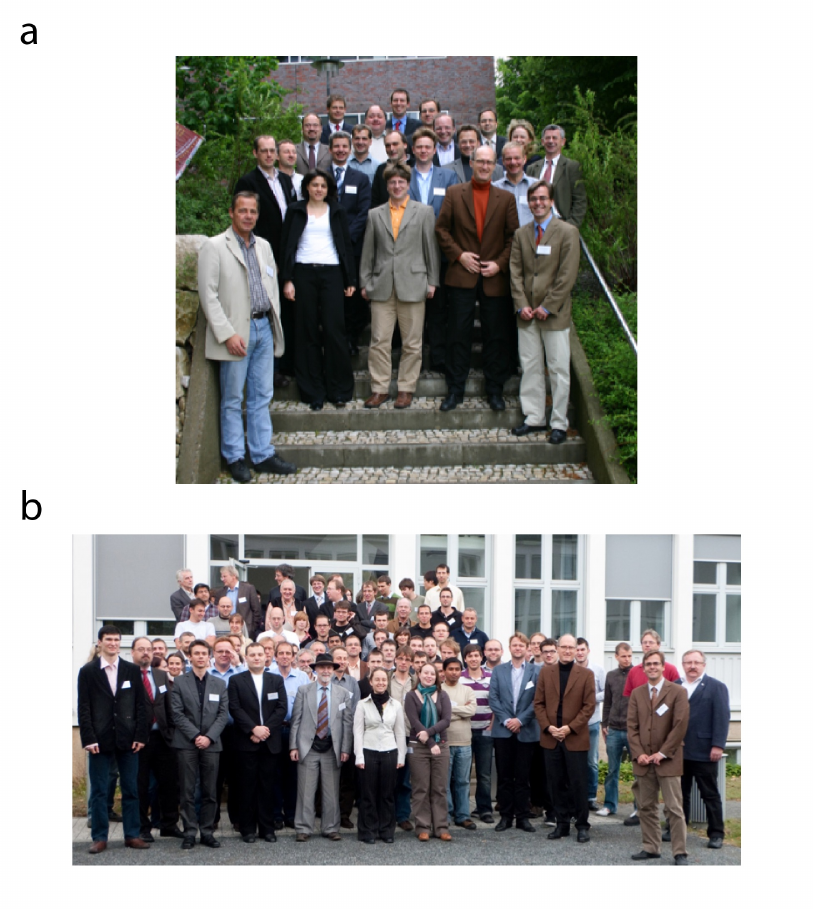}
\caption{(a) Participants of the first round table discussion and initiators of the German MOF research program SPP 1362 (2006, TU Dresden). (b) Participants of the German MOF Program (2008, Dresden).}
\label{fig:SPP1362}
\end{center}
\end{figure}

In Germany, an early promoter of MOF research was BASF (U. Müller), who initiated industrial research, upscaling and tested the early catalytic applications of MOFs.
BASF also filed several patents quite early in the field, discouraging competing enterprises to enter the field. F. Schüth promoted MOF research by including an early review chapter on MOF structures in 2002 in the Handbook of Porous Solids \cite{10.1002/9783527618286.ch19}.
Early German reports in the field of porous coordination polymers appeared by R. Kempe, mostly focused on the structural aspects, C. Janiak, and N. Stock \cite{10.1002/1521-3749(200108)627:8<1711::AID-ZAAC1711>3.0.CO;2-D,10.1039/B303498D,10.1039/B305705B,10.1016/j.micromeso.2003.12.026}.
A few groups stressed quite early the potential of these highly functional materials for applications in catalysis, for example the metal@MOF concept or Lewis acid catalyzed reactions \cite{10.1002/anie.200462515,10.1016/j.micromeso.2003.12.027}.
An important driver for the promotion of German MOF research was the DFG priority program SPP 1362 (``Porous Metal-Organic Frameworks'') initiated in 2006 by S. Kaskel (coordinator), T. Bein, S. Ernst, R. A. Fischer, J. Kärger, E. Klemm, J. Lercher, J. Sauer, and F. Schüth \cite{SPP1362}. In May 2006, 17 principal investigators presented their ongoing research in an informal meeting to identify new directions in MOF research in Dresden (Figure~\ref{fig:SPP1362}).

At that early stage the highly interdisciplinary character of this research field was expeditiously identified ranging from high throughput synthesis (N. Stock), advanced characterization methods (A. Pöppl, J. Kärger), simulations (R. Schmid, G. Seifert), nanomorphologies (T. Bein) towards applications in hydrogen storage (M. Hirscher), catalysis (E. Klemm), and technical chemistry (M, Hartmann).  However, it took until 2008 to get 13 consortia funded, in total involving 39 investigators distributed all over Germany. The highly interdisciplinary character of the consortia was an important driver to promote cross-disciplinary cooperation at an early stage of MOF research in Germany. SPP 1362 was in force from 2008-2014 supporting the increasing interest in MOF research, organizing network meetings and promoting cooperative MOF projects and supporting more than 100~PhD students financially, in total over the 6 years.

More than 160 scientists from 14 different countries (Europe, Asia, North- and South America) attended the ``International MOF Symposium'' organized by SPP 1362 in 2013 \cite{MOFSymposiumDFG}.
The most important highlights generated either within this program or achieved by German groups outside the program are discussed in this review. Some of the outcomes are also reflected in a special issue in \textit{Microporous and Mesoporous Materials} dedicated to the ``New Generation of Metal--Organic Frameworks'' published in 2015 \cite{10.1016/j.micromeso.2015.06.045}. Leading research groups from the program also contributed significantly to a recent monograph on the ``Chemistry of Metal-Organic Frameworks'' \cite{10.1002/9783527693078}.
Two large cooperative EU projects (Macademia and NanoMOF \cite{MACADEMIA,nanoMOF}) further promoted the development of industrial applications of MOFs in various fields, also involving partners in Germany. The European training network DEFNET \cite{DEFNET} further interlinked PhD students in Germany in MOF research with those in other European countries. F. Schüth initiated the first discussions with G. Férey for the first international MOF conference organized by DECHEMA in Augsburg 2008. He motivated G. Férey to act as the chair of this conference with M. Hartmann and S. Kaskel in the local organizing committee.
In recent years, the need to integrate MOFs into electronic or photonic device architectures motivated efforts towards MOF thin film preparation and characterization targeting electrically conductive or semiconducting MOFs. To this end, a new DFG priority program (SPP 1928, COORNETs, 2016--2021) was initiated by R. A. Fischer in 2014 and the first projects were funded in 2016 \cite{COORNET}.
Over more than one decade, MOF research in Germany has significantly benefited from excellent international cooperation with groups widespread all over the world in an amazingly diverse and interdisciplinary field. In this regard, we hope our current review and summary of MOF science achievements in Germany may serve as fruitful source and guidance for partnering, initiation of future international cooperation and prospective scientific exchange.

\section{Chemistry and Materials}

\subsection{Zirconium MOFs and related compounds}

MOFs based on the tetravalent \ce{Zr^{4+}} cation are probably the most intensively investigated subclass of this group of materials in recent years.
The main reasons for this are their outstanding stability due to strong metal-ligand bonding \cite{10.1039/C4CS00081A} and their chemical tunability using chemically different, but topologically identical ligands \cite{10.1039/C0CC02990D}.
These advantages combined with substantial porosity were demonstrated for the first zirconium carboxylate MOF denoted as {UiO-66} (\ce{[Zr_6O_4(OH)_4(BDC)_6]}) reported in 2008, where UiO stands for University of Oslo and BDC is benzenedicarboxylate \cite{10.1021/ja8057953}.
The hexanuclear inorganic \rev{structural} building unit (SBU) observed in this MOF has subsequently been encountered in most of the other {Zr-MOFs} reported.
The contribution from the German community to this nevertheless diverse structural chemistry is summarized below.
It should be noted that several of the MOFs mentioned herein can be as well prepared employing hafnium instead of zirconium salts due to the similar chemistry of these two elements.

This topic emerged in the German research community more clearly after the group of {P. Behrens} elaborated the concept of modulated synthesis for zirconium MOFs \cite{10.1002/chem.201003211}.
Initially the synthesis attempts for {Zr-MOFs} only yielded microcrystalline compounds and the use of linker molecules other than terephthalic acid was often hampered by a low degree of reproducibility.
Using monocarboxylic acids like benzoic, acetic or formic acid in excess and accurately controlling the water content of the synthesis mixture allowed for the reproducible synthesis of Zr-MOFs based on benzene-, biphenyl- and terphenyldicarboxylic acids with controllable crystal size.
Eventually, crystals even suitable for single crystal X-ray diffraction could be prepared.
Today it is believed that modulators induce the formation of the \rev{SBU} in the synthesis solution, making the preformed \rev{SBU} readily accessible.
However, modulators also act as competing ligands for the linker molecules and thus the nucleation process is sufficiently slowed to allow the growth of large single crystals, or at least highly ordered microcrystals.

This concept was further extended by the {P. Behrens} group using fumaric acid \cite{10.1016/j.micromeso.2011.12.010} or azobenzenedicarboxylic acid \cite{10.1002/ejic.201101151} as linker molecules and resulted in the discovery of a series of compounds denoted as {PIZOFs} (porous interpenetrated zirconium--organic frameworks) \cite{10.1002/chem.201101015}.
PIZOFs are formed when the length of the linear dicarboxylic acid exceeds the size of terphenyl building units.
Hence, molecules with alternating phenylene (P) and ethynylene (E) fragments, like a dicarboxylic acid with PEPEP sequence, but also other linker molecules like quaterphenyldicarboxylic acid and others \cite{10.1021/acs.inorgchem.6b01814} are able to form PIZOF structures.
The topology of these remarkable frameworks is based on two interpenetrating \textit{fcu} nets, the prototypical {UiO-66} and its derivatives exhibit a non-interpenetrated \textit{fcu} topology.
Using modulator assisted synthesis, developed in Hannover, the field of research dealing with new Zr-MOFs has prospered, mostly since the formation of large single crystals sufficient for conventional X-ray diffraction allows for a comparably facile way of structure determination.
Consequently, following this study several dozens of {Zr-MOFs} have been reported, of which the vast majority were obtained using modulating monocarboxylic acids as additives \cite{10.1039/c5cs00837a}.

Particularly in Germany this topic was further extended by the group of {S. Kaskel}.
In very thorough studies the influence of the amount and nature of the modulating agent was investigated, using predominantly bent dicarboxylic acids as linker molecules in which the angle between the acid groups is smaller than \ang{180}.
This led to the discovery of a remarkable number of new {Zr-MOFs} with differing and thitherto unreported topologies, different than the archetypical \textit{fcu} framework.
As a result, the first {Zr-MOF} with a slightly bent linker molecule (dithienothiophenedicarboxylic acid) exhibiting the \textit{reo} topology could be obtained, denoted as {DUT-51}, where DUT stands for Dresden University of Technology \cite{10.1039/C2CC34246D}.
Employing 2,5-thiophenedicarboxylic acid as the linker, three new compounds ({DUT-67}, -68 and -69) were synthesized, all crystallizing with different framework topologies (\textit{reo}, \textit{bon} and \textit{bct}, respectively) \cite{10.1021/cg301691d}.
\rev{The different framework topologies exhibit very different pore systems with a pore diameter of \SI{5}{\angstrom} for the \textit{bct} ({DUT-67}) topology and a \SI{27.7}{\angstrom} mesopore for the \textit{reo} ({DUT-68}) topology.}
Very recently, the compound DUT-126 was added to this series of 2,5-thiophenedicarboxylates, exhibiting a \textit{hbr} topology \cite{10.1098/rsta.2016.0027}.
Employing 9-fluoreneone-2,7-dicarboxylic acid as linker, the typical \textit{fcu} structure commonly occurring for linear linker molecules was observed for a bent organic building block in {DUT-122} \cite{10.1002/ejic.201600261} and the MOFs {DUT-80} and {DUT-98} were reported which incorporate 9-(4-carboxyphenyl)-9H-carbazole-3,6-dicarboxylate as the linker.

Other than bent linker molecules, the linear 2,6-naphthalenedicarboxylic acid was as well investigated.
This study yielded an analogue of {UiO-66} denoted as {DUT-52} (\textit{fcu}), a compound with \textit{bcu} topology denoted as {DUT-53} and a layered MOF named {DUT-84}.
Notably, these frameworks are based on identical building units and synthesized by varying only the concentration of modulator.
It is worth mentioning that at least some of the aforementioned observed topologies (\textit{bcu}, \textit{reo}, \textit{bct}) can be constructed starting from the \textit{fcu} topology by systematically removing inorganic and/or organic building units while mostly preserving the packing mode of the \rev{SBU}s.
Hence the face centered cubic arrangement, resembling the \textit{fcu} topology, can be effectively manipulated into other, related structures by adjustment of linker geometry and the use of modulators.

In further studies, MOFs exhibiting specific functionalities were generated to induce a certain property of the solids.
For example using enantiopure proline-functionalized linker molecules, {Zr-MOFs} with \textit{fcu} framework topology could be obtained, which are highly active catalysts for diastereoselective aldol additions \cite{10.1021/acs.chemmater.5b04575}.
Similarly, the use of a tetrazine-based analogue of terephthalic acid resulted in a {UiO-66} analogue which showed a clear color change upon reaction with oxidizing probe molecules \cite{10.1039/C4CC08136F}.

\begin{figure}[tbhp]
\begin{center}
\includegraphics[width=0.8\linewidth]{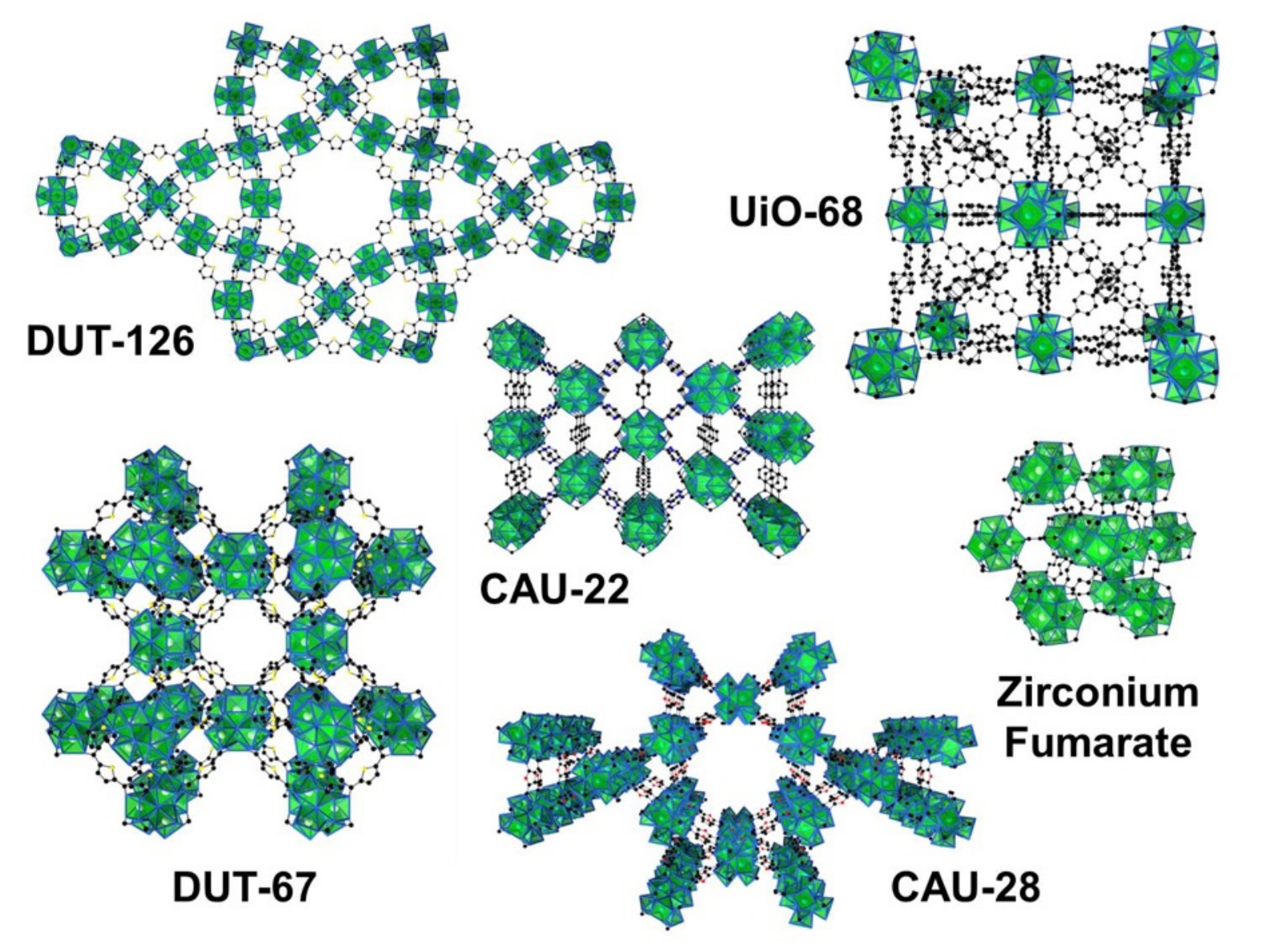}
\caption{Some of the described MOFs based on \ce{Zr^{4+}}, \ce{Hf^{4+}} or \ce{Ce^{4+}} that were discovered or first characterized in Germany.}
\label{fig:zirconia_sum}
\end{center}
\end{figure}

Recently, the group of {N. Stock} further extended the diversity of {Zr-MOFs}.
For example, the first Zr-MOF exhibiting \textit{scu} topology based on 1,2,4,5-tetrakis(4-carboxyphenyl)benzene and denoted as {CAU-24} was reported \cite{10.1039/C6DT03852B}.
In efforts to broaden the structural chemistry of {Zr-MOFs} further, the use of water instead of the most often employed solvent dimethylformamide (DMF) was investigated.
This was inspired by the report of the synthesis of zirconium fumarate in mixtures of water and modulator \cite{10.1016/j.micromeso.2014.10.034} by the group of {P. Behrens}.
As a consequence, the new compound {CAU-28} based on 2,5-furandicarboxylic acid was reported in Kiel \cite{10.1021/acs.inorgchem.6b02969}.
While the linker molecule is very similar to the 2,5-thiophenedicarboxylic acid used for the DUT-series, the resulting structure is not similar to those reported for the DUT-compounds.
This is possibly a consequence of the different solvent employed.
Furthermore, {CAU-22} based on 2,5-pyrazinedicarboxylate could be obtained in water as solvent in combination with formic acid \cite{10.1039/C6CC06287C}.
In contrast to all aforementioned Zr-MOFs, this framework does not incorporate the typical \ce{[Zr_6O_4(OH)_4(O_2CR)_12]} building unit, but a condensed version which represents a 1D polymer of edge-sharing hexanuclear clusters.
Under very similar conditions, 2,5-pyridinedicarboxylic acid allows for the synthesis of analogues of {UiO-66} \cite{10.1039/C7DT03641H}.

In addition to compounds based on \ce{Zr^{4+}}, the synthesis of \ce{Ce^{4+}}-based MOF was also investigated in the {N. Stock} group.
The first reported examples were analogues of the {UiO-66} structure incorporating (functionalized) terephthalate, fumarate or 4,4’-biphenyldicarboxylate as linker molecules \cite{10.1039/C5CC02606G}.
Similarly, isoreticular versions of {Zr-MOFs} with 5,5’-bipyridinedicarboxylate forming the {UiO-66} analogue, with trimesate forming the {MOF-808} analogue and with 2,5-thiophenedicarboxylate or 3,5-pyrazoledicarboxylate forming the {DUT-67} analogue were reported using cerium \cite{10.1021/acs.cgd.6b01512}.
The compounds {CAU-24} and {CAU-28} mentioned above are also accessible as {Ce-MOFs}.
In general the reactions are carried out in water/DMF/modulator mixtures and in order to obtain {Ce(IV)-MOFs} synthesis times must be restricted to minutes.
The reactive tetravalent cation tends to be reduced under these conditions upon prolonged heating, eventually forming trivalent cerium formate.
The \ce{Ce^{4+}}-based compounds are often catalytically active due the redox behavior of \ce{Ce^{4+}} \cite{10.1002/cphc.201700967}.
Unfortunately this also leads to a high reactivity and thus lower stability compared with their Zr-counterparts.
In order to achieve higher stability without compromising the redox activity, mixed metal MOFs based on \ce{Zr^{4+}} and \ce{Ce^{4+}} were also synthesized, exhibiting the {UiO-66} or {MOF-808} structure \cite{10.1039/C7DT00259A}.

\subsection{Mesoporous MOFs}

The modular approach to the synthesis of MOFs clearly allows for a variety of inorganic and organic building blocks to be combined in many different topologies.
This unique approach allows for MOFs to exhibit pore size distributions and textural properties almost without
limitations \cite{10.1039/C4CC00524D}.
Notably, MOFs have the unique potential of having a hierarchical pore structure ranging from the microporous to mesoporous size regime or singly microporous or mesoporous \cite{10.1039/c1cs15196g,10.1016/j.ccr.2018.05.001}.
Though most of the MOFs are microporous materials there are numerous mesoporous examples with extremely high specific surface areas above \SI{7000}{\square\metre\per\gram} and Germany-based researchers have reported several examples and provided important insight into their synthesis.

One strategy to extend the pores of a MOF to the mesoporous range is using the isoreticular principle, systematic enlargement of the ligand in a fixed framework topology \cite{10.1038/nature01650}.
The first mesoporous MOFs were introduced by {O. M. Yaghi} et al. using this approach \cite{10.1126/science.1067208}.
However, there are two main problems with using this method to produce mesoporous materials, interpenetration and polymorphism, which often appear with increased ligand length.
The group of {S. Kaskel} reported a route of avoiding interpenetration in {MOF-14} \cite{10.1126/science.1056598} by adjusting the synthesis conditions to produce {DUT-34} \cite{10.1002/chem.201101383}.
Though this did require screening a large number of alternative synthetic routes in order to obtain a phase pure non-interpenetrated sample.
Alternatively, a similar non-interpenetrated structure can be obtained by overcoming the energetic stabilization provided by interpenetration.
This was also demonstrated for {MOF-14} by inserting a linear neutral ligand (bipyridine (bipy)) between the the \rev{SBU} paddlewheels to produce DUT-23 possessing an impressive pore volume of \SI{2.03}{\cubic\centi\metre\per\gram} \cite{10.1002/chem.201101383}.

In a similar strategy, by using two or more ligands during the synthesis a number of mesoporous materials with impressive porosity can be obtained.
In particular the use of tritopic and ditopic carboxylic ligands has been most successful.
For example, {UMCM-1} and {DUT-6} ({MOF-205}) were synthesized using this approach \cite{10.1002/anie.200705020,10.1002/anie.200904599}.
Most examples employ benzene-1,3,5-tribenzoate (btb) as the
tritopic ligand and \ce{Zn4O^{6+}} as the metal node.
For successful copolymerization of MOFs, there are geometric considerations for the length ratio of the ligands and their connectivity to the inorganic unit.
Analysis suggests for MOFs constructed using di- and tritopic ligands ($L_{D}$ and $L_{T}$) the $\frac{L_{D}}{L_{T}}$ ratio must be within the range of 0.44 to 0.66 \cite{10.1021/ja1065009}.

Alternatively ligands with a bent shape can be employed to produce large mesoporous cavities.
{H.-C. Zhou} et al. successfully demonstrated this concept to produce the porous framework {PCN-21} with \SI{4}{\nano\metre} cavities and a surface area of \SI{4485}{\square\metre\per\gram}, which was constructed using angular tetratopic ligands and Cu paddlewheels \cite{10.1039/c0cc00779j}.
The group of {S. Kaskel} also used this approach by combining a bent thiophene dicarboxylic ligands (2,5-thiophenedicarboxylic) and Zr-clusters to produce the mesoporous framework {DUT-68} \cite{10.1021/cg301691d}.
The complex hierarchical pore structure of {DUT-68} contains: a rhombicuboctahedral mesopore, a cuboctahedral cage, a square antiprism and small octahedral pores.
Notably this framework is also observed to be robust, hydrophilic, chemically, and thermally stable.

\begin{figure}[tbhp]
\begin{center}
\includegraphics[width=0.6\linewidth]{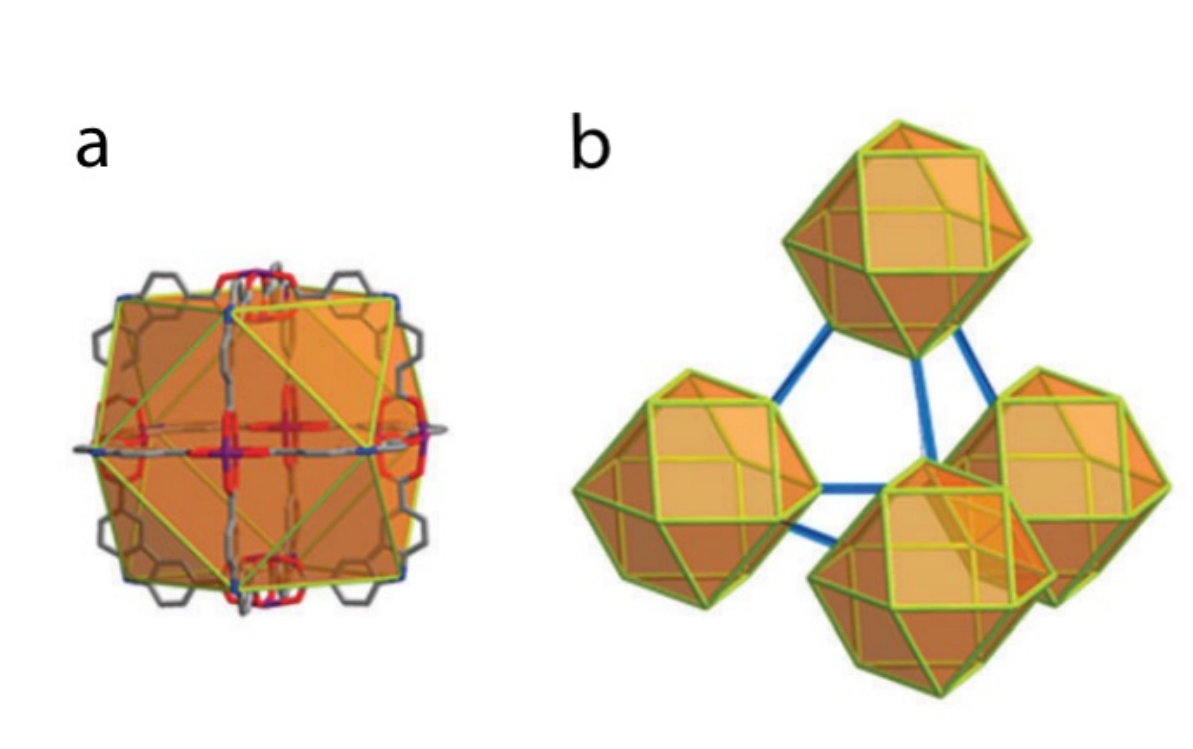}
\caption{(a) A carbazol-based MOP is linked by linear moieties to form (b), a three-dimensional framework. This is typified by {DUT-49}. \cite{10.1039/C4CC00524D} - Reproduced by permission of The Royal Society of Chemistry.}
\label{fig:MOP2MOF}
\end{center}
\end{figure}

Finally, using discrete metal--organic polyhedra (MOPs) as building units to construct highly porous three-dimensional materials is an excellent approach as it provides a high degree of control over the resulting pore structure and topology \cite{10.1039/B807086P}.
The polyhedra act as supramolecular building blocks that can subsequently be linked using multi-topic ligands to form a three-dimensional structure as demonstrated in Figure~\ref{fig:MOP2MOF}.
The smallest pore of the resulting framework can be straightforwardly tuned by changing the size of the initial polyhedra.
In addition by designing the shape, size and symmetry of the ligand connecting the polyhedra the number of resulting pores and their size and shape can be controlled.
Both {S. Kaskel} and {H.-C. Zhou} groups used this approach to produce mesoporous MOFs from cuboctahedral MOPs comprising of copper paddlewheels and carbazole-3,6-dicarboxylate ligands \cite{10.1039/C2DT32479B,10.1039/C2CC34840C}.
For example a linear moiety (biphenylene) is used to connect the  MOPs in {DUT-49} to produce a framework. which can be  described as the extended cubic closed packing of cuboctahedral building units, with exceptional gravimetric methane adsorption capacity \cite{10.1039/C2CC34840C}.

\subsection{Chiral MOFs}

Another interesting aspect of porous MOFs consists of their chiral nature.
On a compulsory prerequisite, chiral environment of a MOF is reflected from its crystal packing. Thus packing in a chiral space group is the essential condition for exhibiting chiral property of the structure.

Introduction of chirality has been made possible through four different approaches.
The first approach begins with achiral constituents, however self-resolution is induced to the framework structure during crystallization, thus bringing chirality to the final MOF structure.
This chiral environment can also be induced into the framework composed of chiral components, through use of a chiral template, thereby directing the framework to crystallize in chiral environment as a second approach.
The third approach begins with chiral ligands and thereby forming a chiral framework with a stereocenter.
Additionally a fourth approach consists of post synthetic modification to introduce a chiral moiety into an otherwise achiral MOF \cite{10.1039/B802258P}.
The chirality of the MOF structures has been proven useful in multiple aspects of asymmetric catalysis and chiral separation \cite{10.1002/cite.201000188}.

Of these approaches, starting with chiral ligand provides more control of the resulting framework chirality and thus German researchers have preferred this approach over the others for producing chiral MOFs for desired purposes.
The BDC linker was replaced with a chiral linker (enantiopure (S)-oxazolidinone) decorated BDC in the synthesis of UMCM-1 \cite{10.1002/anie.200705020}; a chiral version of the MOF was obtained \cite{10.1039/c1cc14893a}.
The obtained chiral MOFs (iPr-Chir-UMCM-1 and Bn-Chir-UMCM-1) differ by the absence of an inversion center in the framework structures, which was present for UMCM-1.
This generated chirality in the new frameworks allows for chiral separation applications.
A stationary bed formed with the chiral MOFs was employed as a HPLC column and different chiral analogues were passed through, an immediate change in the retention time was observed over the non-functionalized achiral MOF.
An effective enanatioselective separation was observed for 1-phenylethanol, with selectivity ($\alpha$) and resolution (R$_{\mathrm{S}}$) for the enantiomer as 1.6 and 0.65, respectively.
Interestingly, the porous nature of the MOF was retained, even after 200 injections of analytes through the MOF made HPLC column.

\begin{figure}[tbhp]
\begin{center}
\includegraphics[width=1\linewidth]{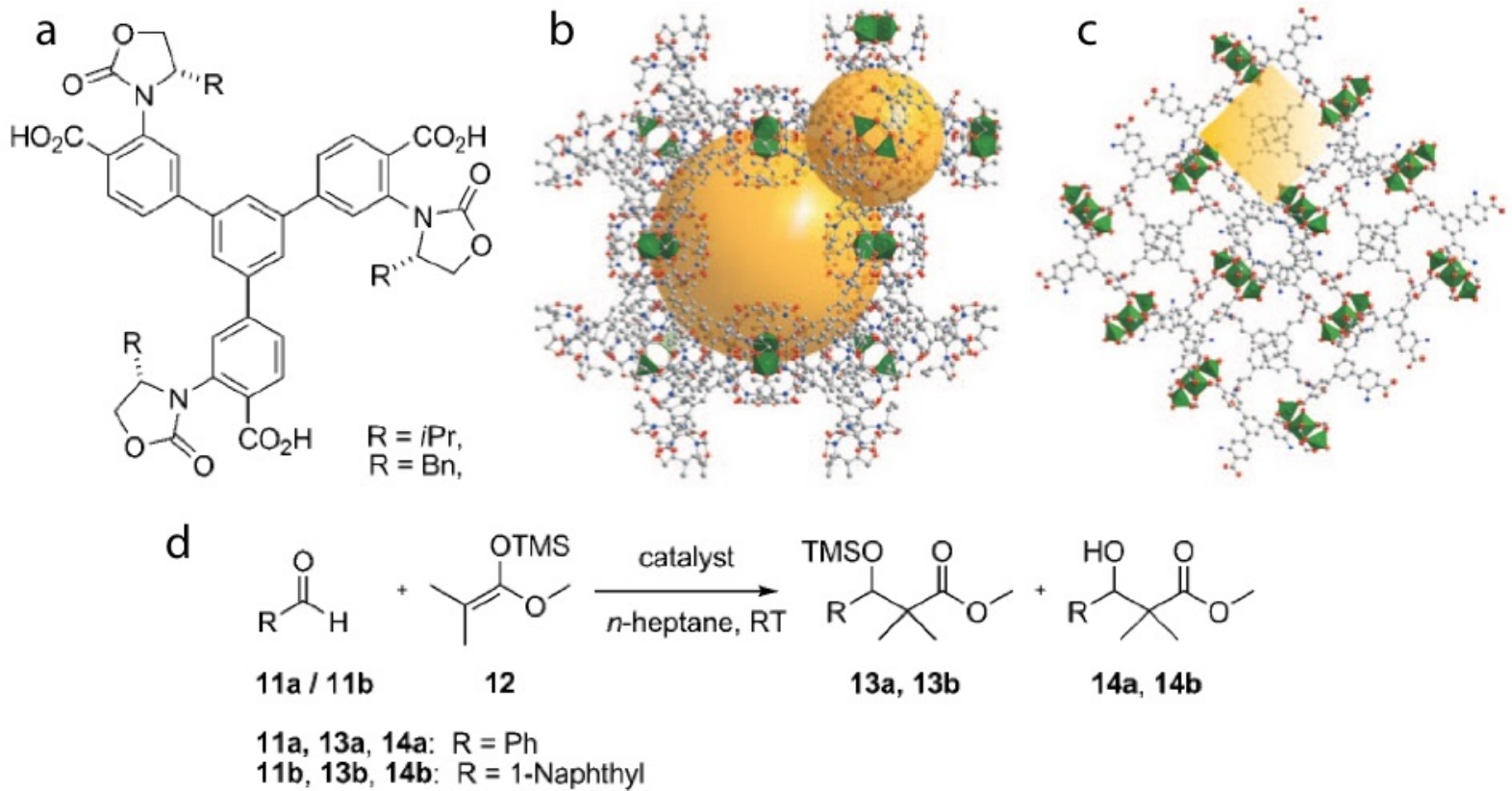}
\caption{a) chirBTB-n linker containing the chiral moieties and representative framework structures of the chiral MOFs (b: \ce{Zn_3(ChirBTB-1)_2} and c: \ce{Zn_3(ChirBTB-2)_2)}. d) Scheme of the enantioselective Mukaiyama aldol reaction reaction performed using the chiral MOFs as catalyst. Adapted from \cite{10.1002/chem.201002568}, with permission.}
\label{fig:ChirBTB1}
\end{center}
\end{figure}

Another family of chiral MOFs, bearing a chiral oxazolidinone group, was strategically synthesized by {S. Kaskel} and coworkers.
The basic design of the chiral linkers consists of a tripodal unit suitable for a large pore aperture a with distinct topological feature \cite{10.1002/chem.201002568}.
The chiral oxazolidinone group was then attached to the linker through chemical functionalization in an attempt to develop chiral environment near to the metal-linker junction.
This functionalization put the chiral oxazolodinone moiety in close proximity to the paddlewheel \rev{(SBU)}, which is known for catalytic activity through possible open metal site generation.
On reaction of these chiral linkers with Zn(II) centers, two different frameworks (\ce{Zn3(ChirBTB-1)_2} and \ce{Zn3(ChirBTB-2)_2}) were observed as displayed in Figure~\ref{fig:ChirBTB1}.
The resulting frameworks were observed to be structurally different, owing to the difference on their steric bulk near to the coordinating positions.
Thus, the less hindered moiety containing linker (ChirBTB-1) gives rise to a highly porous form with 3 different types of pores in the structure with pore diameter ranging from  \SI{13.7}{\angstrom} to \SI{33.7}{\angstrom}.
The MOF formed from the other linker (ChirBTB-2) also bears the chiral characteristic but with a smaller pore size of \SI{18}{\angstrom}.
The observed difference in pore size has been used to encapsulate large sized dye molecules into the wider pores of the MOF.
Being located near to the catalytically active sites, the chiral moiety has clearly influenced the tested Mukaiyama aldol reaction.
This leads to an elevation of the enantiomeric excess of the product formed by catalysis from the neighboring open metal sites of the paddlewheel SBU.

Insertion of such chiral functionality into the linker for MOF synthesis has proven beneficial in bringing in chirality into the MOF framework.
However, such approaches are often difficult to apply when the ligand of the pristine MOF structure is smaller in size.
An alternative approach for functionalizing the terephthalic acid linker requires a new methodology that has been presented in a recent report \cite{10.1039/c7ce00410a}.
A three-dimensional analogue of the BDC linker was chosen for the purpose and the subsequent functionalization leads to equivalent distribution of chiral moieties in its vicinity.
This makes it possible for chiral functionalities of the ligand to be evenly distributed along the structure of {DUT-129}, indicating a possible enhancement in the interaction with other chiral components.

Apart from the asymmetric linker induced chirality, a recent approach from {D. Volkmer} and coworkers highlights self-resolution during MOF formation to obtain the chiral MOF {CFA-1} \cite{10.1039/C3DT50787D}.
Here the achiral triazole ligand, discussed in more detail in the subsequent section, on coordination with Zn(II) gives rise to a chiral MOF in the chiral space group {$P$321}.
This provides a spiral pore channel inside the structure with a maximal width of \SI{14.35}{\angstrom}, accessible through two different pore windows, having effective diameters of 3.43 and \SI{6.18}{\angstrom}.
Thus, there is potential for chiral MOFs to be successfully applied in related applications through an easy, and feasible, synthetic approach.

Furthermore, in an attempt to achieve thin films for more practical application of the chiral MOFs for separation purpose, a MOF thin film (SURMOF) was developed by the {C. Wöll} group \cite{10.1002/anie.201104240}.
The pillar layered structure consisting of camphoric acid, as the chiral moiety, and DABCO as pillars was grown on a quartz crystal microbalance (QCM) substrate to have the SURMOF grown in the desired (110) and (001) directions.
Expected enantioselectivity was observed when the vapor of \textit{R} and \textit{S} form of a chiral probe (2,5-hexanediol) was passed over the resulting SURMOFs.
This serves as an example for application of such materials towards stationary phases to separate racemates by GC.

\subsection{Nitrogen containing ligands}

Metal carboxylate frameworks are ubiquitous in the chemistry of metal--organic frameworks. However, nitrogen containing ligands which form metal--azolate frameworks (MAFs) represent an important and structurally diverse subfamily of (crystalline) coordination polymers \cite{10.1351/PAC-REC-12-11-20}, which have been comprehensively reviewed by J.P. Zhang et al. in 2012 \cite{10.1021/cr200139g} and for which Germany-based researchers have made a significant contribution.

\begin{figure}[tbhp]
\begin{center}
\includegraphics[]{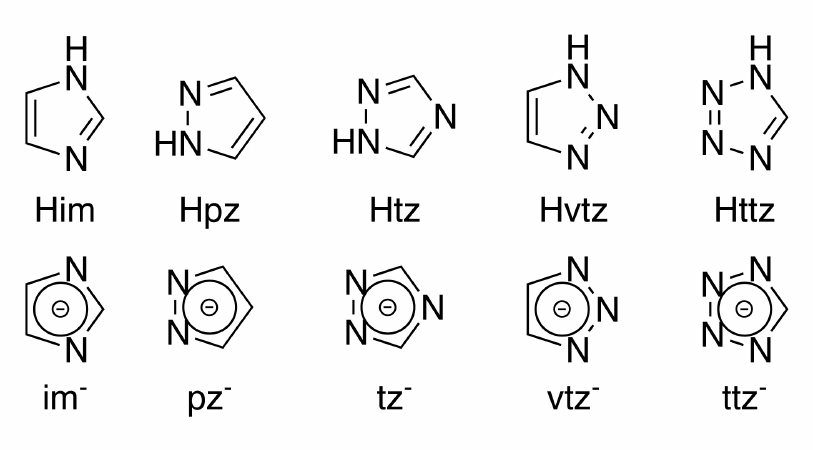}
\caption{Structures of prototypical azoles and their corresponding azolates.}
\label{fig:azolateschemdraw}
\end{center}
\end{figure}

In a broad sense the term MAFs refers to coordination polymers which are constructed from one-, two-, or three-dimensional chains of metal ions linked by anionic \ce{N}-heterocyclic aromatic ligands, most of which are derived from simple five-membered ring systems, such as those displayed in Figure~\ref{fig:azolateschemdraw} (Note that the list is non-comprehensive and does not include fused aromatic ring systems that will be highlighted later in this section).
The term ``MAF'', as yet, has not received wide acceptance in scientific literature, presumably because the compound family itself contains distinguished members of coordination polymers such as the zeolitic imidazolate frameworks (ZIFs) \cite{10.1002/chem.200304957,10.1073/pnas.0602439103}, derived from imidazolate (im$^{-}$) ligands (Figure~\ref{fig:azolateschemdraw}), a first member of which was described by F. Seel and J. Rodrian in 1969 \cite{10.1016/S0022-328X(00)89773-9}.
ZIFs receive constantly growing attention within the scientific community owing to their ease of synthesis and exceptional chemical robustness \cite{10.1021/ar900116g}.
This chapter on MAFs is devoted to the seminal contributions from German research groups to triazolates, with emphasis on continuous developments of the structural chemistry of these compounds rather than their (potential) use in technical applications.

\begin{figure}[tbhp]
\begin{center}
\includegraphics[]{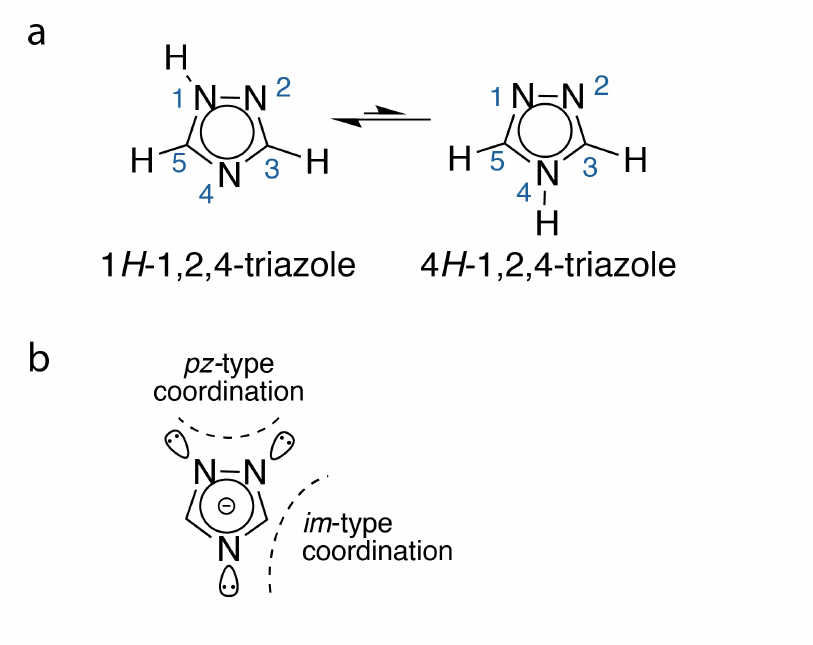}
\caption{(a) Naming and numbering scheme, tautomerism and (b) coordination properties of 1,2,4-triazole and its corresponding anion.}
\label{fig:namingandproperties}
\end{center}
\end{figure}

The structural chemistry of triazolate frameworks based on the most simple basic aromatic heterocyclic ring systems Htz (1H-1,2,4-triazole) and Hvtz (1H-1,2,3-triazole) have been thoroughly investigated.
Systematic investigations on tz$^{-}$ derived framework compounds have been conducted, amongst others, by J. Zubieta and coworkers \cite{10.1021/ic700790h,10.1021/ic061102e,10.1021/ic062269a}.
The synthesis of 1,2,4-triazoles (Figure~\ref{fig:namingandproperties}) has previously been reviewed \cite{Maddila:2013:1570-1786:693,10.1021/cr60210a001}, as well as aspects of their structurally diverse coordination chemistry \cite{10.1016/S0010-8545(00)00266-6,10.1016/j.ccr.2010.10.038}.

In Germany, systematic crystallographic work on coordination polymers derived from 1,2,4-triazolate ligands has been conducted by the {H. Krautscheid} group at the University of Leipzig in close collaboration with {K. V. Domasevitch} and coworkers from Kiev University.
However, the group’s work has focused on N1-functionalized Htz linkers, which lack the ability to form an azolate, i.e. the anionic form of the linker.
Hence, their coordination chemistry formally resembles that of neutral 1,2-diazole ligands and shall not be reviewed here any further.
Illustrative examples of frameworks derived from this approach include bitopic linkers such as 4-(4H-1,2,4-triazol-4-yl)benzoic acid \cite{10.1016/j.tetlet.2009.11.098,10.1021/ic300235s,10.1016/j.micromeso.2010.11.017} or polytopic triazole linkers containing an adamantyl backbone \cite{10.1039/C2DT30362K,10.1021/ic5009736}.

\begin{figure}[tbhp]
\begin{center}
\includegraphics[width=0.5\linewidth]{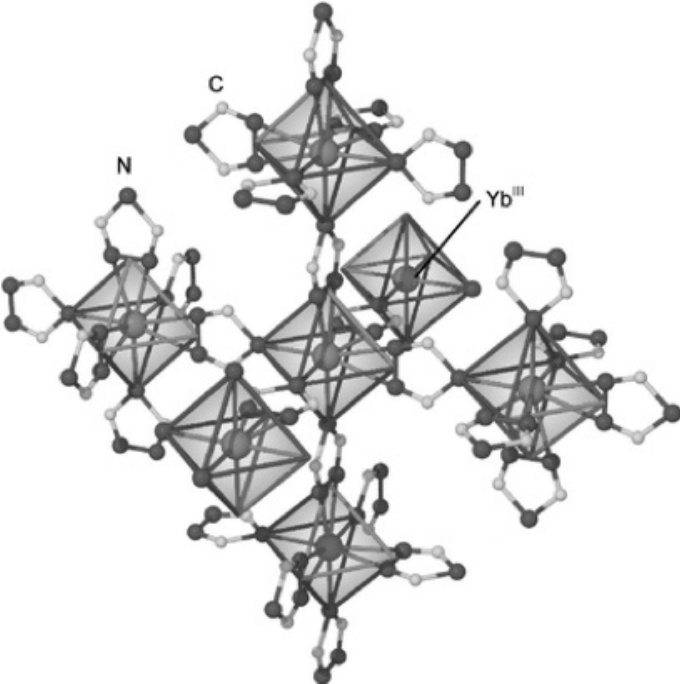}
\caption{The crystal structure of the framework \ce{[Yb(tz)_3]}, showing the \ce{\mu_2-\eta^2} bridging mode of the 1,2,4-triazolato ligand. Reproduced from \cite{10.1039/B601450J}, with permission.}
\label{fig:Ybtz3}
\end{center}
\end{figure}

The coordination chemistry of rare earth (RE) element 1,2,4-triazolates has been explored by the group of {K. Müller-Buschbaum} at Cologne University \cite{10.1039/B601450J}, who presented the first solvent-free synthesis of this linker-type showing exclusive N-coordination of RE metal ions (Figure~\ref{fig:Ybtz3}).

\begin{figure}[tbhp]
\begin{center}
\includegraphics[]{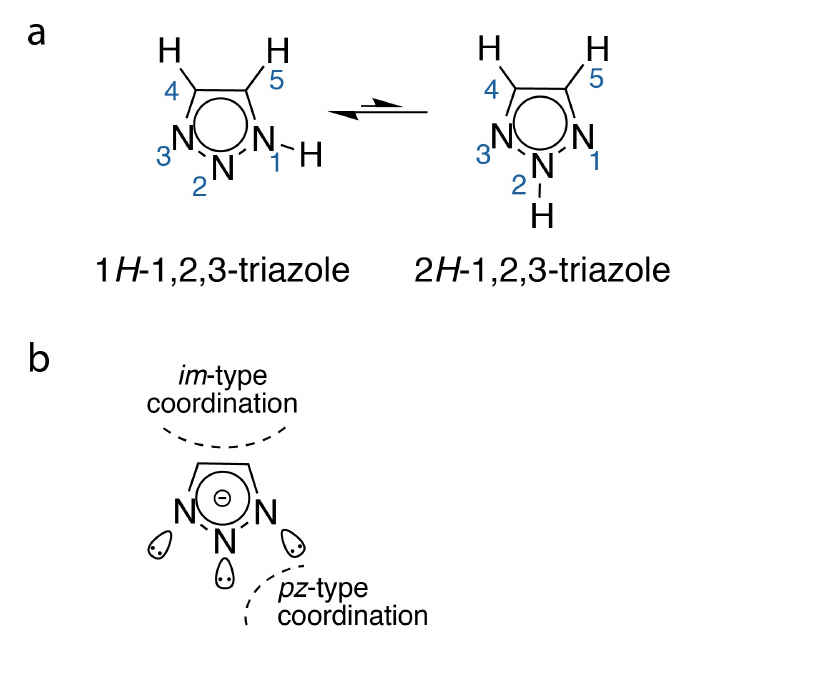}
\caption{(a) Naming and numbering scheme, tautomerism and (b) coordination properties of 1,2,3-triazole and its corresponding anion.}
\label{fig:namingandproperties2}
\end{center}
\end{figure}

\begin{figure}[tbhp]
\begin{center}
\includegraphics[width=0.5\linewidth]{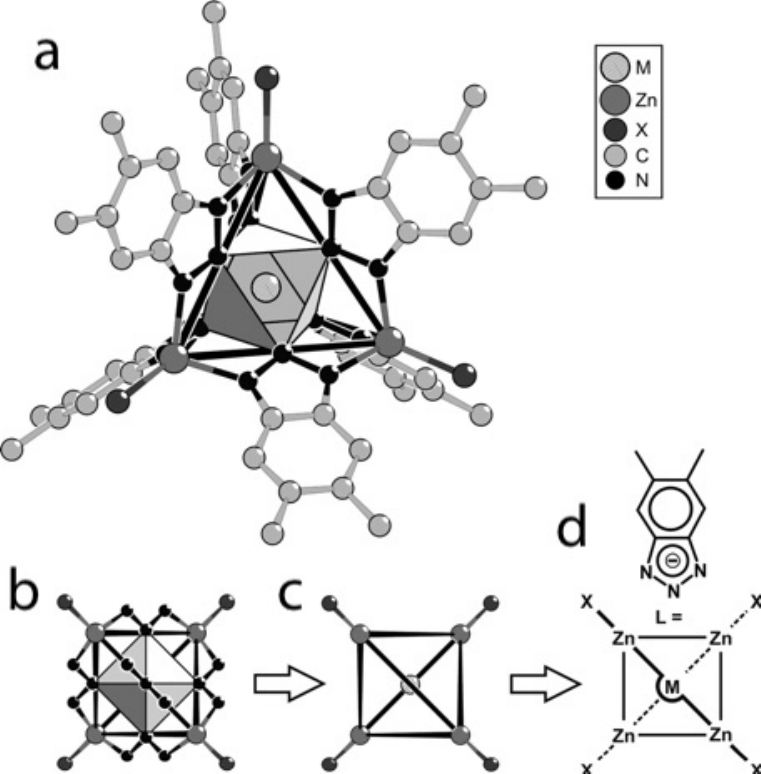}
\caption{(a) Ball-and-stick representation of ``Kuratowski-type'' coordination units, highlighting the central octahedral coordination site that results from six N-donor atoms situated at the centers of the edges of an imaginary tetrahedron, which is spanned by the four Zn(II) ions at the corners. \rev{M represents Zn, Fe, Co, Ni, or Cu and X represents Cl.}  (b) Simplified representation of metal centers and ligand donor atoms. (c and d) Derivation of a rational graphical scheme representing the connectivity of the coordination units. Reprinted with permission from \cite{10.1021/ic100749k}. Copyright 2010 American Chemical Society.}
\label{fig:kuratowski}
\end{center}
\end{figure}

The synthesis of 1,2,3-triazoles (Figure~\ref{fig:namingandproperties2}) has been reviewed previously \cite{0036-021X-74-4-R03} and aspects of the general coordination chemistry of these ligands are covered in \cite{10.1016/j.ccr.2010.10.038}.
The structural chemistry of metal-organic frameworks derived from 1,2,3-triazolate-type ligands has only recently been investigated.
{D. Volkmer} and coworkers at Ulm University have pioneered the field with systematic work on the development of {MFU-4-type} metal--organic framework compounds ({MFU-4} being an acronym for metal--organic framework Ulm University), reported in 2009 \cite{10.1039/B904280F}.
The design of {MFU-4-type} framework compounds rests on unique pentanuclear structural building units (SBUs), for which the term ``Kuratowski-type'' SBUs has been suggested by the authors (Figure~\ref{fig:kuratowski}) \cite{10.1021/ic100749k}.

\begin{figure}[tbhp]
 \begin{center}
 \includegraphics[]{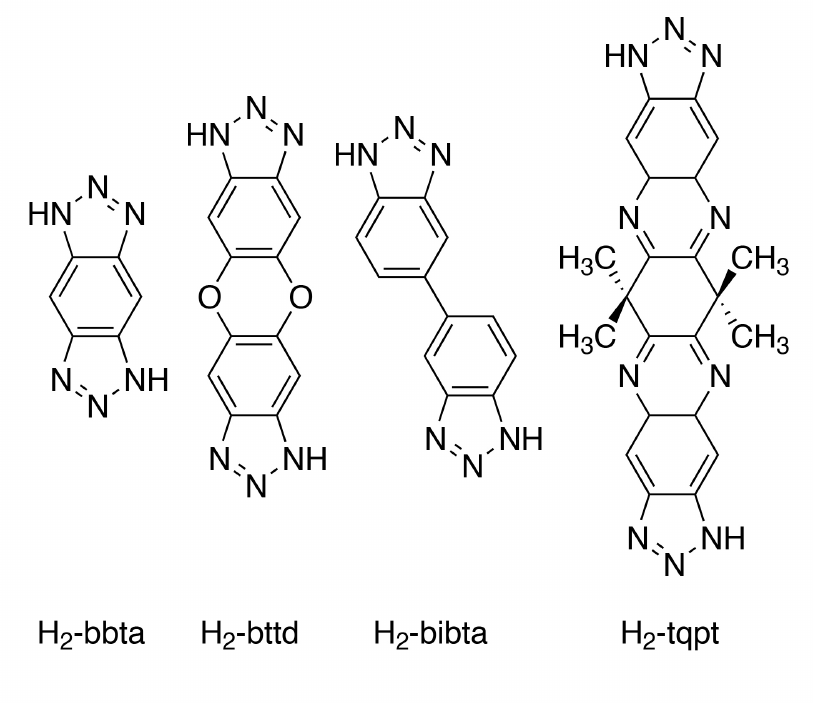}
 \caption{Structural formulas of bis-(1,2,3-triazole) linkers prepared by {D. Volkmer} et al.}
 \label{fig:volkmerligands}
 \end{center}
\end{figure}

 While molecular analogs of such units have been known since 1981 \cite{10.1021/ja00391a049} their use in constructing robust frameworks was realized by D. Volkmer et al. in various examples employing a range of rigid and partly flexible back-to-back bound bis-1,2,3-triazolate ligands (Figure~\ref{fig:volkmerligands}).

\begin{figure}[tbhp]
 \begin{center}
 \includegraphics[width=0.8\linewidth]{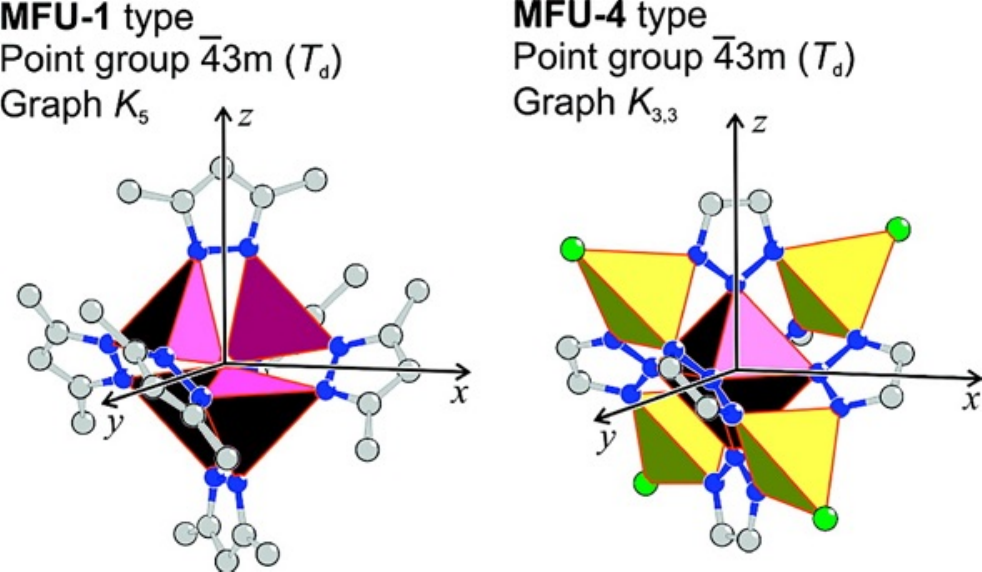}
 \caption{Structural features of SBUs found in MFU‐1 and MFU‐4. Reprinted with permission from \cite{10.1021/ic100749k}. Copyright 2010 American Chemical Society.}
 \label{fig:MFU14}
 \end{center}
\end{figure}

 In clear distinction to the SBU of pyrazolate-based MFU-1 (Figure~\ref{fig:MFU14} left) \cite{10.1002/anie.200901241}, which seems to be limited to the presence of a \ce{Co4O} core, the development of the structurally more flexible 1,2,3-triazolate-based {MFU-4} family, which contains Kuratowski-type pentanuclear SBUs of the general formula \ce{[M^{II}Zn4X_4(L)_6]},  \rev{where M represents Zn, Fe, Co, Ni, or Cu and X represents Cl}, (Figure~\ref{fig:MFU14} right) provides synthetic access to a wide variety of metal--organic frameworks.
 Kuratowski-type SBUs feature a central metal ion coordinated to six triazolate ligands (L) that span a Cartesian coordinate system.

The central metal ion in the octahedral coordination environment can be varied \cite{10.1021/ic100749k} for the opportunity to obtain redox-active SBUs that are also Lewis acidic, with chemical properties which can be fine-tuned by selecting appropriate metal ions \cite{10.1039/C0DT01750G}.
A graph theoretical analysis proves that \ce{[M^{II}Zn4X_4(L)_6]} units contain the nonplanar K$_{3,3}$ graph.
Accordingly, there is no way to draw \ce{[M^{II}Zn4X_4(L)_6]} as a planar graph and thus a pseudoperspective skeletal formula as derived in Figure~\ref{fig:kuratowski} to represent Kuratowski-type coordination compounds was proposed.
The first triazolate-based MOF featuring Kuratowski-type secondary building units, {MFU-4}, was constructed from bbta$^{2-}$ linkers (Figure~\ref{fig:volkmerligands}) and \ce{Zn_5Cl_4^{6+}} inoganic clusters to produce a framework featuring a very high thermal and hydrolytic stability \cite{10.1039/B904280F}.
{MFU-4} also features small pore apertures (\SI{2.5}{\angstrom}) which are highly selective for the adsorption of atoms or small molecules such as \ce{H2}, and it can therefore be applied in molecular sieving applications including some of which are difficult to achieve with other kinds of porous materials, such as the separation of \ce{H2}/\ce{D2} \cite{10.1002/adma.201203383} or \ce{N2}/\ce{CO2} mixtures \cite{10.1039/C4DT00365A}.

In order to separate mixtures of larger molecular adsorbates, or for catalytic transformations, a porous framework featuring large pore apertures is required.
This led to the development of {MFU-4large} ({MFU-4l}) \cite{10.1002/chem.201001872}, an isoreticular large-pore variant of {MFU-4}, constructed from larger bttd$^{2-}$ ligands (Figure~\ref{fig:volkmerligands}).

\begin{figure}[tbhp]
\begin{center}
\includegraphics[width=0.8\linewidth]{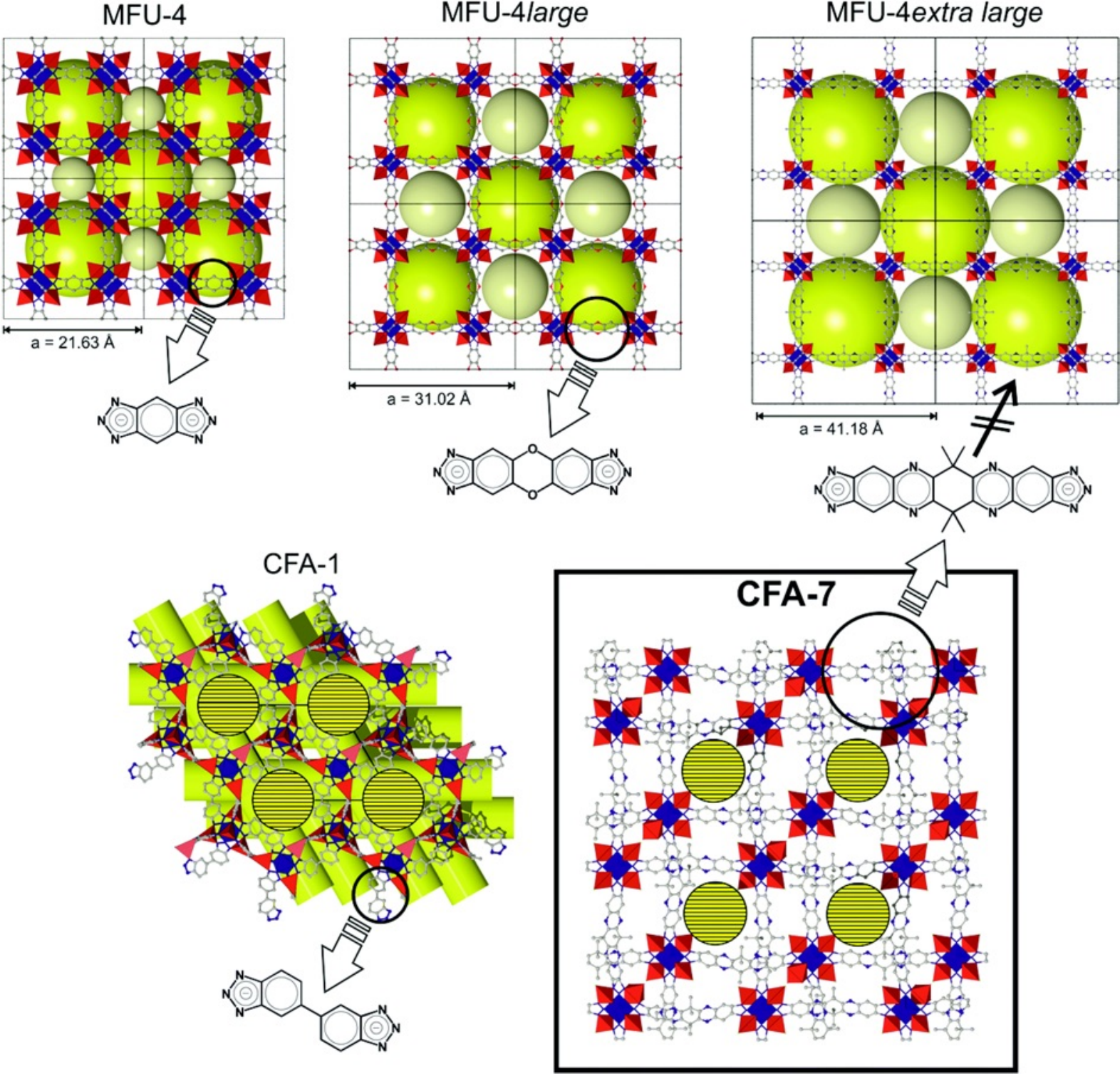}
\caption{Overview of currently known framework types comprising rigid or partly flexible bis-triazolato linkers and Kuratowski-type SBUs. \cite{10.1039/c5dt01673h} - Published by The Royal Society of Chemistry.}
\label{fig:triazolato}
\end{center}
\end{figure}

Attempts to extend the isoreticular series of cubic {MFU-4-type} frameworks with larger ligands, such as bibta$^{2-}$ or tqpt$^{2-}$ (Figure~\ref{fig:volkmerligands}) met with little success.
However, these investigations led to the discovery of crystallographically distinct framework structures (Figure~\ref{fig:triazolato}), in which the fundamental Kuratowski SBU is retained --- clearly demonstrating the robustness of the building block approach.
Employing the easy-to-synthesize \ce{H_2}-bibta linker, D. Volkmer and coworkers reported {CFA-1} (CFA being an acronym for coordination framework Augsburg University), the first chiral metal--organic framework containing Kuratowski-type SBUs \cite{10.1039/C3DT50787D}.
This was soon followed by {CFA-7}, an interpenetrated framework variant of the {MFU-4} family \cite{10.1039/c5dt01673h}.

Amongst the different {MFU-4} frameworks, {MFU-4l} attracted particular attention owing to its versatility with respect to post-synthetic metal- and ligand-exchange and its functional properties.
In a communication from 2012 the \ce{Co(II)}-exchanged derivative of {MFU-4l} ({Co-MFU-4l}) was reported to demonstrate reversible gas-phase oxidation properties \cite{10.1039/C2CC16235K}.
Subsequently, in 2014, {D. Volkmer} et al. showed the remarkable characteristics of {\ce{Cu(I)}-MFU-4l} toward binding and activation of small molecules, such as \ce{O2}, \ce{N2}, and \ce{H2}, with corresponding isosteric heats of adsorption of 53, 42, and \SI{32}{\kilo\joule\per\mol}, respectively, determined by gas-sorption measurements and confirmed by density functional theory (DFT) calculations \cite{10.1002/anie.201310004}.
The \ce{H2} complex of trigonal pyramidal coordinated \ce{Cu(I)} metal ions in this framework ranks among the strongest molecular hydrogen complexes of a 3D transition metal ion described in literature to date, and it has been used in boosting the efficiency of \ce{H2}/\ce{D2} separation by means of quantum sieving \cite{10.1038/ncomms14496}.

\begin{figure}[tbhp]
\begin{center}
\includegraphics[width=0.8\linewidth]{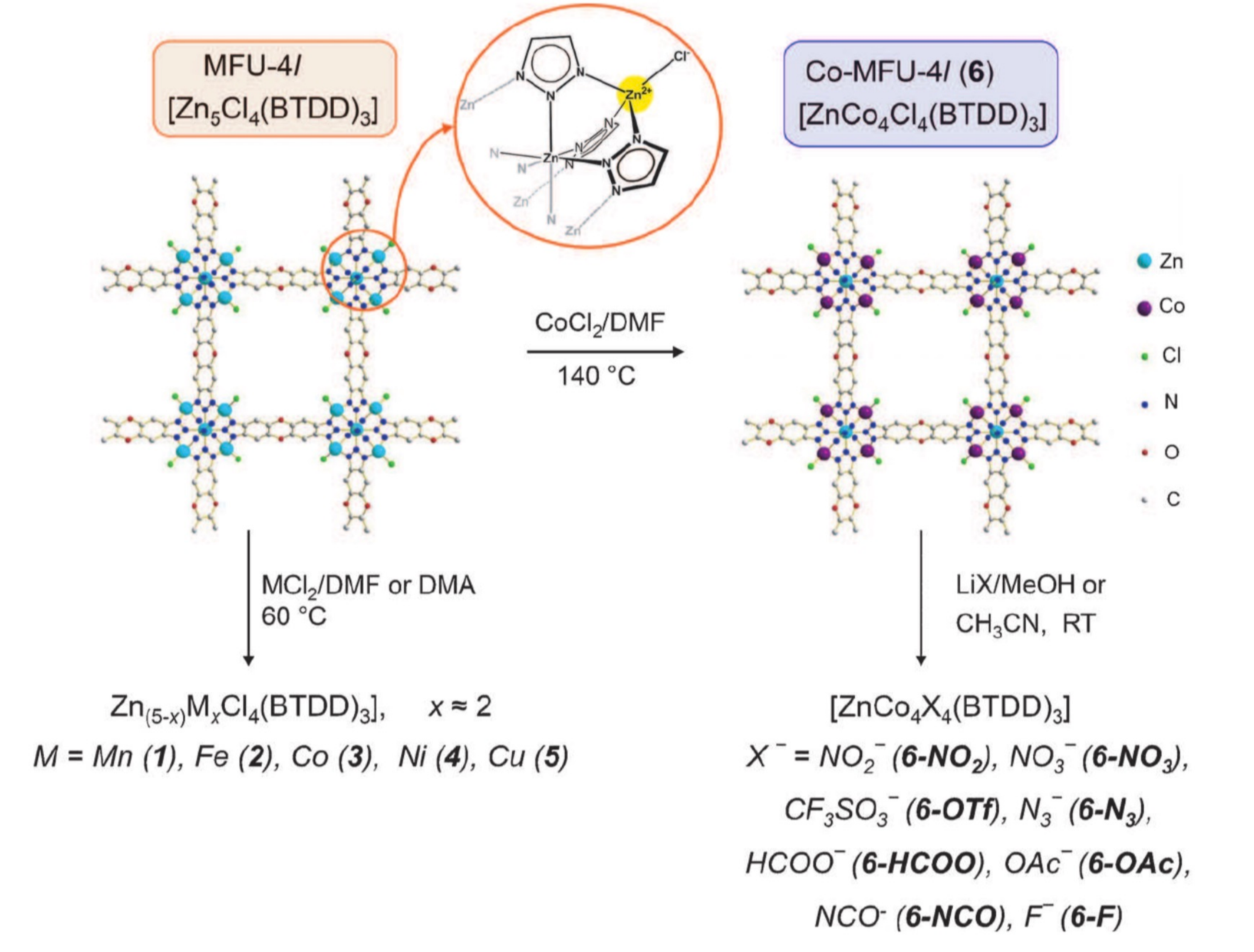}
\caption{Post‐synthetic modifications of MFU‐4l (DMF=N,N‐dimethylformamide, DMA=N,N‐dimethylacetamide). Reproduced from \cite{10.1002/chem.201406564}, with permission}
\label{fig:exchangeMFU}
\end{center}
\end{figure}

Exchanging chloride with hydride anions, {MFU-4l} exhibits hydride transfer to electrophiles \cite{10.1039/C2CC16235K}, which suggests these novel, yet robust, metal--organic frameworks are promising single-site catalytic materials, comprising earth-abundant metal elements.
The concept of single-site catalysts requires the peripheral position in the SBU of {MFU‐4‐type} frameworks zinc ions (``scorpionate-type coordination units''; Figure~\ref{fig:exchangeMFU}, left) to be exchanged by coordinatively unsaturated (``open'') metal sites.
A variety of such exchange reactions has been demonstrated by the groups of {M. Dincă} \cite{10.1002/chem.201402682} and {D. Volkmer} \cite{10.1002/chem.201406564} (Figure~\ref{fig:exchangeMFU}, right).
Structures based on hetero-metal derivatives of {MFU-4l} exhibit a range of interesting catalytic reactions, recently demonstrated by {M. Dincă} and coworkers.
These encompass the dimerization of ethylene to 1-butene \cite{10.1021/acscentsci.6b00012} and the dimerization of propylene \cite{10.1021/acs.organomet.7b00178}, both reactions being catalyzed by \ce{Ni(II)}-exchanged {MFU-4l}.
Additionally, {Co-MFU-4l} displays highly stereoselective heterogeneous diene polymerization \cite{10.1021/jacs.7b06841}.
{\ce{Ni(II)}-MFU-4l}, also is shown to be catalytically active in a gas phase cyclic process with an overall stoichiometry \ce{2NO + CO -> N2O + CO2}, demonstrating a potential use of these frameworks for the removal of highly toxic gases \cite{10.1039/C7FD00034K}.

An interesting metal azolate framework, corresponding to the formula \ce{[M(II)(vtz)_2]}, is derived from the commercially available 1H-1,2,3-triazole ligand and divalent metal cations.
The first member of this series, \ce{[Cd(vtz)_2]}, was described in 2009 \cite{10.1039/B819302A} followed by a report from the {D. Volkmer} group on \ce{[Cu(vtz)_2]} in 2012 \cite{10.1039/C2DT12311H}.
Later in the same year, {O. M. Yaghi} et al. reported the synthesis of the metal triazolate series termed MET-1 to 6, corresponding to \ce{[M(vtz)_2]} with \ce{M(II)} = Mg, Mn, Fe, Co, Cu, and Zn \cite{10.1002/chem.201103433}, for which they elucidated a cubic structure based on X-ray powder diffraction investigations of microcrystalline powders.
For the \ce{Cu(II)}-containing framework this assignment is at odds with the single crystal structure determination for \ce{[Cu(vtz)_2]}, which was shown to crystallize in the tetragonal crystal system, from which it reversibly transforms into the cubic phase within the temperature range of 120--\SI{160}{\celsius} \cite{10.1002/anie.200902274}.

Among the {MET} frameworks, \ce{[Fe(II)(vtz)_2]} showed unexpectedly high electrical conductivity, with a conductivity value of \SI{0.77E-4}{\siemens\per\centi\metre} which the authors could increase to a value of \SI{1.0E-3}{\siemens\per\centi\metre} by exposing the sample to iodine vapor.
Although the given explanation, namely the introduction of \ce{Fe(III)} sites by partial oxidation, seems unlikely with regard to commonly reported standard redox potentials of \ce{Fe(II)}/\ce{Fe(III)} vs. \ce{I2}/\ce{I^-} redox couples, the study had a seminal influence on the development on electrically conductive frameworks.
A more detailed theoretical analysis for the high electrical conductivity of {MET-3} (\ce{[Fe(II)(vtz)_2]}), based on accurate band gap calculations, was recently given by {M. Dincă} and coworkers. \cite{10.1039/C7SC00647K}.

It is noted that up to now the structures of intermediate oligomeric species are largely unknown if the simple vtz$^{-}$ ligand is employed.
However, using larger benzotriazolate ligands, discrete nonanuclear coordination compounds can be isolated representing oligomeric species from the fusion of Kuratowski-type building units \cite{10.1002/ejic.200900156,10.1039/C0DT00556H}.
Lastly, the review on metal--triazolate frameworks would be rather incomplete without mentioning a series of europium-containing triazolate framework structures reported by {K. Müller-Buschbaum} et al. \cite{10.1016/j.solidstatesciences.2007.12.003,10.1002/zaac.201000077,10.1002/anie.200603682}, among which \ce{[Eu(II)(bta)_2]}, is topologically equivalent to the cubic metal--triazolates.

\subsection{Phosphorus containing ligands}

In contrast to the well-developed chemistry of porous metal carboxylates and azolates, the number of porous metal phosphonates is limited to a few dozens of compounds \cite{10.1002/ejic.201600207,10.1021/cr2002257,10.1039/9781849733571,10.1039/B802423P}.
This is due to the larger coordination flexibility of the phosphonate groups and the possibility to form \ce{\bond{-}PO3H^-} and \ce{\bond{-}PO3^{2-}} groups, which in turn leads to more dense, often layered, structures and makes the formation of suitable and easily available \rev{SBU}s, prevalent in carboxylate MOFs, more difficult.
In comparison to the carboxylate group, the higher number of coordinating atoms and the higher charge of the phosphonate group often results in a more chemically stable MOF, which is highly desirable for possible applications  \cite{10.1039/B802423P}.
On the other hand, these properties also explain the difficulty in formation of large single crystals suitable for single crystal X-ray structure determination \cite{10.1039/C4QI00011K}.
Hence, scientists working in the field of metal phosphonates have relied on the structure elucidation using powder X-ray diffraction (PXRD) or more recently electron diffraction data, often in combination with structure modelling.

\begin{sloppy}
The chemistry of porous metal phosphonates has, at the beginning of the 1970s, been dominated by zirconium phosphonates \cite{10.1039/9781849733571,10.1016/0022-1902(78)80520-X}.
Their structures can be derived from the ones of \ce{[Zr(HPO4)2]*H2O} and \ce{[Zr(PO4)(O3POH)2]*H2O} by replacing the \ce{\bond{-}OH} by organic moieties \cite{10.1021/ic50073a005,10.1021/ic00148a036}.
Mostly dense non-porous structures are formed and porosity was introduced using a mixed-ligand approach through a) post-synthetic hydrolysis, b) use of sterically demanding phosphonic acids or c) topotactic ligand exchange.
Metal phosphonates with defined regular pore structures, highly crystalline metal phosphonates, have been reported by less than a dozen of groups worldwide including the groups of {T. Bein}, {H. Krautscheid} and {N. Stock} in Germany.
The first porous metal phosphonate was obtained using small methyl phosphonic acid \ce{CH3PO3H2}, but subsequently such compounds were synthesized by employing polyphosphonic acids with N-containing groups, such as secondary or tertiary amines or triazoles.
From the reported structures it can be deduced, that either coordination of the N atoms, the presence of protonated amines or the formation of \ce{N\bond{-}H\bond{...}O\bond{=}P} hydrogen bonds seem to be helpful to form porous 3D framework structures.
\end{sloppy}

\begin{sloppy}
Nitrogen-containing polyphosphonic acid molecules with \ce{\bond{-}NCH2PO3H2} or \ce{\bond{-}N(CH2PO3H2)_2} groups have been mainly employed, since these can be easily prepared starting from primary or secondary amines.
Thus the groups of {T. Bein} and {N. Stock} were able to hydrothermally synthesize two alkaline earth phosphonates \ce{Ca[(HO3PCH2)_2NHCH2C6H4CH2N2(CH2PO3H)_2]*2H2O} \cite{10.1016/j.micromeso.2003.12.026}, and \ce{Ba3[O3PCH2NH2CH2PO3]2*7H2O} \cite{10.1021/ic050935m} starting from the respective phosphonic acids.
Both compounds exhibit reversible water uptake and are rare cases, even for porous metal phosphonates, since they do not exhibit a one-dimensional \rev{SBU} in their crystal structure.
In the Ca-based compound two-dimensional nets of 8-rings are observed which are composed of alternating corner-linked \ce{CaO6}- and \ce{PO3C}-polyhedra.
These nets are connected to a three-dimensional structure through the organic part of the linker molecules leaving space for non-coordinating water molecules.
The crystal structure of the Ba compound is even more exceptional as the \ce{BaO} polyhedra are connected to a three-dimensional \ce{BaO} framework and the ligands ``line'' the inner walls of the pores.
\end{sloppy}

\begin{figure}[tbhp]
\begin{center}
\includegraphics[width=1\linewidth]{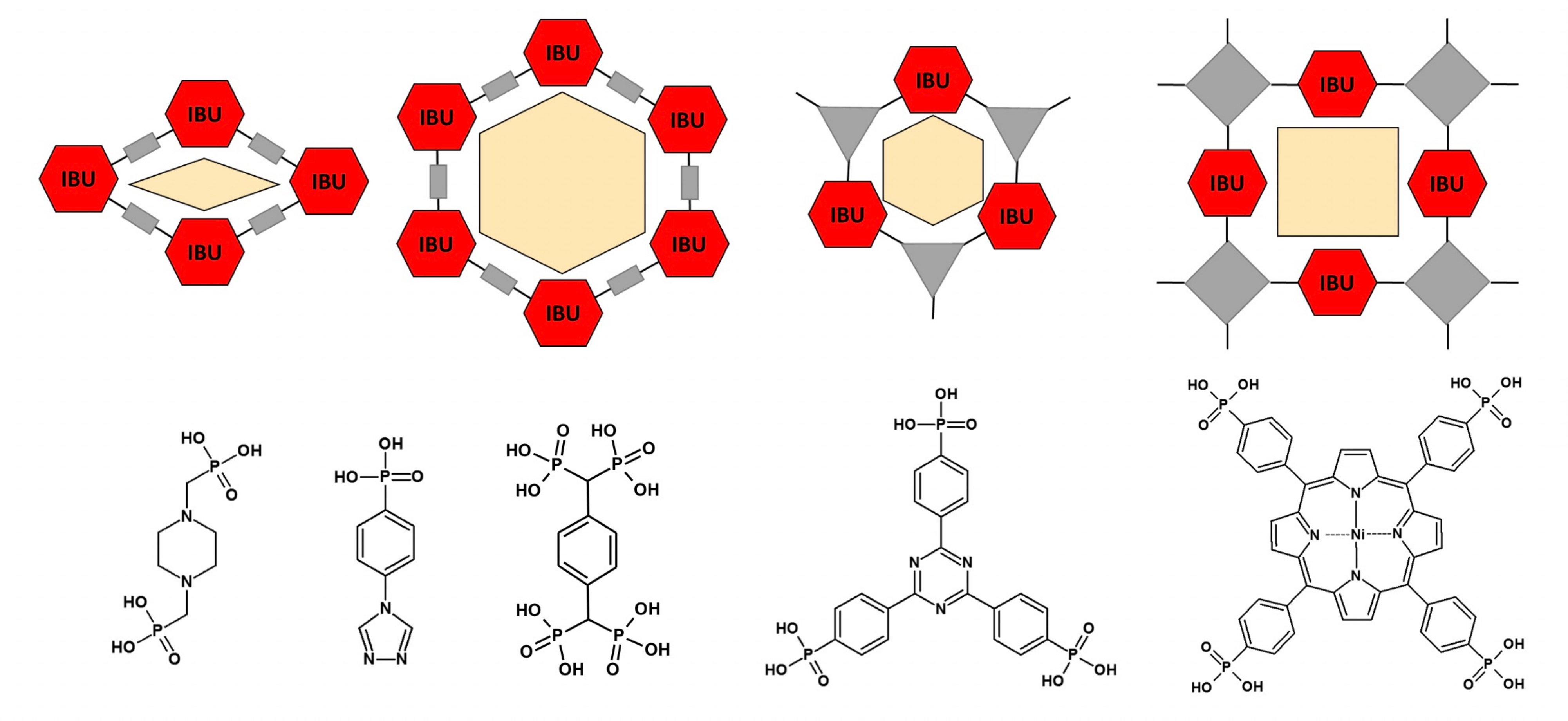}
\caption{Scheme representing the structures of the majority of porous metal phosphonates. Red hexagons: inorganic structural building unit (\rev{SBU}), grey rectangle, triangle and square: phosphonic acids. Examples for polyphosphonic acids which have been successfully employed are shown at the bottom.}
\label{fig:IBU_phosphonates}
\end{center}
\end{figure}

The most common structural motive in porous metal phosphonates is the presence of one-dimensional \rev{SBU}s (Figure~\ref{fig:IBU_phosphonates}).

\begin{sloppy}
The connection of one-dimensional \rev{SBU}s to porous metal phosphonates seems to be especially effective using N,N-bis(phosphonmethyl)piperazine as the linker, which was extensively studied by the groups of {G. Férey}, {C. Serre}, {N. Stock} and especially {P. A. Wright} \cite{10.1021/cm052149l,10.1039/B605400E,10.1016/j.micromeso.2011.12.003}.
Thus, two series of compounds have been described {M-MIL-91} (\ce{[MOH_n(HO3PCH2\bond{-}C4H8N2\bond{-}CH2PO3H)]} with M = Al, Ti and n = 1, 0)  \cite{10.1021/cm052149l}, the well-known {STA-12} series (\ce{[M_2(H2O)_2(O3PCH2NC4H8NCH2PO3)]*xH2O} with M = Mg, Mn, Fe, Co, Ni) \cite{10.1039/B605400E,10.1016/j.micromeso.2011.12.003} and STA-16, which displays the same topology as {STA-12} but contains a larger linker and thus is the only known example for isoreticular structures in porous metal phosphonates \cite{10.1021/ja1097995}.
Important structural data was obtained by single crystal X-ray diffraction of {STA-12(Mg)} \cite{10.1016/j.micromeso.2011.12.003} and high resolution PXRD of Al-MIL-91 \cite{10.1002/zaac.201600358}.

Using a linear tetraphosphonic acid containing two bisphosphonate groups, the group of {T. Bein} obtained the only known flexible open-framework metal phosphonate of composition \ce{[NaLa((HO3P)_2CH\bond{-}C6H4\bond{-}CH(PO3H)_2)]}, \ce{[NaLa(H_4L)]}, \cite{10.1021/ic802294e} in a high-throughput study.
The one-dimensional \rev{SBU}s are formed by alternating bisphosphonate \ce{(HO3P\bond{-}CHR\bond{-}PO3H)^{2-}} and \ce{La^{3+}} ions, which result in the formation of an anionic framework which is highly selective towards monovalent metal cations.
Coordination by the phosphonate and the N-containing groups was also observed by {H. Krautscheid} et al. in metal phosphonates of composition \ce{[La3L_4(H2O)_6]Cl*xH2O} and \ce{[Co4L_3($\mu$_3\bond{-}OH)(H2O)_3](SO4)_{0.5}*xH2O} which were synthesized using the ligand {4-(4H-1,2,4‐triazol‐4‐yl)phenyl} phosphonic acid (\ce{H_{2}L}) (Figure~\ref{fig:IBU_phosphonates}) \cite{10.1039/C5DT02651B,10.1002/chem.201402886}.
In these compounds also one-dimensional \rev{SBU}s are observed which are linked by the organic component of the ligand.
The La phosphonate structure contains large pores with a diameter of \SI{1.9}{\nano\metre} and has been studied for proton conductivity due to its high water stability.
Diffusivities close to that of liquid water were observed and thus conductivity via the vehicle mechanism was proposed \cite{10.1002/chem.201402886}.
\end{sloppy}

The formation of porous structures can also be accomplished using tri- and tetraphosphonic acids with trigonal or tetragonal planar or tetrahedral geometry, respectively: most porous metal phosphonates belong to this class of compounds \cite{10.1002/ejic.201600207,10.1038/ncomms15369,10.1002/slct.201700573}.
Especially triphosphonic acids containing aromatic rings have been employed since this ensures both rigidity and robustness \cite{10.1002/ejic.201600207}.
One-dimensional \rev{SBU}s are predominantly present in these compounds and their connection often leads to honeycomb-like arrangements.

Notably, a structure containing one-dimensional channels between dense, corrugated, hydrogen bonded layers has been reported \cite{10.1039/C6CE01580H}.
Using this information of tetratopic phosphonic acid chemistry, the {N. Stock} group recently used a planar geometrically demanding tetraphosphonic acid, Ni(4-tetraphosphonophenyl)porphyrin (Ni-\ce{H_8}TPPP), in combination with \ce{Co^{2+}} and \ce{Zr^{4+}}/\ce{Hf^{4+}} ions to construct porous metal phosphonates with one-dimensional \rev{SBU}s (Figure~\ref{fig:IBU_phosphonates}).
Importantly, electron diffraction data was necessary to elucidate their crystal structures.
In \ce{[Co_2(Ni\bond{-}H_4TPPP)]*2DABCO*6H2O} even the guest molecules and individual hydrogen atoms could be directly located from the electron diffraction data \cite{StockSubmitted1}.
The crystal structure of \ce{[M_2(Ni\bond{-}H_2TPPP)(OH/F)_2]*xH2O} (M= Zr, Hf), denoted M-CAU-30 where CAU stands for Christian-Albrechts-University, was determined by combining electron diffraction tomography for structure solution and PXRD data for subsequent structural refinement \cite{10.1039/c8sc01533c}.
In addition to very high thermal and chemical stabilities, these compounds show exceptionally high specific surface areas for metal phosphonates.

The chemistry of porous metal phosphonates still relies to a large extend on serendipity, although important structural trends have already been established.
One-dimensional \rev{SBU}s have been shown to lead to porous metal phosphonates but a higher variability in framework structures is expected when well-defined \ce{M\bond{-}O} clusters are accessible.
Unfortunately no robust synthesis conditions for such clusters are yet known, partially due to the lack of commercially available phosphonic acids including a variety of additional functional groups, which would allow systematic investigations and the possibility to establish additional and crucial structural trends.

\section{Towards an Industrial Scale}

\subsection{High-throughput Methods}

The synthesis of MOFs is the most important step in their possible commercialization, since there is no application without a suitable material.
An important synthetic approach, which has proved to be highly valuable for MOF science, are high-throughput (HT) methods using reactors for the solvothermal synthesis of materials.
These were first developed in the 1990s \cite{10.1002/(SICI)1521-3773(19981231)37:24<3369::AID-ANIE3369>3.0.CO;2-H} for the investigation of zeolites but reactor development by German groups led to their use for MOF discovery and synthesis optimization in 2004 by the group of {T. Bein} \cite{10.1002/anie.200351718} and later in 2008 by the groups of {O. M. Yaghi} \cite{10.1126/science.1152516} and {N. Stock} \cite{10.1021/ic800538r}.
Subsequently, HT methodology has been established in many groups and is routinely employed in the investigation of metal carboxylates, phosphonates, imidazolates and pyrazolates.
Unfortunately, the use of HT methods is often not mentioned in the main manuscript of a publication but ``hidden'' in the supporting information, although it can contain a wealth of information on reaction trends.

\begin{figure}[tbhp]
\begin{center}
\includegraphics[width=1\linewidth]{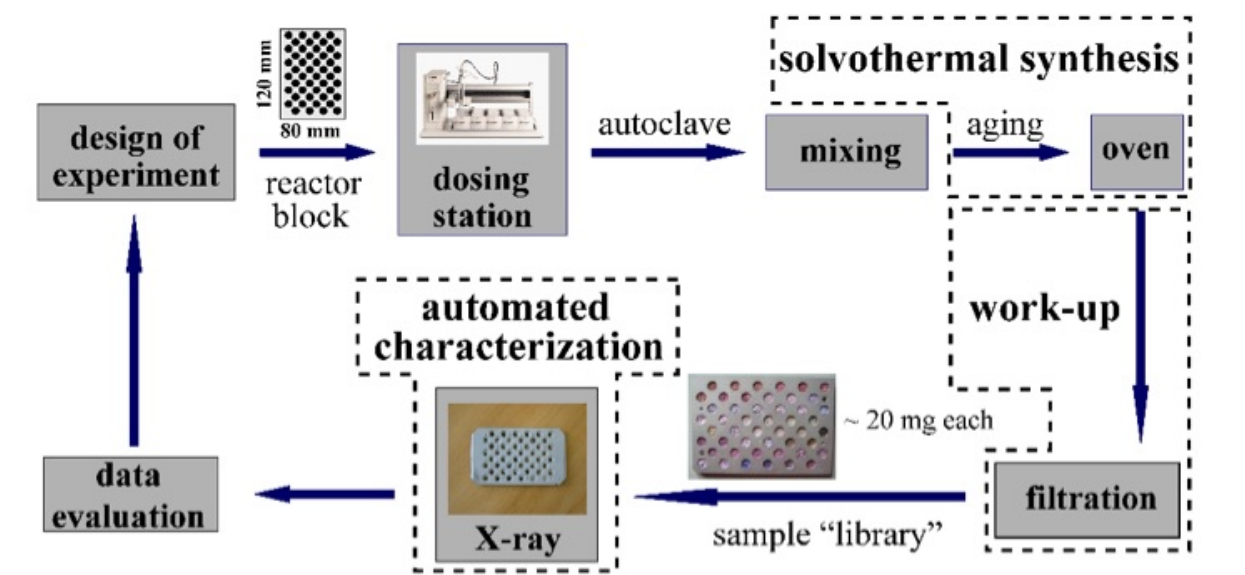}
\caption{Typical HT workflow as used in the discovery of new crystalline materials from solvothermal synthesis. \rev{Reproduced from \cite{10.1016/j.micromeso.2009.06.007}, with permission.}}
\label{fig:HTworkflow}
\end{center}
\end{figure}

HT methodology is based on the concepts of parallelization and miniaturization (of reactors) and automation (of synthesis and characterization).
The combination of such concepts allows for the efficient and accelerated investigation of the complex parameter space frequently observed for solvothermal synthesis \cite{10.1016/j.micromeso.2009.06.007}.
Parameters, such as compositional (molar ratios of starting materials, pH, solvent, etc.) and process parameters (reaction time, temperature and pressure), are varied and their influence on product formation is monitored.
As a result, new compounds can be discovered, their synthesis optimized and, importantly, fields of formation and synthesis-structure trends are established.
HT methods can be regarded as important tool for bringing MOFs to an industrial scale \cite{10.1016/j.micromeso.2012.12.024}.
A typical HT workflow comprising the design of experiment, the solvothermal synthesis, the work-up and automated characterization followed by data evaluation is shown in Figure~\ref{fig:HTworkflow}.

\begin{figure}[tbhp]
\begin{center}
\includegraphics[width=1\linewidth]{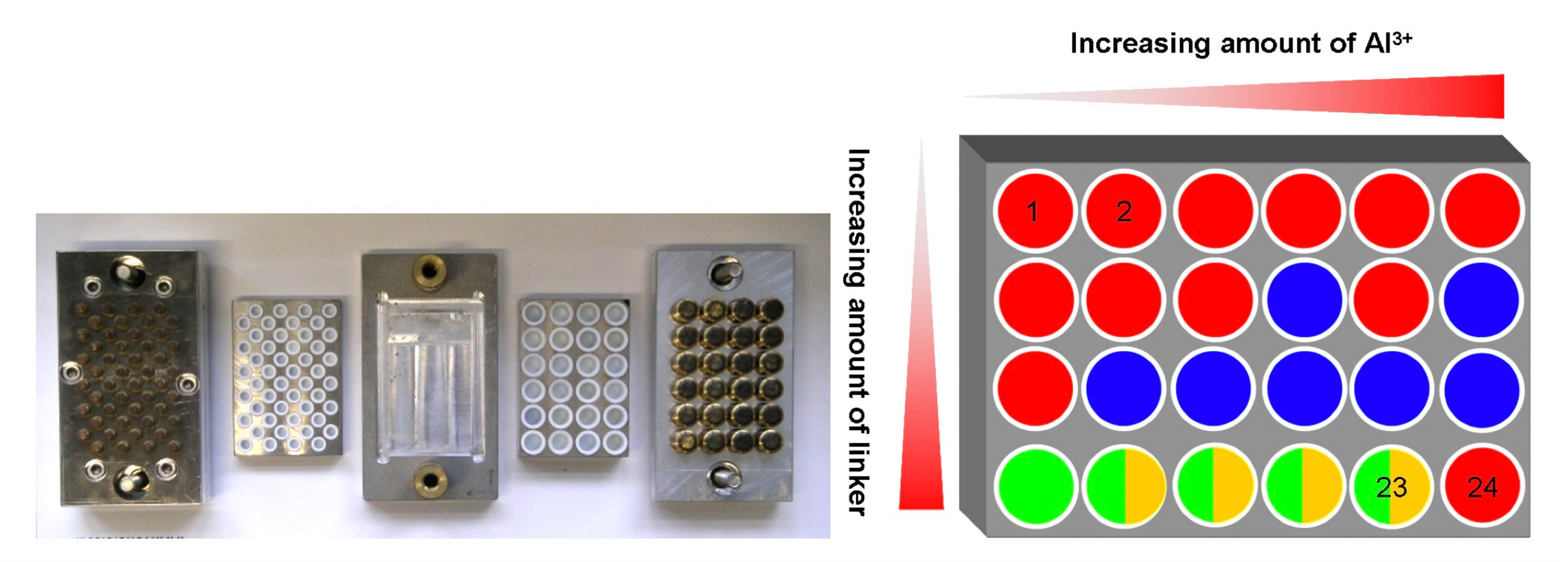}
\caption{Left 24 and 48 eight multiclaves typically used in the HT investigation under solvothermal reaction conditions. Right: HT screening of the system \ce{Al3+} / isophthtalic acid (m-\ce{H2}BDC) / DMF-water that led to the discovery of CAU-10-H. In the first three rows \ce{AlCl3*6H2O} was used as the metal source and in the last row \ce{Al2(SO4)3*18H2O} was used. Product characterization was carried out using PXRD. Green: CAU-10-H; blue: m-\ce{H2}BDC; Red: mixture of CAU-10-H and \ce{H2}BDC; Yellow: unknown phase. \rev{Reprinted with permission from \cite{10.1021/cm3025445}. Copyright 2012 American Chemical Society.}}
\label{fig:HTsolvothermal}
\end{center}
\end{figure}

For the synthesis of MOFs, parallel reactors, often based on the 96 well-plate format, have been employed which can hold reaction mixtures ranging from several \si{\micro\liter} to a few \si{\milli\liter} (Figure~\ref{fig:HTsolvothermal}, left).
In addition to conventional heating of the reactors, the group of {N. Stock} has introduced microwave irradiation \cite{10.1016/j.micromeso.2012.12.024,10.1039/C3DT50413A,10.1021/ic200381f} and ultrasound \cite{10.1021/cg401459a,10.1039/C3DT52576G} as energy sources and also a temperature gradient oven in the HT workflow \cite{10.1002/anie.200701575}.
In order to facilitate the synthesis and rapid characterization of products, at least a small degree of automation is necessary.
Fully automated HT set-ups have been described including automated liquid handling and solid dosing in the \si{\milli\gram}-range \cite{10.1016/j.micromeso.2004.09.018,10.1039/C0SC00179A}.
HT characterization is routinely carried out using PXRD, but also high throughput adsorption measurements have been described, the latter being important for the fast screening of sample porosity \cite{10.1021/co300128w,10.1039/C1CC10674K}.

The advantages of HT-methods in the discovery and synthesis optimization of CPs and MOFs has been summarized in various publications \cite{10.1016/j.micromeso.2012.12.024,10.1021/cr200304e,10.1039/9781849733571-00087}.
The following three examples, by Germany-based researchers, demonstrate the advantages of this methodology and the resulting impact to MOF research.

One of the first HT studies of MOFs in 2008, the functionalization of prototypical materials MIL-53, MIL-88 and MIL-101 was systematically studied for the first time and the importance of the amino functionalization for possible modification reactions and applications was stressed \cite{10.1021/ic800538r}.
The fields of formation of all the studied functionalized materials were identified by solvothermal reactions of \ce{FeCl3} and 2-aminoterephthalic acid (\ce{H_2BDC-NH2}) in protic and aprotic reaction media, which allowed extraction of reaction trends.
Surprisingly the molar ratio \ce{FeCl3}:\ce{H_2BDC\bond{-}NH2} was found to be less important.
The solvent was identified to have the most profound impact on the product formation and the other key parameters include reaction temperature and, remarkably, the overall concentration of the reaction mixture, an important consideration for synthesis scale-up.

The limited amount of data obtained for Ni-based MOFs in 2010 inspired another HT-investigation \cite{10.1021/ic200381f}.
Ni salts were reacted in DMF with different aromatic polycarboxylic acids, varying in size and geometry, and a number of reaction parameters were varied including  metal source (counter ion), reaction temperature, time and addition of bases.
In this HT study microwave-assisted heating was employed and isoreticular compounds of HKUST-1 and MOF-14 were obtained with \ce{Ni^{2+}} ions.
Interestingly, by changing the ligand from tricarboxylic acid to terephthalic acid and increasing the complexity of the solvothermal system by the addition of a diamine, two framework polymorphs were obtained.
One is based on a square-grid of Ni-paddlewheel units and terephthalate ions, while the other is based on a kagome-grid.
The layers in these polymorphs are interconnected by diamine molecules to form highly porous framework compounds.

As a last example the results of the study on Al-MOFs is presented, which led to the discovery of the aluminium hydroxide isophthalate CAU-10-H \cite{10.1021/cm3025445}.
Due to the interesting water sorption properties a green synthesis route was subsequently established employing HT methods and the synthesis scale-up was carried out which allowed the use of CAU-10 in adsorption driven chillers, as discussed later in this review \cite{10.1002/adma.201705869}.
CAU-10-H was originally discovered employing dimethylformamide (DMF) as a solvent.
The investigation of the system \ce{Al^{3+}} / isophthtalic acid (m-\ce{H_2BDC}) / DMF-water mixture was carried out in a 24-reactor HT set-up.
The metal source, the molar ratio \ce{Al^{3+}} : linker and the solvent, a mixture of DMF and \ce{H2O}, were varied.
After reaction at \SI{145}{\celsius} for 12~h, the product was filtered off and characterized by PXRD (Figure~\ref{fig:HTsolvothermal}) \cite{10.1021/cm3025445}.
The evaluation of the PXRD data resulted in the following information and trends.
As seen in the first three rows, under the given reaction conditions using aluminium chloride in the reaction mixture resulted in the formation of mixtures of crystalline CAU-10-H and m-\ce{H_2BDC} or pure m-\ce{H_2BDC}.
In the last row the results using aluminium sulfate as the metal source is given. In almost all reactions CAU-10-H is detected.
Starting from this discovery library the synthesis was further optimized and the decisive role of the solubility of m-\ce{H_2BDC} was seen as a key factor since a mixture of DMF and water led to a phase pure product.
This knowledge has been very important in establishing a green synthesis route of CAU-10-H.
In further studies, the DMF could be replaced by using ethanol and the pH was adjusted using sodium aluminate (\ce{NaAlO_2}) and m-\ce{Na_{2}BDC} as starting materials.
Via high throughput screening suitable reaction conditions were identified at the \SI{2}{\milli\liter} level and the synthesis was scaled to at least the \SI{0.5}{\kilo\gram}-scale with a yield of up to 95\% (based on the amount of linker), using a \SI{10}{\liter} round bottom flask.

The speed of a HT investigation is determined by the time-limiting step in the workflow.
The HT characterisation of new compounds is routinely carried out by PXRD measurements within a few minutes.
While PXRD data only permits the identification of crystalline phases, no information on the porosity is obtained, which is a key property for many applications.
An effective and rapid characterisation tool for the characterisation of up to 12 porous materials at ambient conditions has been reported by the {S. Kaskel} group in 2011 \cite{10.1039/C1CC10674K}.
The InfraSORP technique measures the thermal response upon adsorption of a test gas in real time using an optical sensor and it was used to screen various micro- and mesoporous materials for their adsorption capacity and specific surface area.
It also allows for the investigation adsorption kinetics, pore structure and isotherm characteristics within a few minutes.
Up to now, mainly physisorption of test gases such as butane, propane or carbon dioxide were carried out, but the InfraSORP was also used to investigate chemisorption of \ce{H2S} and \ce{NH3}.
By carrying out two consecutive adsorption/desorption runs, the contributions of physisorbed and chemisorbed guest molecules could also be distinguished \cite{10.1016/j.micromeso.2018.03.032}.

\subsection{Industrial Development and Green Synthesis}
The scale-up of common research laboratory parameters yielding amounts \SI{<1}{\gram} of MOF up to conditions which allow the production at industrial scale can be considered a crucial challenge for the transfer of MOFs to applications as adsorbents in real life.
This was discussed by industrial researchers from BASF already in 2006 along with the potential fields of applications for MOFs \cite{10.1039/B511962F}.
This challenge is closely related to the development of synthesis conditions for the respective MOF that result in a reasonable pricing while being --- if possible --- environmentally benign and of course preserving performance.
Hence such ``green'' synthesis conditions for MOFs were already postulated very early as a likely prerequisite for MOF production at large scale by BASF researchers \cite{10.1016/j.micromeso.2011.08.016} and therefore the topics ``green'' synthesis and industrial development will be discussed side by side herein.
\rev{Recently, {M. R Hill} and coworkers produced an excellent review detailing new and general synthetic routes for MOF production at a commercial scale \cite{10.1039/C7CS00109F}.}

Green synthesis conditions comprise mild heating up to reflux only, and rely on employing mostly or only water as solvent and have been recently described by {H. Reinsch} in more detail elsewhere \cite{10.1002/ejic.201600286}.
Already in 2006, researchers at BASF filed a patent in which the synthesis of the MOF aluminium fumarate was described using DMF as solvent \cite{US8734652B2}.
The precise structure of this compound could be elucidated only nine years later and the framework was identified as an analogue of the archetypical {MIL-53} structure \cite{10.1002/anie.201410459}.
The compound shows particularly good performance for heat pumps \cite{10.1039/C4RA03794D} and other applications.
Due to these remarkable properties and the commercially promising reactants, a truly green synthesis using only sodium fumarate and aluminium sulfate as reactants under reflux was subsequently developed, again by the very same research group \cite{US8524932B2}.
This currently serves as the benchmark for industrially feasible, mild and cost-efficient synthesis conditions, as the synthesis uses only harmless reactants at very mild conditions after short reaction times, producing minimal waste.

Especially in the group of N. Stock, these synthesis conditions have served as inspiration for the development of similar processes yielding several other Al-MOFs under mild conditions.
An interesting example is the synthesis of the aluminium isophthalate {CAU-10}, which originally required the use of DMF in closed autoclaves under pressure \cite{10.1021/cm3025445}.
This potentially hazardous procedure could be optimized into an aqueous process, using only ethanol in small amounts as additive, and carried out under reflux \cite{10.1002/adma.201705869}, thus giving access to \si{\kilo\gram}-scale amounts of {CAU-10}.

In further studies by {H. Reinsch} the linker molecules mesaconic and citraconic acid were employed, separately, under conditions very similar to the production process for aluminium fumarate.
This resulted in the synthesis of MOFs {Al-MIL-68-Mes} \cite{10.1002/chem.201704771} and CAU-15-Cit \cite{10.1039/C7DT04221C}.
These compounds are of particular interest since the organic building units are derivatives of citric acid, which are readily available at industrial scale.
Furthermore, the ``green'' synthesis of {Zr-MOFs} were also investigated by the group in Kiel.
By using aqueous conditions suitable for the synthesis of zirconium fumarate, as reported by the {P. Behrens} group, as a starting point several other compounds based on zirconium ions could be obtained in water/acetic acid mixtures under reflux \cite{10.1002/ejic.201600295}.
While this necessitates zirconylchloride as metal source, using the less corrosive sulphate allowed for the synthesis of zirconium trimellitate with {UiO-66} framework structure \cite{10.1039/C7RE00214A}.
However, MOFs usually cannot be obtained under completely arbitrary conditions and thus the improvement of very harsh to milder synthesis conditions can be considered a relevant approach to ``green'' synthesis.
In addition, the scalability is definitely not limited to mild conditions as various products of interest can be only obtained under harsh conditions, or employing hazardous reactants.

Examples of desired improvements for challenging synthesis conditions were recently outlined by the group of {C. Janiak}.
The mesoporous chromium terephthalate {MIL-101(Cr)} is a versatile and extremely stable MOF, usually synthesized employing hydrofluoric acid as additive at very high temperatures, up to \SI{220}{\celsius} \cite{10.1126/science.1116275}.
In a different approach, the original synthesis of iron trimesate, {MIL-100(Fe)}, was improved.
This synthesis procedure also involves hydrofluoric acid and requires high temperatures, using elemental iron as metal source \cite{10.1039/B704325B}.
Employing a mixture of water and dimethylsulfoxide, the {C. Janiak} group succeeded in the synthesis of {MIL-100(Fe)} at ambient pressure \cite{10.1039/C6DT01179A}, again preserving the high porosity reported for the harsher synthesis conditions.
Investigating the synthesis of another iconic MOF, zirconium terephthalate {UiO-66}, the amount of employed solvent could be drastically reduced by dry-gel conversion \cite{10.1039/C7DT01717K}.
The respective reactants were employed as solid mixtures and kept spatially separated from the comparably small amount of solvent used in a closed reactor.
Under heating, the solvent evaporates and induces the crystallization of {UiO-66} in the solid mixture.
This procedure is also applicable for some derivatives of this MOF and the solvent is easily recycled and reused.

Despite all these academic and industrial efforts, it must be stated that the availability of MOFs in amounts larger than few grams is still limited and the commercial exploitation of MOFs is still in its infancy.
A few industrially prepared solids known by the ``basolite'' product name are commercially available in small quantities and marketed via Sigma Aldrich.
While several start-ups in other countries try to develop suitable procedures to make MOFs accessible in large amounts, BASF only produces Aluminium fumarate at larger scale.
Notably, shaping is an important requirement for any industrial transfer \cite{10.1111/j.1551-2916.2010.03824.x} and shaped granules and spheres of MOFs are available from Materials Center Dresden \cite{MaterialsCenterDresden}.

\section{Advanced Characterization}

\subsection{In situ characterization during synthesis}

The traditional synthesis of MOFs usually occurs in solution, often at temperatures above the boiling point of the respective solvent.
Under these conditions slight changes of reaction parameters, for example the replacement of the counter ion of the metal source, can have a strong impact on the product formation.
As previously described, variation of reactant concentration, temperature or time may lead to interpenetration and completely different framework structures.
Thus the parameter space of a chemical system is usually screened to find new MOFs.

In situ techniques, which very often necessitate the use of synchrotron radiation, allow the non-invasive observation of chemical processes within a sealed reaction vessel without the need for quenching and ex situ characterization \cite{10.1021/cr200304e,10.1021/cr200304e,10.1002/9783527693078.ch24}.
Crystalline intermediate phases may be observed and crystallization kinetics can be deduced by applying physical models.
An overview on in situ monitoring of the formation of crystalline solids in general and in situ studies on MOF syntheses have been summarized by {N. Pienack} and {W. Bensch} \cite{10.1002/anie.201001180}, {N. Stock} and {S. Biswas} \cite{10.1021/cr200304e}, {R. Walton} and {F. Millange} \cite{10.1002/9783527693078.ch23} and {I. Senkovska and V. Bon} \cite{10.1002/9783527693078.ch24}.
In situ studies carried out by the groups of {R. A. Fischer} and {C. Wöll} on layer-by-layer growth of SURMOFs \cite{10.1002/9783527693078.ch17} will be covered in Section~5.8.

Early in situ XRD studies of microporous materials were carried out on zeolites and AlPOs.
In 2010 the first study on MOFs was reported by the groups of {G. Férey} and {R. Walton} \cite{10.1002/anie.200905627} and this as well as following work demonstrated the wealth of information obtained by such investigations.
At this time energy dispersive X-ray diffraction (EDXRD) studies were employed and in addition to the evaluation of crystallization kinetics, ``windows of stability'' for certain MOF syntheses were established and crystalline intermediates were discovered that were further characterized after quenching.
Shortly after, the German groups of {P. Behrens}, {N. Stock} and {M. Wiebcke} have made extensive use of the available equipment and subsequent reactor development was carried out by the groups of {R. E. Dinnebier}, {F. Emmerling} and {N. Stock}.
Some of the results of their work is summarized in the following paragraphs.

The group of {M. Wiebcke} has used in situ EDXRD often in combination with other techniques, such as time resolved static light scattering, to shed light on the role of modulators in the synthesis of ZIF-8 and the formation and transformation of zinc imidazolate polymorphs \cite{10.1039/C1CE06002C,10.1515/zkri-2016-1968,10.1021/cm103571y}.
While {P. Behrens} and coworkers studied the modulated synthesis of zirconium fumarate MOF with UiO-66 structure \cite{10.1039/C4CE01095G}.
They were able to elucidate the role of modulators as competitive ligands in the coordination equilibria and bases in the deprotonation equilibria during nucleation and crystal growth.
These important results has allowed for the fine adjustment of size and morphology of ZIFs and Zr-MOFs \rev{, an area which has seen continued interest and investigation by researchers outside of Germany \cite{10.1039/C7NR07949D,10.1039/C002088E}.}

\begin{figure}[tbhp]
\begin{center}
\includegraphics[width=1\linewidth]{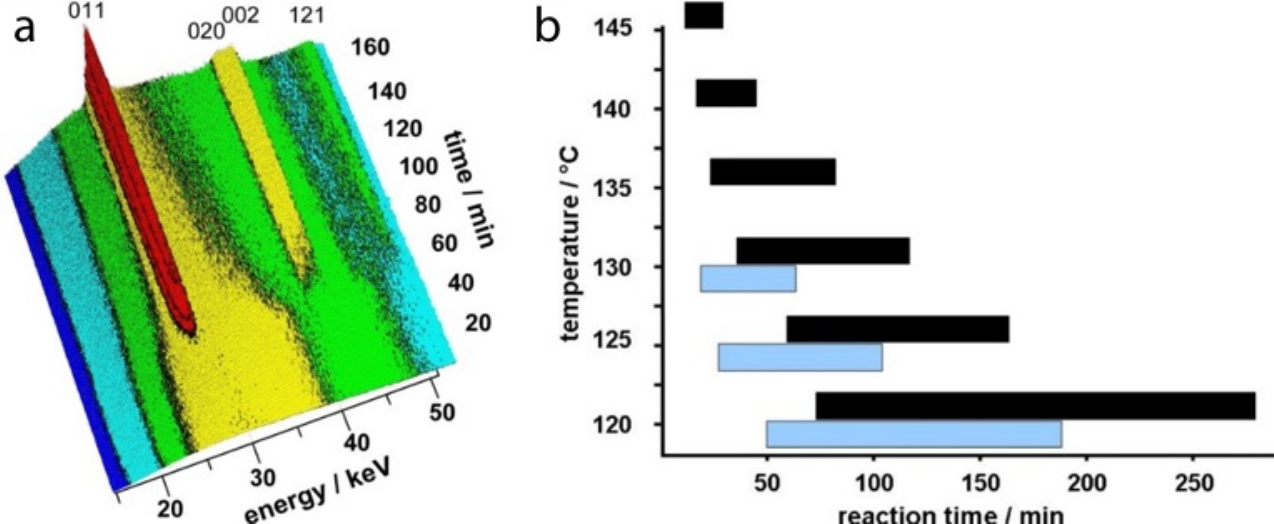}
\caption{(a) Time-resolved in-situ EDXRD data measured during the crystallization of CAU-1-\ce{(OH)_2} at 125 °C using conventional heating. (b) Comparison of induction and reaction times between MW-assisted (blue) and conventional (black) heating for the crystal growth of CAU-1-\ce{(OH)_2} in the range 120-\SI{145}{\celsius}. Reproduced from \cite{10.1002/chem.201003708}, with permission.}
\label{fig:EDXRD_CAU}
\end{center}
\end{figure}

The chemistry of Al-MOFs has been extensively studied by the group of N. Stock using in situ EDXRD and high-energy monochromatic X-rays.
This group also introduced energy input through MW-assisted heating \cite{10.1021/ic101786r,10.1002/chem.201003708} and ultrasonication \cite{10.1039/C3DT52576G}, for in situ crystallization studies.
MW-assisted heating in the synthesis of lanthanide phosphonates led to shorter crystallization times in comparison to conventional heating, although no significant influence on induction time was observed.
In these studies intermediate crystalline phases were also discovered and in some cases these could be isolated by quenching.
Further characterization of the crystalline intermediates established the structural evolution of the final product \cite{10.1021/ic301976s}.
In a study of the Al-MOF denoted {CAU-1}, the role of ligand functionalization on product formation was elucidated and conventional and MW-assisted heating were compared (Figure~\ref{fig:EDXRD_CAU}).
Here, MW-assisted heating led to shorter induction times and faster crystallization and the amino-modified compound ({CAU-1-\ce{NH2}}) showed faster crystallization kinetics compared to the dihydroxy modified compound {CAU-1-\ce{(OH)_2}} \cite{10.1002/chem.201003708}.
Surprisingly, using MW-assisted heating at higher temperatures no product formation of CAU-1-\ce{(OH)_2} was observed, while conventional heating resulted in this desired product.

A combination of in situ infrared radiation (IR) and EDXRD measurements on the formation of a nickel phosphonate under ultrasonication revealed faster crystallization upon increase of the ultrasound amplitude under constant temperature \cite{10.1039/C3DT52576G}.
Temporal evolution of the signals in the IR spectra were attributed to enhanced crystallization kinetics for faster dissolution of the ligand.

Development of third generation synchrotron sources and improved detectors have dramatically increased the scope of possible in situ analyses \cite{10.1002/9783527693078.ch24}.
Far better time resolution and signal quality, in combination with new in situ reactors, has recently allowed the study of more complex reaction systems in unprecedented detail.
It was first demonstrated by the groups of {T. Friščić} and {R. E. Dinnebier} that mechanochemical syntheses can be followed in situ within a ball mill \cite{10.1038/nchem.1566}.
Fast data collection and full pattern-fitting allowed the study of sequential crystallization of zinc 2-ethylimidazolates and the role of small amounts of liquids (liquid-assisted grinding) or additional salts (ion and liquid-assisted grinding).
The group of {F. Emmerling} even combined in situ XRD analysis with Raman spectroscopic measurements to investigate the formation of four model compounds and demonstrated information of crystallization at the molecular and the crystallite scale \cite{10.1002/anie.201409834}.

The combination of small- and wide-angle X-ray scattering (SAXS/WAXS) is also useful when short acquisition times are necessitated, although such studies are still very rare.
The synthesis of ZIFs \cite{10.1002/anie.201102071,10.1002/chem.201204638} and Al-MOFs \cite{10.1002/anie.201101757} were investigated using these methods to provide insight to nucleation and crystal growth processes.
The groups of {M. Wiebcke} and {K. Huber} studied the formation of {ZIF-8} with one second time resolution.
SAXS provided information on the presence and size of nano‐sized clusters during the nucleation and early growth of nanocrystals while WAXS followed the larger crystal growth.
Based on these results two possible pathways on the {ZIF-8} formation at room temperature were established.

Recently a new in situ reaction cell SynRAC (synchrotron-based reaction cell for the analysis of chemical reactions) has been reported which was jointly developed by the groups of {N. Stock}, {W. Bensch} and the beamline staff at P08, PETRA III, DESY \cite{10.1063/1.4999688}.
In addition to using glass vessels that are routinely employed in laboratory synthesis, the reaction cell also allows for precise control of reaction temperature, including fast heating rates, which is important for rapid crystallization processes.
Complex reaction schemes with multiple temperature ramps can be investigated, fully computer-controlled, and the option to add liquids and solids at the beginning or during the synthesis can be chosen.
Thus reactions can be quenched or chemical parameters (concentration, pH, etc.) can be changed to study their influence on product formation.

\begin{figure}[tbhp]
\begin{center}
\includegraphics[width=0.8\linewidth]{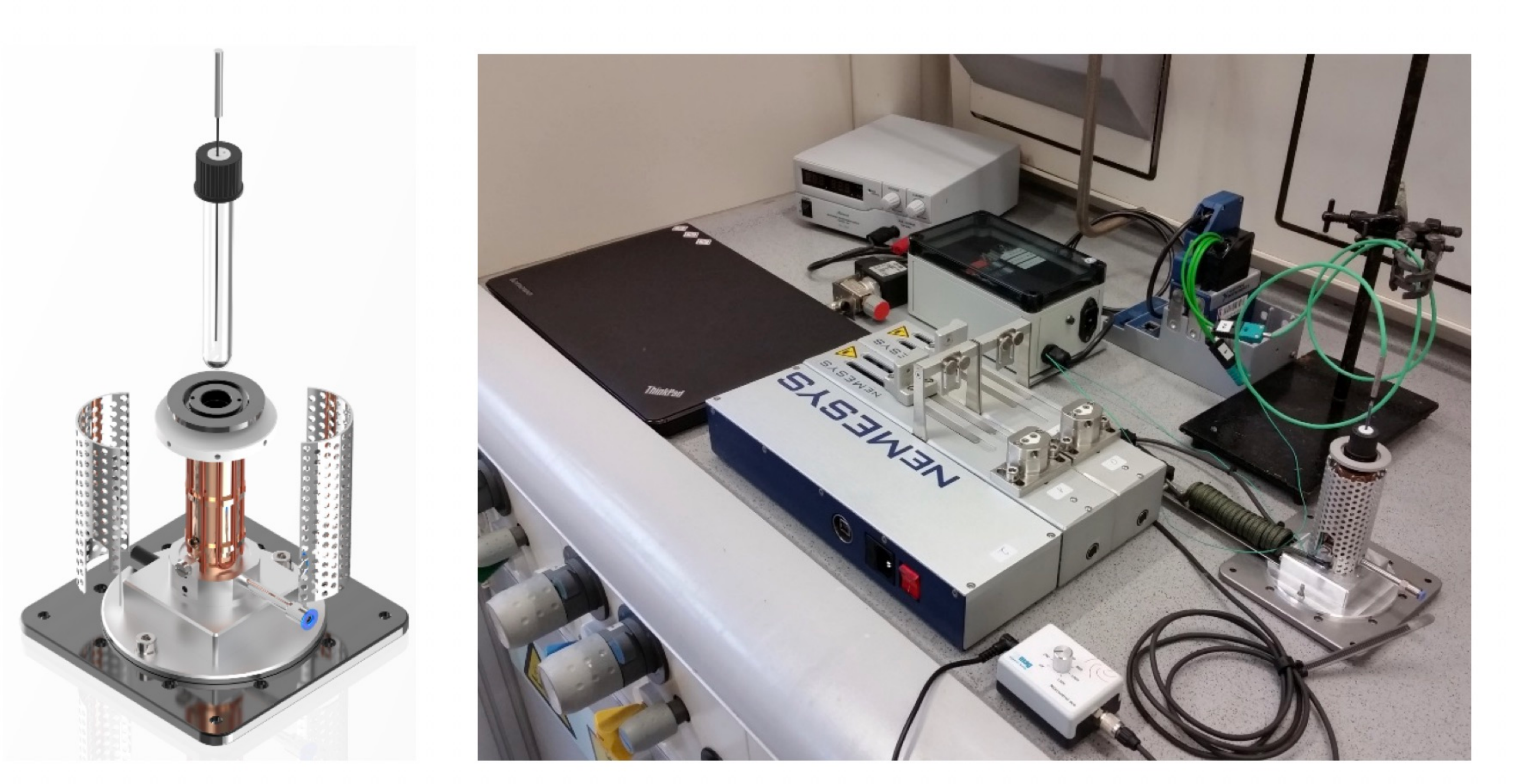}
\caption{Left: picture of the recently developed fully computer-controlled SynRac reaction cell, which allows the precise control of the reaction temperature, including fast heating and cooling as well as the possibility to add liquids and solids during the reaction. Right: complete setup with computer control and liquid dosing.}
\label{fig:SynRac}
\end{center}
\end{figure}

In proof of concept studies, this reactor has been employed to study the synthesis of Ce-UiO-66, which forms rapidly and decomposes upon extended reaction times under the formation of cerium formate \cite{10.1039/C5CC02606G}.
Pre-heating of the solvent/linker solution to the targeted reaction temperature and injection of the linker solutions allowed the reactor to reach the targeted reaction temperature within $\approx$\SI{20}{\second}, which allowed for determination of kinetics of this fast crystallization process.
In addition, a crystalline intermediate was isolated and identified during the formation of [\ce{Bi(HIDC)(IDC)}] \cite{10.1021/cg4000654} through quenching of the reaction mixture by direct cooling and subsequent characterization.
The versatility of the SynRAC has been proven in other in situ PXRD investigations and in a combined in situ XRD / XAS study of the evolution of palladium species in Pd@MOF catalysts during the Heck coupling reaction \cite{Stock_JACSrev}.
In one of the first full scale studies using the SynRAC the linker conformation controlled flexibility of {CAU-13} upon interaction of different pyrazines in solution at various temperatures was investigated \cite{10.1039/c6dt03998g}.
The results showed that stronger host-guest interactions required milder adsorption conditions while harsher conditions nevertheless accelerated the conversion.
Surprisingly the kinetic parameters for the intercalation of pyrazine indicated that dependent on intercalation temperature the rate limiting step differs.

As a final example, the green synthesis of {Al-MIL-68-Mes} is mentioned.
This compound is readily obtained upon heating a reaction slurry, which is rapidly formed directly after mixing the solutions of the starting materials, after short reaction times and under mild aqueous reaction conditions.
Surprisingly, the reaction temperature has only a slight influence on the induction time for crystallization, but the shape of the crystallization curves are substantially different demonstrating that a temperature increase does not simply accelerate the reaction, as expected.
To explain the crystallization kinetics, the homogeneity of the slurry and the dissolution properties of the intermediate amorphous phase were taken into account and based on the extracted Avrami exponents, a plausible, but rather complex formation process was postulated.

The impressive developments at synchrotron sources in combination with newly developed reactor designs open up a wide variety of experiments that will allow to study MOF formation with unprecedented precision.
Especially the combination of characterization methods (scattering, spectroscopic etc.), which allow for the study of crystallisation processes at different length scales, will help us in the future to understand the formation and properties of MOFs.
This may allow us to tune the reaction conditions in order to obtain materials with desired properties.
It should be considered that many reactors are well established at different synchrotron facilities and scientists interested in such investigations are strongly urged to contact beamline scientists or the reactor developers.

\subsection{In situ characterization during gas adsorption}

Adsorption-induced flexibility in MOFs, discovered in early 2000s by {K. Kaneko}, {S. Kitagawa} and {G. Férey}, required new characterization techniques, allowing for the detailed characterization of structural changes during the adsorption of guest molecules.
The presence of long range order in MOFs make X-ray diffraction techniques preferential and most exact, for following the phase transitions.
Since most of flexible MOFs could be synthesized as fine powders or show large amplitude of structural changes, PXRD is the appropriate technique for detection of the structural changes.
Hence, {R. Matsuda} and coworkers used synchrotron X-ray powder diffraction for direct observation of the adsorbed hydrogen in the pores of \ce{[Cu_2(pzdc)_2(pyz)]_n} (pzdc = pyrazine-2,3-dicarboxylate, pyz = pyrazine) \cite{10.1002/anie.200461895}.
In Europe, {P. L. Llewellyn} and coworkers designed a similar capillary-based setup for investigations of adsorption-induced transitions in materials of the MIL-53 family \cite{10.1021/ja803899q,10.1021/ja907556q}.

In Germany, groups of {R. A. Fischer}, {S. Kaskel}, {H. Krautscheid} and {N. Stock} recognized early the both academic and industrial potential of flexible MOFs and contributed not only to the synthesis of new flexible materials, but also to the development of new in situ PXRD instrumentation.
In 2006, a flexible ``gate pressure'' pillared-layer MOF {DUT-8(Ni)} was synthesized in Dresden for the first time \cite{10.1039/C003835K}.
At that time, there was no experimental setup at German large scale facilities for measurement of PXRD patterns under the required gas loading and temperature for {DUT-8(Ni)} flexibility.
Therefore in order to explain the structural behavior of DUT-8(Ni) during gas adsorption experiments, the group at {TU Dresden}, with a support of the sample environment group of Helmholtz-Zentrum Berlin für Materialien und Energie (HZB), developed an automated instrumentation for in situ PXRD measurements during adsorption and desorption for all non-corrosive gases \cite{10.1016/j.micromeso.2013.12.024}.
The instrumentation is commissioned at the KMC-2 beamline of HZB and is available for the whole user community of the BESSY II synchrotron.

This important instrumentation includes an adsorption chamber, mounted on a closed-cycle helium cryostat, which ensures isothermal adsorption / desorption conditions within the temperature range 10--450~$\pm$~0.01~K.
The sample is sealed within a beryllium dome within the adsorption chamber, which is adjusted in the synchrotron beam.
The PXRD patterns could be measured in transmission geometry using $2\theta$ scans.
The adsorption chamber is then connected to the adsorption instrument (BELSORP-max), allowing to measure low-pressure isotherms within the range {$10^{-6}$--1~bar}.
After reaching an adsorption equilibrium conditions, PXRD patterns can be measured at each adsorption/desorption point.
However, this setup allows the measurement of only low-pressure isotherms because of the limitation of the adsorption instrument and beryllium dome.
Small modifications also allowed the use of this same setup for in situ X-ray absorption spectroscopy measurements on the same beamline and temperature/pressure conditions.

In order to test flexible MOFs under conditions of pressure and temperature swing adsorption another capillary-based setup was combined with a BELSORP-HP instrument at the same facility.
The instrumentation allows for the monitoring of adsorption in the pressure range of up to 80 bar and temperatures between 200--300~K.
Recently the setup was successfully used in the in situ study of methane hydrate formation in the confined pore space of porous carbons \cite{10.1039/C6CP03993F}.

One of the potential applications of chemically stable MOFs is adsorptive heat pumps.
To shed the light on the adsorption mechanism of water and other vapors, an in situ vapor cell is now constructed and will be commissioned at the same facility.
With three different types of in situ cells, KMC-2 beamline at BESSY II synchrotron became a multipurpose instrument for the study of crystalline porous materials at defined gas / vapor loading in broad pressure and temperature ranges.
In the recent five years, adsorption induced flexibility in a number of flexible MOFs namely, {DUT-8(Ni)} \cite{10.1039/C5CP02180D}, {ELM-11} \cite{10.1016/j.micromeso.2013.12.024}, {CAU-13} \cite{10.1021/ic500288w} solid solutions of \ce{[Zn_2(BME-bdc)_x(DB-bdc)_2-xdabco]_n} \cite{10.1016/j.micromeso.2015.02.042}, {SNU-9} \cite{10.1021/ic4024844} and {DUT-49} \cite{10.1038/nature17430,10.1038/s41467-018-03979-2}, were studied using these experimental setups.

\begin{figure}[tbhp]
\begin{center}
\includegraphics[width=0.8\linewidth]{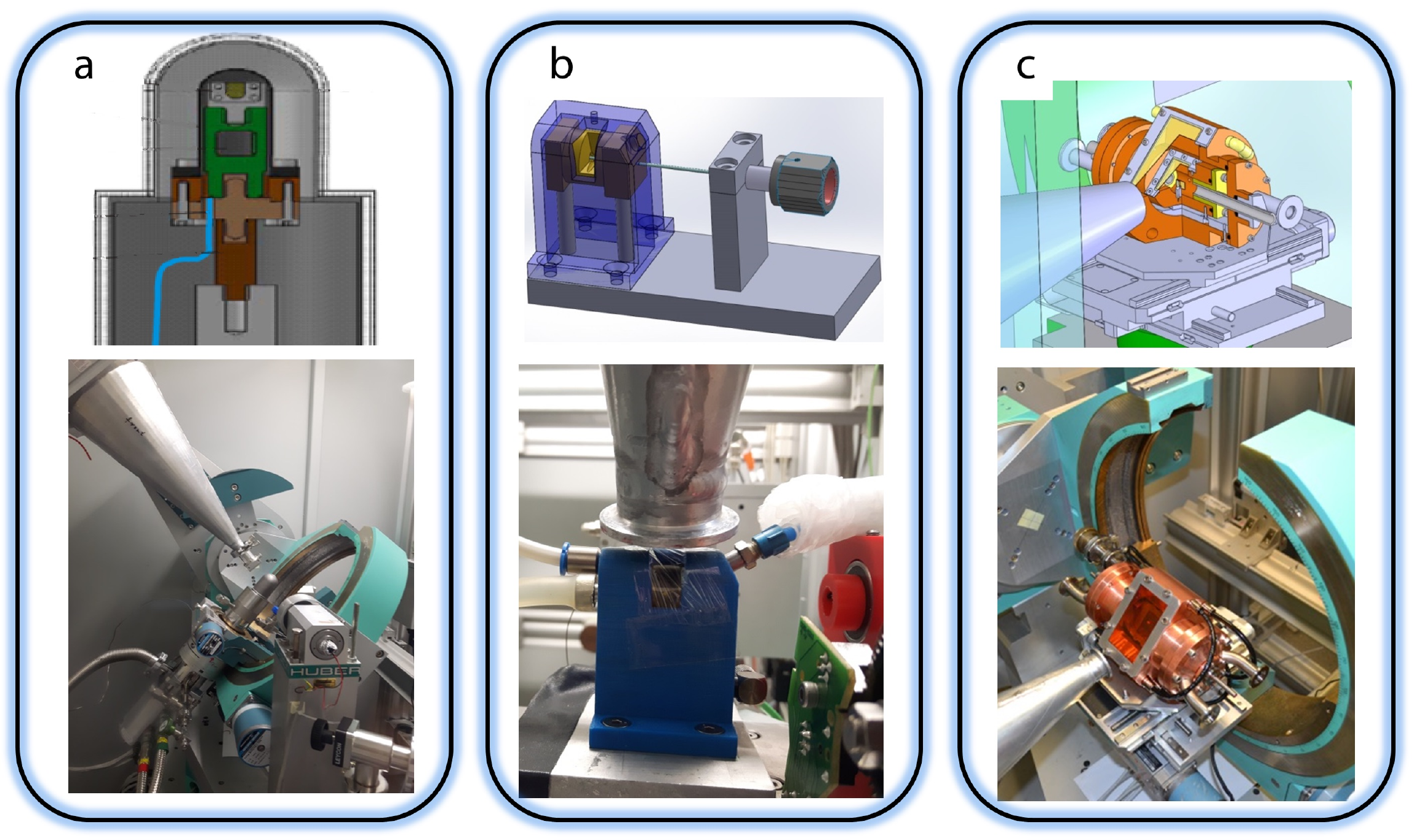}
\caption{In situ PXRD cells for characterization of MOFs during gas and vapor adsorption: a) low-pressure cell; b) high-pressure cell; c) vapor cell.}
\label{fig:insitu_ads}
\end{center}
\end{figure}

A similar experimental setup was described by {R. A. Fischer} and coworkers \cite{10.1039/c1ce05446e}.
The setup was used to investigate the multi-step breathing behavior in a pillared-layer system \ce{[Zn_2(BME-bdc)_2(dabco)]_n} at the beamline of the BL9 of the synchrotron radiation facility DELTA.
The cell uses a helium cryostat for tempering of the sample and the measurements proceed in reflection mode.
However no isotherm can be measured with this setup, only PXRD patterns at defined \ce{CO2} loadings.

All above mentioned setups are available only at large scale facilities, which do not imply frequent user access.
Therefore {H. Krautscheid} and coworkers designed a setup, which can be mounted on the laboratory powder diffractometer STOE STADI P.
The cell consists of glass or quartz capillary, glued on the VCR-support and connected to a ``home-made'' gas dosing system.
The adsorption temperature can be controlled by nitrogen cryostream.
Using this setup, PXRD patterns at different \textit{n}-butane and 1-butene loadings were measured on \ce{[Cu_4(\mu_4-O)(\mu_2-OH)_2(Me_2trz-pba)_4]} at room temperature \cite{10.1039/c3ta15331b}.

In summary, development of in situ PXRD instrumentation for characterization of MOFs at variable guest loading is extremely important for elucidating framework dynamics and adsorption mechanisms.
This was one of the crucial factors, which propelled the German flexible MOF community to the level of the leading groups in the field.

\subsection{Characterization of Defects}

Imperfections or defects naturally occur at temperatures above 0~K in every crystalline material and are responsible for several fascinating properties.
The most stunning examples are precious gemstones like Emerald \ce{Be3Al2(SiO3)6} or Sapphire \ce{Al2O3} that acquire their impressive color from chromium impurities.
For synthetic systems, only the precise control of defect concentration, allows for the fine tuning of certain properties of the material, which has led to relevant discoveries and even novel technological applications of such materials.
One example that originates from the precise control of defect chemistry is doped semiconductors that are used in many different devices.
MOFs being crystalline porous materials are no exception.
For this reason, it is exciting to see that defects are now recognized to be similarly useful in tailoring MOFs properties.

The first attempt to gain control over defect formation in MOFs was achieved with the modulation approach.
It is established that small amounts of monocarboxylic acids, or modulator, slow down the speed of crystallization by impacting the equilibrium reaction (the formation of the framework).
In contrast, large modulator concentrations facilitate framework incorporation and in turn the formation of defects.
The first report using this approach was given by {U. Ravon et al.} in 2010, using 2-toluic acid as modulator in the synthesis of MOF-5 \cite{10.1016/j.micromeso.2009.06.008}.
Since then, many research groups focused on the synthesis and characterization of defective MOFs \cite{10.1021/ja405078u,10.1021/ja308786r,10.1016/j.ces.2014.09.047}.
{R. A. Fischer} \cite{10.1002/anie.201311128}, {K. P. Lillerud} \cite{10.1021/acs.chemmater.6b00602} and {A. L. Goodwin} \cite{10.1038/ncomms5176} were one of the first to strongly believe that defects can be exploited to enhance MOFs properties.

In 2014, {R. A. Fischer} and coworkers used the so-called mixed linker approach, where the linker of the parent framework is partially substituted by a linker having one different coordinating group, e.g. 1,3,5-tricarboxylate (\ce{BTC^{3-}}) by pyridine-3,5-dicarboxylate (\ce{pydc^{2−}})  \cite{10.1002/anie.201311128}.
The material showed high activity in the hydrogenation of 1-octene and \ce{CO2} adsorption.

\begin{figure}[tbhp]
\begin{center}
\includegraphics[width=1.0\linewidth]{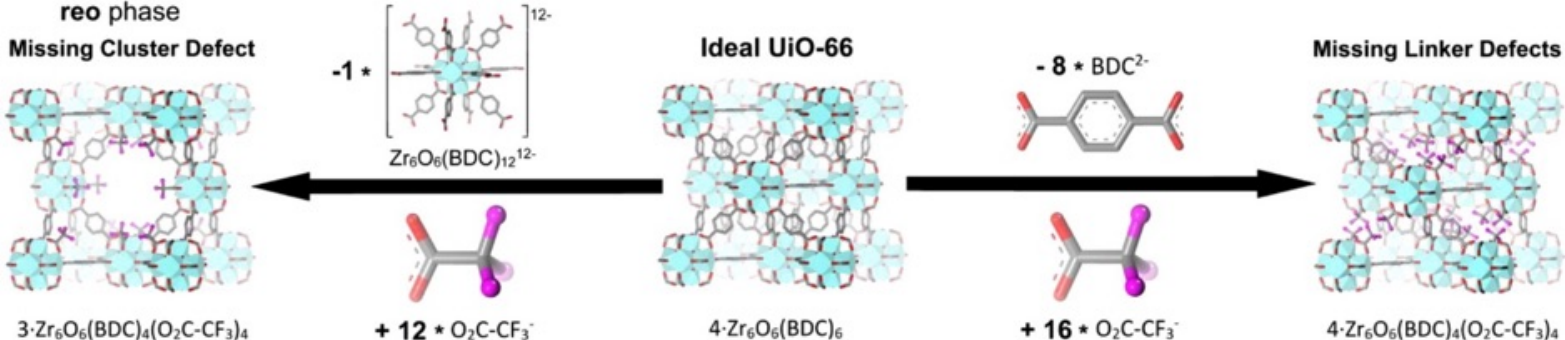}
\caption{Schematic representation of the modulation approach showing the generation of two types of defects, missing linker on the right and missing cluster defects on left. Reprinted with permission from \cite{10.1021/acs.chemmater.6b00602}. Copyright 2016 American Chemical Society.}
\label{fig:defect1}
\end{center}
\end{figure}

In 2016, {K. P. Lillerud} and coworkers systematically studied defective {UiO-66} (Figure~\ref{fig:defect1}) by applying several different characterization techniques such as PXRD, BET (\ce{N2} physisorption), and high-resolution nuclear magnetic resonance (NMR) after digesting the defective MOF \cite{10.1021/acs.chemmater.6b00602}.

In another study, the group of {A. L. Goodwin} achieved a milestone for defect characterization where a combination of several techniques such as (anomalous) powder X-ray diffraction (PXRD) and pair distribution function analysis (PDF) combined with computational modelling was used to access the defect chemistry of {UiO-66} \cite{10.1038/ncomms5176}.
These contributions are just few examples of the many other important studies and it is not surprising that this area of research has gained much attention during the past few years.

In a recent review {R. A. Fischer} and coworkers outlined recent progress in the field \cite{10.1002/adma.201704501}.
One highlighted concept of the review is the intrinsic difficulty to characterize defects (such as the elucidation of defect concentration and spatial distribution).
For instance, diffraction patterns of defective MOFs differ from their respective pristine counterpart by minor changes which are difficult to detect by a laboratory X-ray diffraction due to the instrument resolution limitations.
Therefore, experimental techniques like X-ray absorption fine structure (EXAFS) which can probe the local structure should be used.
This technique requires the use of synchrotron sources which are not readily accessible; therefore, the structural insight into defects in MOFs is limited.

Further insight from computational modeling of defective MOFs has appeared since 2015.
So far the systems of choice are mainly {UiO-66} \cite{10.1039/C5DT04330A,10.1021/acs.chemmater.6b01956} and its derivatives and some other examples HKUST-1 \cite{10.1002/ejic.201600566}.
Nevertheless, useful insight on how defects may affect MOFs mechanical properties have been achieved.
In particular, the group of {V. Van Speybroeck} provided a thermodynamic characterization of the high-pressure behaviour of UiO-66 as a function of missing linker defects and linker expansion in the absence of guests.
Indeed, for the defect-containing and/or expanded linker samples, a reduced mechanical stability is observed \cite{10.1021/acs.chemmater.6b01956}.

It is clear that the in depth characterization of defective MOFs bear many problems which equally challenges experimental and computational chemists.
Presently, the community has demonstrated many properties which are closely linked to the defect concentration and chemistry.
However, the defect structure (including local and correlated defects) in MOFs is barely investigated.
Recently some unconventional characterization techniques were used in an attempt to solve this problem.
Positron annihilation lifetime spectroscopy (PALS) \cite{10.1039/C7DT00350A,10.1002/slct.201601205}, ultra-high vacuum infrared spectroscopy (UHV-IR) with probe molecules \cite{10.1002/chem.201602641}, electron paramagnetic resonance (EPR) \cite{10.1039/C5TA07687K}, acid-base titration \cite{10.1039/C5TA07687K} and water adsorption \cite{10.1039/C4CS00078A} are a few techniques recently applied to MOFs.

\begin{figure}[tbhp]
\begin{center}
\includegraphics[width=0.6\linewidth]{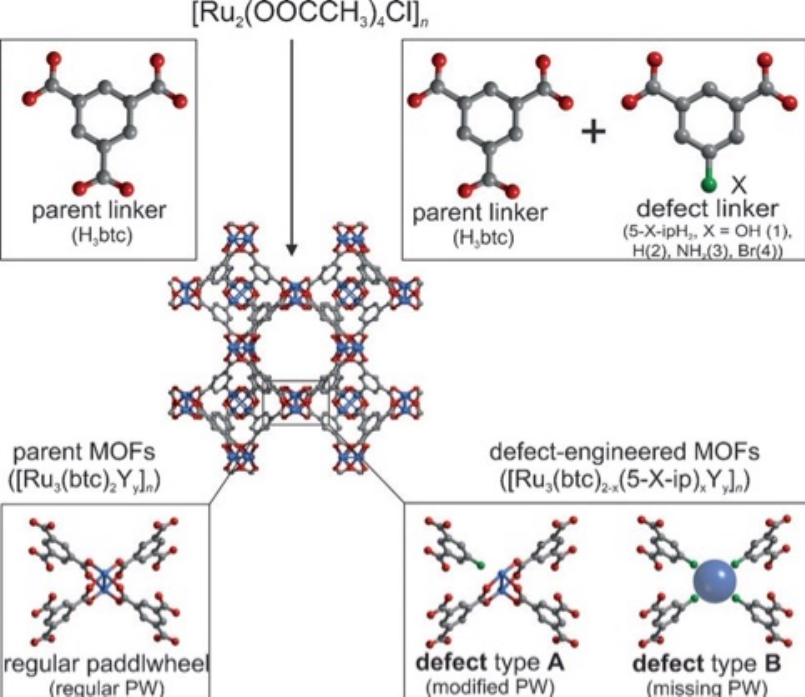}
\caption{Schematic description of the mixed linker approach used to produce type A and B defects in HKUST-1(Ru). Reproduced from \cite{10.1002/chem.201602641}, with permission.}
\label{fig:defect2}
\end{center}
\end{figure}

One attempt to obtain a complete and clear picture of defects which are present was made by {R. A. Fischer} and coworkers in 2016. The authors combined a set of spectroscopic characterization data (XANES, XPS, UHV-FTIR with \ce{CO} and \ce{CO2}) to solve a complicated picture of the defects present in {HKUST-1(Ru)} \cite{10.1002/chem.201602641}.
In their work the authors identified two types of defects obtained with the use mixed-linker approach (Figure~\ref{fig:defect2}): Type A which corresponds to a structural and electronically modified Ru paddlewheel and Type B which corresponds to a complete missing paddlewheel defect.
In type A defects, one out of the four bridging carboxylate groups of the parent 1,3,5-tricarboxylate (BTC) linker in the regular paddlewheels was partly substituted by the (neutral) functional group of the defect linker (H, OH, \ce{NH2} and Br).
Consequently, for charge compensation, the metal sites can be partly reduced (mixed-valent state $0<\delta<2$).
According to XANES and XPS results the application of defect linkers with coordinatively inactive functional groups like H is likely to induce missing-node type B defects which are, accompanied by lower abundance of \ce{Ru^{2+}} and \ce{Ru^{\delta+}} centers in the framework.
On the contrary, the material with coordinatively active functional groups like OH, \ce{NH2} and Br can induce preferentially defects of type A, which are accompanied by an increase of the \ce{Ru^{\delta+}} centers, especially at moderate degrees of incorporation.
However, the authors could not exclude the simultaneous generation of both defects in these samples.
Indeed, {UHV-IR} with CO as a  probe molecule confirmed the presence of both sites, in particular at high concentration of defects for the samples that have linkers with coordinatively active functional groups.

A new fascinating technique that was recently used to characterize defects is PALS.
This method can detect the release of gamma rays that are produced by the interaction of Positrons (Ps) (the antiparticles of electron) with electrons.
In a MOF, the main electron density is located at the framework, coupling the Ps lifetime to the pore size.
Already in 2010 PALS was used by {M. Liu} et al. for this purpose \cite{10.1002/adma.200903618}.
However, in 2016 {S. S. Mondal} et al. \cite{10.1002/slct.201601205} applied PALS to study mesopore formation in hydrogen-bonded imidazolate framework ({HIF-3}).
The authors assigned missing building blocks in {HIF-3} as reason for mesopores which are responsible for structural flexibility during gas uptake.

In a more recent work the authors were able to detect the small differences in pore-sizes of imidazolate framework Potsdam (IFP) structures and variations of gas adsorptions properties, between microwave-assisted (MW) materials and conventional electrical (CE) heating conditions \cite{10.1039/C7DT00350A}.
In particular MW-synthesized materials, due to the fast crystallization process, free linker molecules can be trapped and therefore, reduce the pore sizes from micropores to ultramicropores.
According to the authors these ultramicropores are responsible for the enhancement in the gas uptake capacities since the gas molecules could interact more efficiently with the pore walls \cite{10.1039/C7DT00350A}.
This example demonstrated that PALS has some inherent advantages over common \ce{N2} physisorption measurements.
Most importantly, interconnected pore space is not required to determine pore size from PALS.
Therefore, this method can be applied for open and closed-pore systems.

EPR is also a useful technique to access the chemistry of defects in MOFs due to its high sensitivity.
{A. Pöppl} and coworkers used this technique to precisely characterize the ligand environment of \ce{Ni^{2+}} ions in flexible DUT-8(Ni) \cite{10.1021/acs.jpcc.6b04984}.
In their work they used the small nitric oxide (NO) as a magnetic probe to investigate the adsorption properties as well as the local adsorption sites of NO in DUT-8(Ni).
The NO molecule is accessible by EPR since it has one unpaired electron in an antibonding molecular $^{2}\Pi$ state.
In particular, the lowest rotational level of the $^{2}\Pi_{3/2}$ state of the free NO molecule allows the detection of even small amounts of desorbed NO gas in adsorption experiments \cite{10.1021/acs.jpcc.6b04984}.
The authors were able to distinguish up to five different signals coming from defective site of the flexible and rigid material which can be attributed to NO molecules forming paramagnetic complexes with the \ce{Ni^{2+}} ions.
These might be defective species since in principle defect free {DUT-8(Ni)} offers no open coordination sites for NO.
Furthermore, the density of these species is one order of magnitude higher in the rigid material than in the flexible material.
In particular, only for the rigid defective {DUT-8(Ni)} material two \ce{Ni^{2+}-NO} adsorption species were observed.
For these two species, the unpaired electron resides in a \ce{dz^2} orbital of the \ce{Ni^{2+}} ion instead of the \ce{dx^2-y^2} orbital, where it resides for the flexible material.
A model where one NO molecule bonds in the equatorial plane of a defective paddlewheel unit to the \ce{Ni^{2+}} ion can explain the two species.
This implies that at least one NDC linker (2,6-naphthalenedicarboxylate) does not coordinate to these units.
The absence of some linkers in the rigid DUT-8(Ni) might decrease the total attractive force between the ligands originating from $\pi$−$\pi$ stacking.
This could also explain why rigid {DUT-8(Ni)} stays porous even in the absence of adsorbates \cite{10.1021/acs.jpcc.6b04984}.

Another useful technique to access the chemistry of defects is water adsorption measurements.
As described by {J. Canivet} et al. \cite{10.1039/C4CS00078A}, this technique can be rationalized using a simple set of parameters: the Henry constant (which is the slope of the adsorption pressure in the low pressure range), the pressure at which pore filling occurs, and the maximum water adsorption capacity.
The first two parameters are correlated, both containing information related to the hydrophilicity of the material.
These parameters, as shown by S. Dissegna et al., can be used to access the chemistry of coordinatively unsaturated sites in {UiO-66} \cite{10.1039/C7CE00224F}.
As expected, an increase of defective sites in {UiO-66} leads to an increase in both hydrophilicity and the maximum water capacity.
Both were then correlated with an increased activity for a Lewis acid catalyzed reaction.

The techniques described briefly in this section are selected to give the reader an overview of the rapid evolving field in the defect characterization by a number of German and international groups.
It is apparent, that to obtain in depth structural information of defects, standard analysis and unconventional techniques for MOFs, (such as PALS and EPR) should be combined in order to reach a higher level of accuracy.

\subsection{Nuclear Magnetic Resonance}

Nuclear magnetic resonance (NMR) has proven to be a powerful tool in characterizing adsorption of porous materials.
Two types of NMR techniques are popularly used for extracting required information about the MOF frameworks \cite{10.1002/9783527693078.ch20}.
The first method uses regular solid state NMR, where chemical shifts of the nuclei of interest are used to determine the chemical environment of constituents within the MOF lattice.
The technique of magic angle spinning (MAS) is used generally to remove the spectral broadening otherwise observed for solid samples.
Further enhancement of the spectral sensitivity is gained from application of hyperpolarization techniques, such as cross polarization (CP) and dynamic nuclear polarization (DNP).
Such solid state NMR spectra provide crucial information about the linker conformation (mostly using \ce{^{13}C}) and also  possibly the metal cluster (\ce{^{67}Zn}, \ce{^{27}Al}, etc.).

\begin{figure}[tbhp]
\begin{center}
\includegraphics[width=0.5\linewidth]{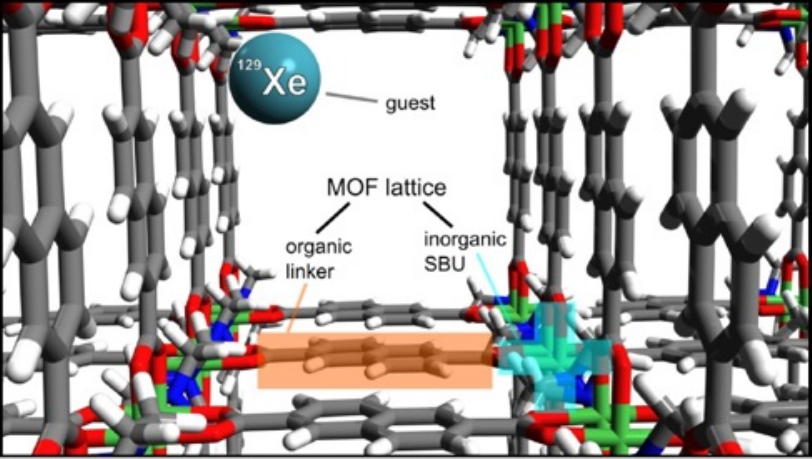}
\caption{Components of MOF structure and interaction with Xe atom to generate shift in NMR spectroscopy. Reproduced from \cite{10.3390/ma5122537}.}
\label{fig:nmr1}
\end{center}
\end{figure}

The second and more beneficial application of NMR spectroscopy involves the use of NMR active guest molecules (like Xe, \ce{CO2}, etc.) within the MOF pores.
Chemical signal of the active nuclei differ significantly with interaction of the environment, thus presenting a description of the local surroundings of the guest molecules (Figure~\ref{fig:nmr1}).
The location of active gases into the pores and subsequently the pore environment can be easily detected by monitoring the change in its chemical shift.
Subsequently investigation by the group of {E. Brunner} using Xe adsorption in MOFs has provided a deep insight to understand pore geometry, pore filling mechanism and most importantly kinetic features, such as the breathing of flexible MOFs.
The basic equation for \ce{^{129}Xe} chemical shift from MOF pores is constructed from the addition of multiple factors: basic reference shift ($\delta_0$), shift from interaction with the pore ($\delta_{\mathrm{S}}$), shift contributed from paramagnetic sites ($\delta_{\mathrm{M}}$), shift contributed from the electric field caused by the metal ions of the framework ($\delta_{\mathrm{E}}$) and moreover, shift from intermolecular collision of the adsorbate molecules ($\delta_{\mathrm{Xe-Xe}}$), Equation~\ref{eqn:nmr1}.

\begin{equation}
\delta = \delta_0 + \delta_{\mathrm{S}} + \delta_{\mathrm{M}} + \delta_{\mathrm{E}} + \delta_{\mathrm{Xe-Xe}}
\label{eqn:nmr1}
\end{equation}

Thus the observation of the chemical shift in Xe NMR provides the location and interaction of the Xe molecules with respect to the MOF pores \cite{10.3390/ma5122537}.
Theoretical support to the observed difference in chemical shift for the Xe NMR in different MOFs has been established by K. Trepte et al.
Theoretical models constructed from the experimental observation during NMR spectroscopy has been applied for two MOFs ({UiO-66} and {UiO-67}) with varying pore sizes \cite{10.1039/c7cp00852j}.
It has been evidenced that at a given pressure for larger pore sizes the inserted Xe molecules have less interaction with the surface, thereby lowering the chemical shift.
The predicted model has potential for predicting the chemical shift for similar scenario.

Application of Xe NMR to study the pore environment of \ce{Cu_3(BTC)_2(H_2O)_3} (Cu-BTC) MOF was reported by {W. Böhlmann} et al. \cite{10.1021/jp063074r}.
The presence of two different peaks confirmed the presence of two types of pores within the framework.
This concludes with lower value of $\delta_{\mathrm{S}}$ for larger sized pores, thus a lower chemical shift indicated adsorption into the larger pore of Cu-BTC framework.
The chemical shift was also found to vary in a non-linear fashion with change in Xe pressure.
Furthermore, co-adsorption of Xe with other species like water and ethylene provides an insight into the pore filling mechanism for such case.
Thus, the smaller pores of Cu-BTC frameworks were observed to be adsorbed with water and ethylene, preferentially over Xe.
The Xe NMR for the ethylene coadsorption also revealed a strong interaction between the gas and \ce{Cu^{2+}} sites.

Application of Xe NMR for studying the kinetic behavior of flexible MOFs has been extensively investigated by the collaboration of the {S. Kaskel} and {E. Brunner} groups.
The well-studied breathing behavior of the DUT-8(Ni) has been examined using Xe NMR to measure and accurately deduce the gate opening pressure \cite{10.1039/C003835K,10.1021/ja201951t}.
In addition to the broad peak from adsorbed Xe into the pore, another narrow peak at lower chemical shift was also observed, which was attributed to the Xe molecules trapped inside the void between MOF particles.
Retention of the peak intensity corresponding to the adsorbed \ce{N2} revealed the complete filling of the pore, in correlation with the adsorption isotherm.
Isotherms collected at different temperature further demonstrated the kinetic behavior, with pronounced hysteresis as observed in the adsorption experiment.
Precise determination of the gate opening pressure, as observed from the NMR analysis further supported the observations from other experimental techniques.

A thorough investigation of the Xe NMR reveals an absence of Xe peak prior to the gate opening pressure for DUT-8(Ni) \cite{10.1021/ja201951t}, with good correlation to both adsorption isotherm as well as XRD pattern.
The pore opening, accompanied with a color change was observed beyond 12 bar, through the gradual appearance of a peak in the desired region.
The observed chemical shift remained pressure independent beyond the gate opening pressure, indicating a complete saturation of the pores.
Presence of the peak corresponding to Xe residing outside the MOF pore shows liquid like behavior under pressure.
These Xe atoms can undergo rapid exchange with the adsorbed atoms within a very short time span (in tens of milliseconds), characterized through cross-peaks at a mixing time of \SI{25}{\milli\second} in the 2D \ce{^{129}Xe} EXSY spectrum.

\begin{figure}[tbhp]
\begin{center}
\includegraphics[width=0.9\linewidth]{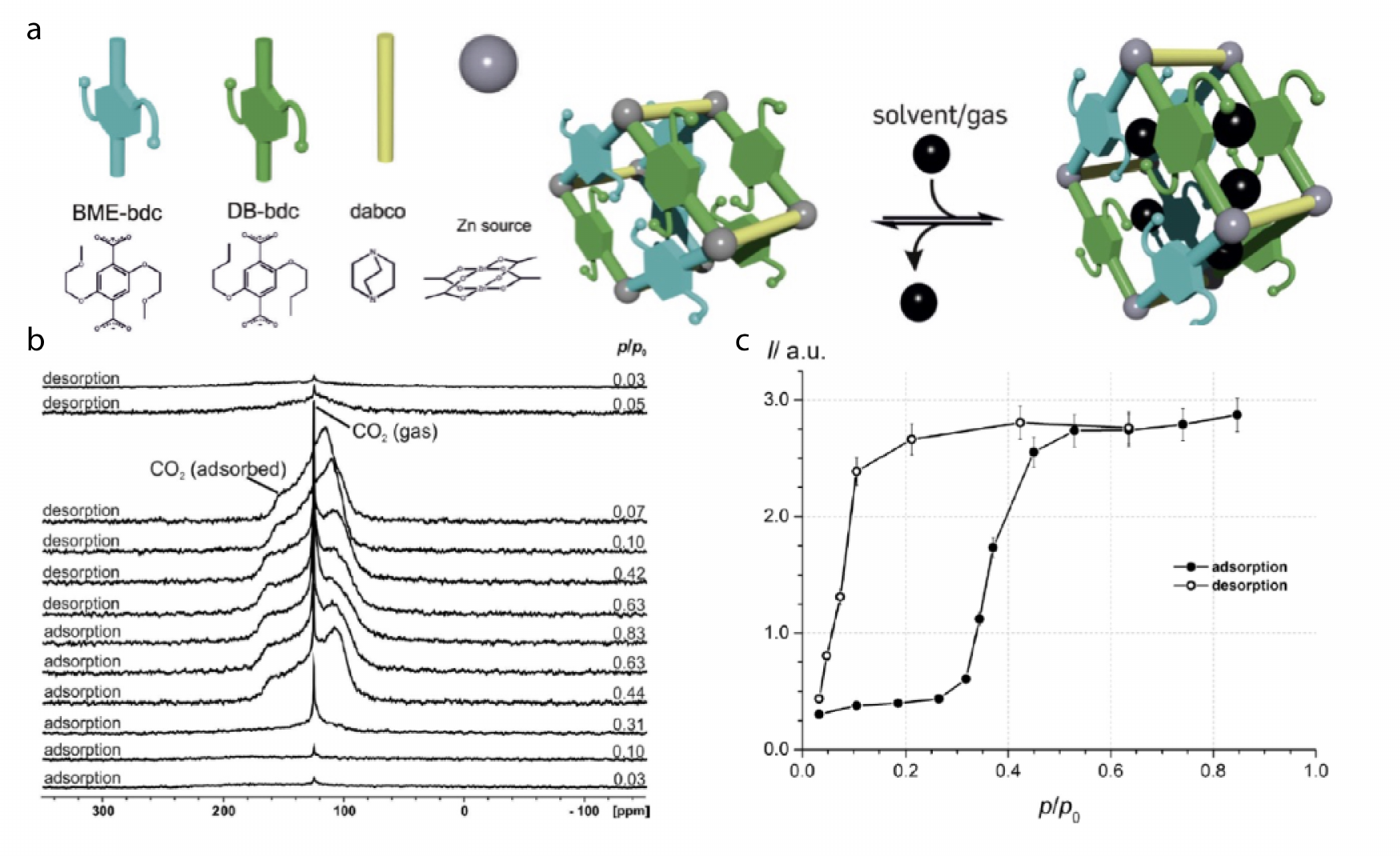}
\caption{Different components their chemical compositions involved for formation of the flexible MOF and its pore opening in present of guest. b) $^{13}$C solid state NMR of the MOF in presence of different relative pressure of \ce{CO2} and c) its plot. Adapted from \cite{10.1016/j.micromeso.2015.02.042}, with permission.}
\label{fig:nmr2}
\end{center}
\end{figure}

Further studies for the breathing behavior were reported for a series of MOFs containing pendant groups, and their solid solutions \cite{10.1016/j.micromeso.2015.02.042}.
Application of in-situ \ce{^{13}C} NMR spectroscopy to \ce{CO2} adsorption in MOFs provide an insight to the process of gate opening (Figure~\ref{fig:nmr2}).
The narrow pore (np) form of the synthesized and activated MOF allows for few \ce{CO2} molecules to be adsorbed within the pores, thereby keeping all the introduced \ce{CO2} outside and a single signal as that of Xe NMR.
On application of the threshold pressure required for phase transition and subsequent opening of the pores, the MOF switches to the large pore (\textit{lp}) form. Conversion into this \textit{lp} form allows more and more \ce{CO2} molecules to be adsorbed, which causes a broadening of the peak.
This chemical shift of the MOF with \ce{CO2} filled pores remains constant indicating a complete filling of the pores, immediate to the breathing phenomenon.
The characteristic desorption profile is clearly observed from the NMR as well, accounting for the hysteresis step.

The unique feature of the recently developed flexible MOF (DUT-49) with negative gas adsorption (NGA) property has also been studied by {E. Brunner et al.} using in-situ Xe NMR \cite{10.1021/acs.jpcc.7b01204}.
The phenomenon of NGA is unique with a feature of pressure amplification in the surrounding atmosphere.
As discussed further in the flexible MOF section, the structural deformation of the carbazole biphenyl linker causes shrinkage in the unit cell of {DUT-49} structure and release of the excess gas molecules resulting in NGA.
Such a unique feature has made DUT-49 framework of acute interest and its study through in-situ NMR provides excellent insight.
At higher temperature, like 273 K, no flexibility was observed from the Xe isotherm. Similar pattern was observed in the case of observed chemical shifts.
However, adsorption isotherms collected at lower temp (200~K) shows the signature change of NGA and the same was observed from drastic change in the difference of chemical shift.
As the open pore (\textit{op}) phase containing the mesopores of the framework undergoes contraction to generate the closed pore (\textit{cp}) phase with smaller pores, the interaction of the residual Xe atoms with the pore walls increases.
This enhanced interaction thus causes a higher chemical shift in the NMR spectrum.
This change in chemical shift thus can be used to finely locate the conversion of the \textit{cp} phase form to \textit{op} phase form at higher pressure.

 \section{Advanced Function}

 The interdisciplinary approach of German MOF groups targeting specific functions by combining synthesis and characterization, as described earlier in this contribution, has produced many materials with a number of advanced functions.

 \subsection{Gas storage}

The research in the field of adsorptive gas storage was strongly motivated by increasing threat of global warming, depletion of fossil fuels and interest to explore alternative renewable energy resources.
Adsorption technology, in principle, could tackle this problem it two ways: improve the energy storage density for energy reach gases, such as hydrocarbons and hydrogen, capture the greenhouse gases, such as carbon monoxide.
But also storage of noble gases is considered as potential industrial application for MOFs.
MOFs are predestinated for adsorption based applications because of their high specific surface area (as adsorption relies on surface upon which gas molecules can adsorb) as well as tunable nature of surface chemistry.

The high potential of MOFs for gas storage materials was recognized quite early and the first reports for methane adsorption was presented by {S. Kitagawa} and coworkers in 1997--2000 \cite{10.1002/(SICI)1521-3773(19990115)38:1/2<140::AID-ANIE140>3.0.CO;2-9,10.1002/anie.199717251}.
In 2003 {O. M. Yaghi} et al. reported the first hydrogen adsorption measurement in {MOF-5} and {IRMOF-8} \cite{10.1126/science.1083440}.
These pioneering studies initiated great interest in MOFs as physisorption based gas storage materials in the world and in Germany.

\begin{figure}[tbhp]
\begin{center}
\includegraphics[width=0.5\linewidth]{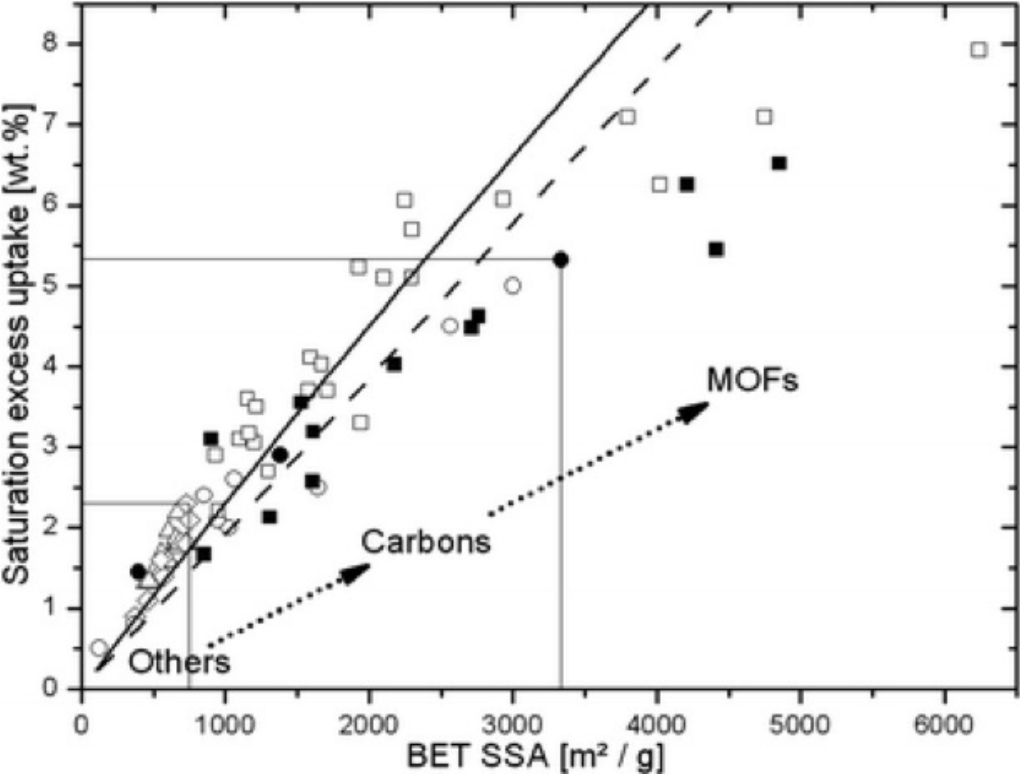}
\caption{Hydrogen storage capacity at 77~K of different materials versus their specific BET specific surface area. The solid line is the hydrogen uptake as predicted from theory, the dashed line is a linear fit to all data points. Reproduced from \cite{10.1039/C2JM15890F} with permission of The Royal Society of Chemistry.}
\label{fig:gasstor1}
\end{center}
\end{figure}

In 2005--2006 {B. Panella} and {M. Hirscher}, based at MPI Stuttgart, presented hydrogen adsorption measurements of MOF-5 and HKUST-1 over a wide range of pressures and at different temperatures ranging from \SI{77}{\kelvin} to \SI{298}{\kelvin} \cite{10.1002/adfm.200500561,10.1002/adma.200400946}.
Further studies were dedicated to the hydrogen storage capacity in other highly porous MOFs, such as MOF-177.
Surprisingly  the hydrogen uptake at 77~K, for all materials (both MOFs, and activated carbons), follows a linear dependence on the surface area with a slope of 1.9 wt\% per \SI{1000}{\square\meter\per\gram} (known as Chahine’s rule, Figure~\ref{fig:gasstor1}) \cite{10.1039/C2JM15890F,10.1002/anie.201006913}.

Isosteric heat of adsorption is an important parameter required to describe the thermal performance of adsorptive storage systems.
It can be calculated from adsorption isotherms measured over wide ranges of pressure and temperature or directly estimated using the coupled calorimetric–volumetric methods.
The {M. Hirscher} group studied the isosteric heats of hydrogen on many prototypical metal–organic frameworks using both methods and find a good agreement between them \cite{10.1007/s00339-015-9484-6}.
{M. Schlichtenmayer} and {M. Hirscher} also determined the storage capacity as a function of operating temperature for the isothermal operation of a storage system.
It shows a maximum at an optimum operating temperature which is at higher temperature for materials with higher enthalpy of adsorption.
The fraction of the total hydrogen stored that can be released at the optimum operating temperature is higher for materials with lower enthalpy of adsorption than for those with higher enthalpy \cite{10.1007/s00339-016-9864-6}.

\begin{figure}[tbhp]
\begin{center}
\includegraphics[width=0.5\linewidth]{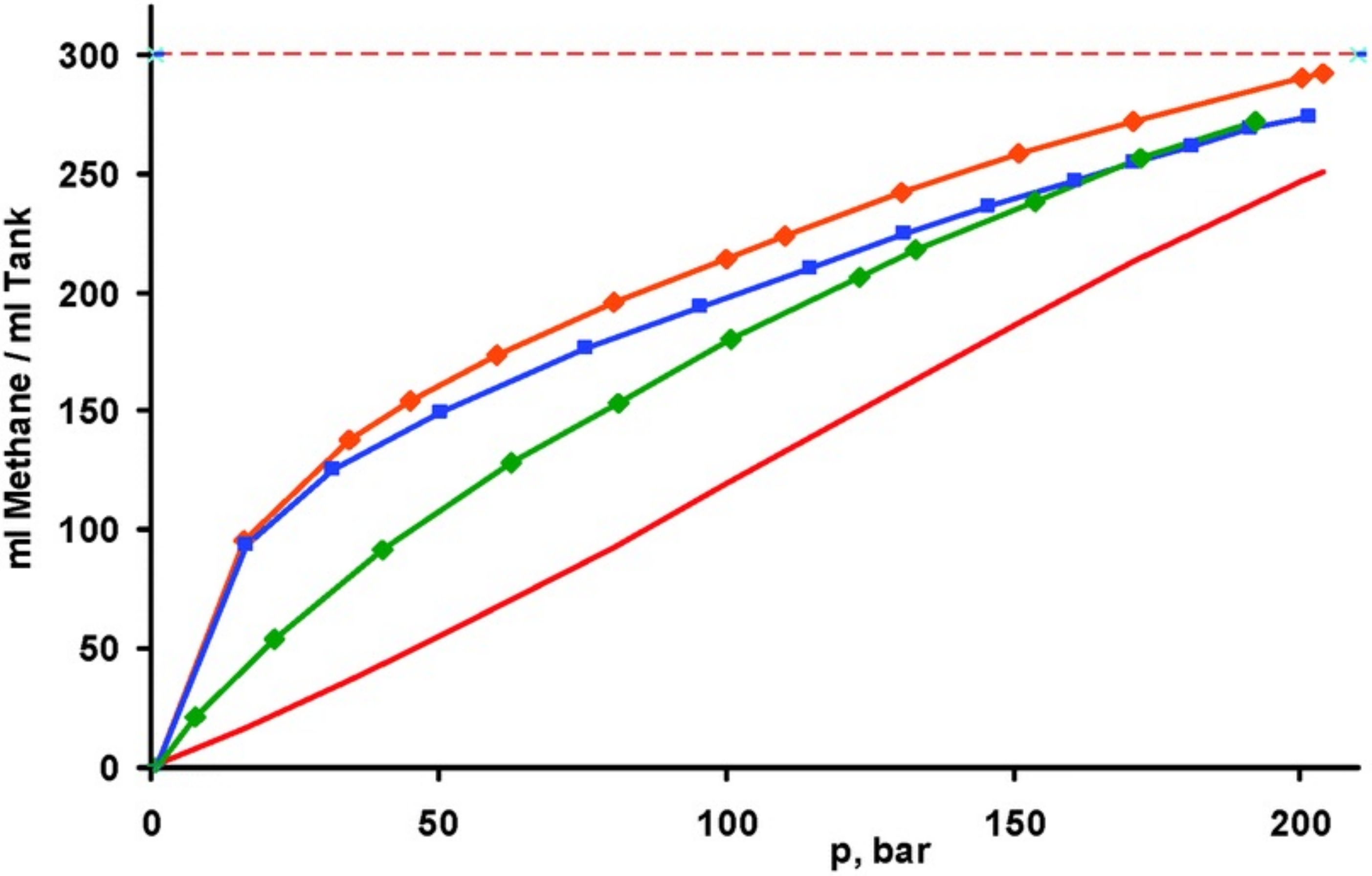}
\caption{Methane desorption isotherms at 293~K for MIL-101(Cr) (green), HKUST-1 (blue), and MIL-101(Cr)–HKUST-1 mixture (orange). The red line represents the amount delivered by a filled empty cylinder. \cite{10.1039/C4CC00524D} - Reproduced by permission of The Royal Society of Chemistry.}
\label{fig:gasstor2}
\end{center}
\end{figure}

In 2008, {I. Senkovska} and {S. Kaskel} reported excess methane storage capacities, and release curves, obtained from a mini-tank prototype operating up to 200~bar at room temperature, for three representative MOFs: HKUST-1, \ce{Zn_2(bdc)_2dabco}, and MIL-101(Cr).
At higher pressure, the specific pore volume dictates the gravimetric storage capacity of a porous material, so mesoporous materials with high pore volume show higher uptake because of higher density of gas compressed inside the pores.
Moreover, to increase the usable capacity of the high pressure tank, where the usable capacity is the amount of gas that can be released at a certain operating temperature between the maximum pressure and the minimum pressure possible for the storage tank \cite{10.1021/ic4024844}, mesoporous materials (such as MIL-101, Figure~\ref{fig:gasstor2}), which adsorb weakly in the low pressure region show better performance.

In the search of such materials with a high pore volume, a plethora of mesoporous MOFs (DUT-6 \cite{10.1002/anie.200904599}, DUT-23 \cite{10.1002/chem.201101383}, DUT-9 \cite{10.1002/anie.201001735}, DUT-32  \cite{10.1039/c4cc00113c}, DUT-49 \cite{10.1039/C2CC34840C}, DUT-76 \cite{10.1039/c4cc07920e}) were reported.
These materials display excellent adsorption characteristics and were developed by the Kaskel group using rational design strategy in 2009--2015.
At the time of the invention, the DUT-23(Co) and DUT-49 were record holders in terms of maximum methane excess adsorption capacity, with \SI{266}{\milli\gram\per\gram} (100~bar, 298~K) and \SI{308}{\milli\gram\per\gram} (110~bar, 298~K), and still belong to the top ten compounds worldwide \cite{10.1016/j.ccr.2017.10.002}.

As previously mentioned, in 2000 the German chemical company BASF started to explore the commercial potential of MOFs.
BASF is focused on commercializing MOF solutions \cite{10.1039/B511962F,10.1039/B804680H} for the transportation industry \cite{10.1016/j.ijhydene.2016.08.153} and is partnering with several technology, development and original equipment manufacturers to pilot the use of MOF materials for enhanced natural gas storage in vehicles in Europe, Asia and North America.
Vehicles equipped with natural gas fuel systems, containing MOF materials, were introduced by BASF in 2013 \cite{10.1002/anie.201410252}.

Not only the gas storage performance but also the understanding of factors influencing the adsorption behavior has been in the focus of research.
Theoretical studies of adsorption properties using grand canonical Monte Carlo (GCMC) simulations and DFT calculations were performed by the {M. Fröba} group \cite{10.1002/cphc.200900459,10.1021/jp1058963,10.1016/j.colsurfa.2009.11.025}, showing that simple force field methods provide a reliable prediction of hydrogen adsorption isotherms in MOFs.
Futher examination of the calculated density fields permits the identification of structural features that create preferential adsorption regions \cite{10.1002/cphc.200900459}.
The same group has also synthesized and measured hydrogen adsorption at 77~K for a number of MOFs, such as PCN-12-Si \cite{10.1021/ic900478z}, and {UHM-6} (UHM: University of Hamburg Materials)  \cite{10.1021/ic201596x}, containing silylene groups.

The group of {H. Krautscheid} studied the adsorption properties of series of isomorphous MOFs in close collaboration with {J. Möllmer} and {R. Staudt} \cite{10.1016/j.micromeso.2015.04.036,10.1021/acs.inorgchem.5b02921}.
For example, Cu paddlewheel based triazolyl isophthalate containing MOFs demonstrated small changes within the MOF structure cause distinct differences in the thermal, adsorptive and catalytic properties, showing the possibility to design the adsorbents with specific gas adsorption properties \cite{10.3390/ma10040338}.

\subsection{Flexible MOFs}

A number of the advanced functions demonstrated by MOFs are caused by dynamic features of the framework structure, such MOFs and similar materials are labelled as ``soft porous crystals'' \cite{10.1038/nchem.444}.
Dynamic features can include flexible moieties extending into the pore network or large volume changes of the framework.
Germany-based researchers have played an important role in the discovery and investigation of flexible MOFs \cite{10.1039/c4cs00101j}.

{S. Kaskel} and coworkers reported the structure and properties of DUT-8 a series of paddlewheel based pillared-layered frameworks containing \ce{Cu^{2+}}, \ce{Zn^{2+}}, \ce{Co^{2+}} and \ce{Ni^{2+}} based metal nodes \cite{10.1039/C2JM15601F,10.1039/C003835K}.
Interestingly, the Cu based framework, DUT-8(Cu) demonstrates rigid behavior with no structural changes
upon solvent removal.
However, DUT-8 frameworks, constructed with Ni, Zn, and Co paddlewheels, indicate structural changes during
evacuation.
Following the adsorption of a number of gases, such as \ce{N2} and Xe, the DUT-8(Ni) structure transforms between closed and open phases, completely, leading to a unique, and large reversible ``breathing'' transition.
The transition in DUT-8(Ni) was investigated using advanced in situ experiments, such as \ce{^{129}Xe} NMR \cite{10.1039/C003835K,10.1021/ja201951t} and in situ PXRD \cite{10.1039/C5CP02180D}, detailed previously.

\begin{figure}[tbhp]
\begin{center}
\includegraphics[]{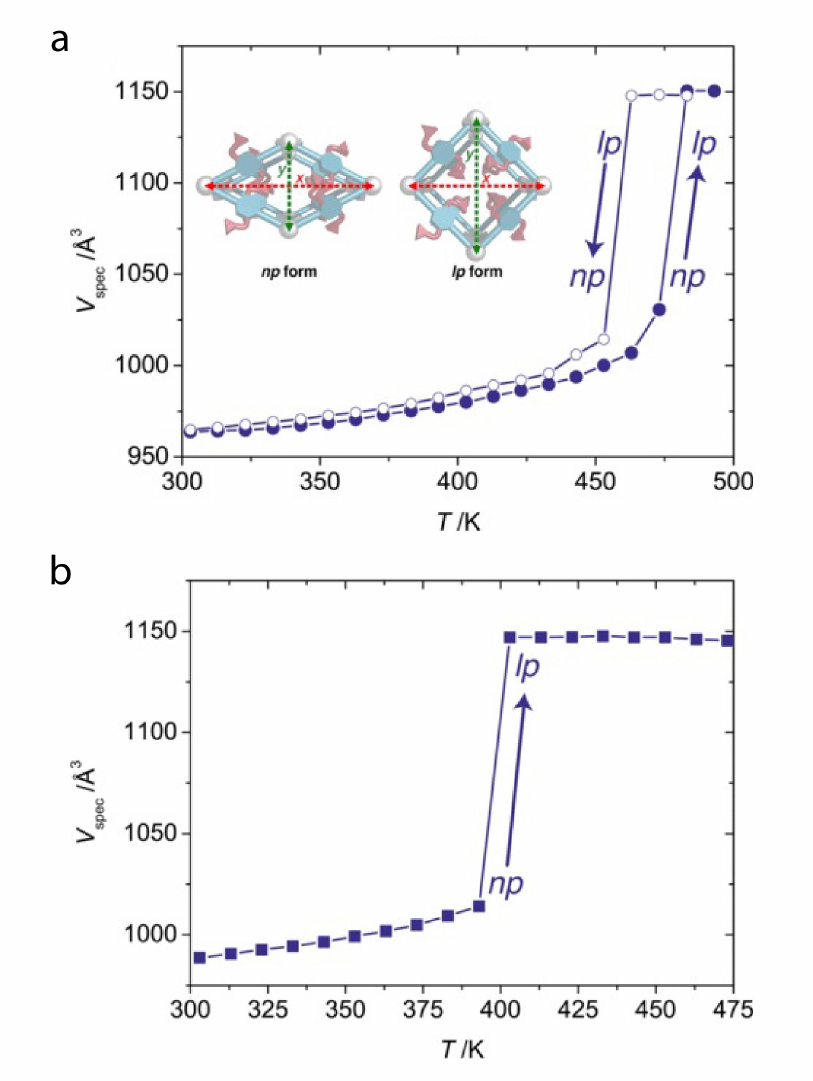}
\caption{Temperature dependence of crystallographic specific volume (Vspec) of fu-MOFs constructed from BME-bdc (a) and DB-bdc (b). Reproduced from \cite{10.1002/adfm.201301256}, with permission.}
\label{fig:flex1}
\end{center}
\end{figure}

\begin{sloppy}
The {R. A. Fischer} group also reported a family of similar pillared-layered frameworks that show a transition between a closed and open phase \cite{10.1002/adfm.201301256}.
The functionalized metal–organic frameworks (fu-MOFs) have the general formula \ce{[Zn_{2}(fu-L)_{2}dabco]_n} and the pores of this material are filled the dangling side chains (L).
In contrast to DUT-8, phase transitions in fu-MOFs can be controlled by temperature.
Upon an increase in temperature, thermal motion of the dangling side chains increase inducing a phase transition from closed pore to open pore, illustrated in Figure~\ref{fig:flex1}.
Interestingly, the temperature of this phase transition is dependent on the type of linker functionalization.
The material based on the ligand BME-bdc (2,5-bis(2-methoxyethoxy)benzenedicarboxylate) shows a transition temperature of \SI{493}{\kelvin} and when exchanged by a larger ligand DB-bdc (2,5-dibutoxybenzenedicarboxylate), a lower transition temperature of \SI{398}{\kelvin} is observed \cite{10.1002/adfm.201301256}.
Materials which contain mixtures of BME-bdc and DB-bdc show transition temperatures between \SI{493}{\kelvin} and \SI{398}{\kelvin}, demonstrating this transition is tunable.
Further studies investigated the influence of chain length, polarity, and grade of saturation of functionalization on the resulting flexibility of fu-MOFs, demonstrating adjustable flexibility of this family of materials \cite{10.1021/ja302991b}.
Recently, {R. A. Fischer} and coworkers also reported different \ce{CO2} induced breathing mechanisms demonstrated by fu-MOF materials constructed with different paddlewheel composition \cite{10.1021/acs.chemmater.7b05052}.
\end{sloppy}

Many of the flexible MOFs reported are chemically fragile thus there has been considerable efforts to produce materials featuring robust building units with similar flexibility.
A flexible, yet very stable, material DUT‐98 was recently reported which features \ce{Zr_6O_4(OH)_4} \rev{SBU}s \cite{10.1002/anie.201702357}.
This material was synthesized using rational supermolecular building block approach to produce a one-dimensional pore architecture.
DUT-98 shows reversible structural transformations upon gas adsorption of specific gases and vapors (\ce{N2}, \ce{CO2}, \textit{n}-butane and alcohols) at characteristic pressures.
This was investigated using in situ PXRD experiments to elucidate the multiple steps in the adsorption isotherm and hysteretic behavior upon desorption.

\begin{figure}[tbhp]
\begin{center}
\includegraphics[width=0.5\linewidth]{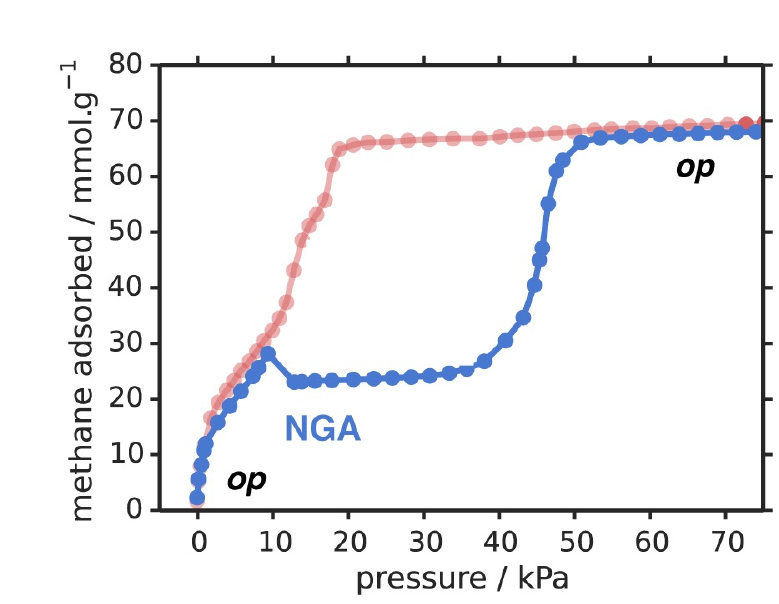}
\caption{The experimental isotherm for DUT-49 for methane adsorption at \SI{111}{\kelvin} and the type 1 isotherm, calculated by simulation, for the static DUT-49 structure (blue and red, respectively). Adapted from \cite{10.1016/j.chempr.2016.11.004}, with permission from Elsevier.}
\label{fig:flex2}
\end{center}
\end{figure}

In 2016, the {S. Kaskel} group reported the behavior of the flexible MOF DUT-49, a material which shows an unexpected negative gas adsorption (NGA) phenomenon \cite{10.1038/nature17430}, Figure~\ref{fig:flex2}.
As DUT-49 fills with gas, it suddenly contracts and releases a significant portion of gas.
This negative gas adsorption process is extremely unique and was not observed in porous materials previously.
Using a combination of advanced in situ experiments this processes was discovered and explained with respect to the disparate relative enthalpies of adsorption of the two phases.

Finally, while this review has focused on the research of experimental groups the work of German simulation groups has provided valuable insight into flexible MOF materials.
The {T. Heine} group has used Born–Oppenheimer molecular dynamics simulations to investigate the low-frequency lattice vibrational modes of DUT-8(Ni) \cite{10.1039/C7CP06225G} and illustrated that they correspond to the oscillation of the breathing mode.
This group has also reported an extension to the universal forcefield \cite{10.1021/ja00051a040} for treatment of MOFs structures that subsequently has been used to describe the flexibility present in these systems \cite{10.1021/ct400952t}.
The group of {R. Schmid} at Ruhr-University Bochum has also reported a forcefield constructed from first principles to describe the dynamics of MOF structures (MOF-FF) \cite{10.1002/pssb.201248460}.
This forcefield has played an important role in elucidating the microscopic mechanism of NGA in DUT-49 \cite{10.1016/j.chempr.2016.11.004}.

\subsection{Membranes and mixed-matrix membranes}

In industrial chemical processes membranes are frequently used, where membranes refer to thin layers of a material with a resistance and selectivity to the passage of different substances thus membranes enable the separation of substances \cite{10.1016/0376-7388(96)82861-4}
Important examples for the application of membranes include the purification of gases, seawater desalination, the purification of sewage and fuel vapor recovery \cite{10.1002/adem.200600032}.
Membrane separation processes are usually advantageous to conventional separation methods, such as distillation, crystallization, adsorption or absorption, as membrane separation processes incur lower costs, lower energy consumption and have simpler process conditions \cite{10.1016/S0376-7388(00)00418-X}.
Notably, energy savings of up to 50\% could be reached with membrane separation processes over other separation technologies \cite{10.1080/01496399308019500,10.3390/membranes2040706}.
\rev{While not detailed here MOF technologies have also demonstrated potential for the separation of industrially relevant liquid mixtures, such as hydrocarbon mixtures \cite{10.1016/j.ccr.2018.04.001}.}

Organic polymers are often the material of choice for commercial membranes since they are inexpensive, easy to manufacture, flexible and sufficiently mechanically stable.
Yet, a major disadvantage of polymeric membranes is their inverse relation, that is, a trade-off between gas permeation and selectivity.  Permeability and selectivity (permselectivity) are the most important membrane parameters that determine the economics of separation processes \cite{10.1016/0376-7388(91)80060-J}.
Permeability ($P$) is the gas flow rate multiplied by the thickness of the membrane, divided by the area and by the pressure difference across the membrane.
Selectivity $S$ is the separation factor for gas mixtures and is calculated from mole fractions of the components produced in the permeate, divided by corresponding mole fractions in the feed \cite{10.1039/C2DT31550E}.
Ideally a membrane should have both a high permeability and a high selectivity.
A low permeability necessitates a larger membrane area or more membrane modules for the given gas volume and time.
A low selectivity requires more steps in the separation with more complex operations, typically leading to higher costs.
Large volume separations, such as \ce{O2}/\ce{N2} for oxy-combustion or \ce{CO2}/\ce{CH4} for natural gas treatments require highly permeable membranes.

High permeability is often the aim in membrane development as optimum membrane selectivity depends on the process and the operating conditions, especially the pressure ratio.
The electricity cost for the compressor to build up the pressure is usually the largest operating expense so an affordably low pressure ratio determines the preferred membrane selectivity \cite{10.1016/j.memsci.2014.03.016}.
Calculations of the optimal design for a membrane post-combustion \ce{CO2} capture process showed a trade-off between membrane area and permeate \ce{CO2} concentration for a given pressure ratio, which results in a narrow optimum selectivity range \cite{10.1016/j.memsci.2009.10.041}.

\begin{figure}[tbhp]
\begin{center}
\includegraphics[width=0.5\linewidth]{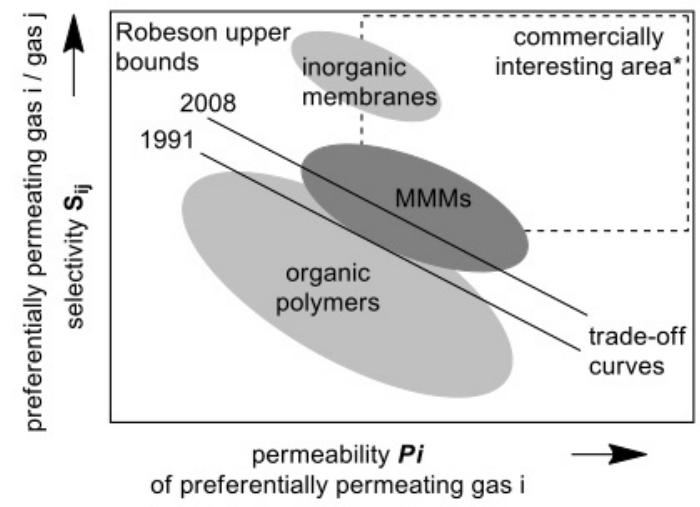}
\caption{Schematic representation of the inverse relation between permeability and selectivity for organic polymer membranes with the 1991 and 2008 Robeson upper bounds\cite{10.1016/0376-7388(91)80060-J,10.1016/j.memsci.2008.04.030}. *The distance or position or the commercially interesting area relative to upper bound depends on the separation problem.}
\label{fig:membr1}
\end{center}
\end{figure}

For polymer membranes high permeability goes together with low selectivity and vice versa.
This inverse relation between permeation and selectivity is illustrated by the Robeson upper bound displayed in Figure~\ref{fig:membr1} \cite{10.1016/0376-7388(91)80060-J,10.1016/j.memsci.2008.04.030}.
A possible way to overcome this trade-off is by using new porous materials or embedding porous fillers as additives into conventional polymer matrices to give mixed-matrix membranes (MMMs) \cite{10.1016/S0376-7388(97)00194-4,10.1039/c1cc13431k,10.1126/science.1069580}.

Zeolitic imidazolate framework (ZIF) based membranes have been studied extensively studied by {J. Caro} and coworkers from Leibniz University Hannover \cite{10.1021/ja907359t}.
These MOF based membranes have been prepared by growing ZIF nanocrystals on a suitable porous support.
The pore window of the ZIF structure thus selectively allows suitably sized gases to pass through, providing molecular sieving.
It has been established that proper orientation of ZIF-8 nanocrystals \cite{10.1021/cm200555s} on the support plays a crucial role in attaining the best separation performance.
Further improvement in separation efficiency has been achieved when the support has been modified with layer double hydroxides (LDH) such as \ce{ZnAl-CO_3} \cite{10.1021/ja507408s} or \ce{ZnAl-NO_3} \cite{10.1002/anie.201411550}.
LDHs provide more sites for heterogeneous nucleation for ZIF-8, resulting in a more intergrown membrane. Additionally, \ce{Zn^{2+}} in the LDH undergoes ``partial self-conversion'' to form a thin continuous layer of ZIF-8 membrane.
Such laminated ZIF-8 membranes have shown high selectivity towards \ce{H2} with a separation factor of 54.1 over \ce{CH4}  \cite{10.1002/anie.201411550}.
Further studies explored the selectivity for \ce{H2} over \ce{CO2} and ZIF-7 was shown to perform better than ZIF-8 \cite{10.1002/anie.200905645,10.1016/j.memsci.2010.02.074}.
The window size of ZIF-8 (\SI{3.4}{\angstrom}) allows few \ce{CO2} molecules (kinetic diameter of \SI{3.3}{\angstrom}) to pass.
However, ZIF-7 has a smaller pore size (\SI{3.0}{\angstrom}) which selectively allows \ce{H2} diffusion over \ce{CO2}, providing an efficient separation (separation factor of 6.5 for ZIF-7 increased from 4.7 for ZIF-8, which follows Knudsen separation behavior).

A second approach using the favorable interaction between the \ce{NH2} groups and \ce{CO2} has also been shown to be useful in reducing \ce{CO2} permeance and thereby enhancing the selectivity for \ce{H2} in the mixture.
Post functionalization of Mg-MOF-74 with \ce{NH2} functionality thus increases the performance of the membrane, demonstrated by an increase in room temperature \ce{H2}/\ce{CO2} selectivity from 10.5 to 28 \cite{10.1016/j.ces.2014.10.037}.

ZIF-22, with similar pore diameter to ZIF-7, also demonstrates a similar enhanced selectivity (separation factor of 7.2) for \ce{H2}/\ce{CO2} \cite{10.1002/anie.201001919}.
Here, ZIF-22 nanocrystals were anchored to the porous \ce{TiO_2} support through chemical functionalization with 3-aminopropyltriethoxysilane (APTES).
The 3-aminopropylsilyl groups from APTES readily coordinate to the available \ce{Zn^{2+}} sites creating a covalent attachment with the support.
This covalent attachment was demonstrated between the amine group of APTES functionality of the modified surface and \ce{-CHO} group of the ZIF linker from ZIF-90 \cite{10.1021/ja108774v}.
The resulting imine condensation thus developed a strong and efficient interaction of the nanocrystal with the alumina support, providing high thermal and hydrothermal stability to the membrane.
However, when APTES functionalization is employed as post-synthetic modification to the pristine membrane of ZIF-90 \cite{10.1002/anie.201204621}, the pore window was narrowed and the defects in the intercrystallite spaces are repaired to produce a less defective membrane.
These changes are observed for the selectivity of \ce{H2} over other tested gases, such as \ce{CO2}, \ce{CH4}, \ce{C2H6} and \ce{C3H8}.

Additionally, the separation of hydrocarbons, such as ethene/ethane has also been achieved using ZIF-8 based membranes \cite{10.1016/j.memsci.2010.12.001}.
Recent reports from {J. Caro} and coworkers has demonstrated the effect of photoisomerization of azobenzene as guests within a membrane based on UiO-67 \cite{10.1021/acs.chemmater.7b00147}.

MMMs are comprised of micro- or nano-sized additives or filler components dispersed within a continuous organic polymer phase.
Various porous or non-porous inorganic additives have been tested as filler materials \cite{10.1039/C2DT31550E,10.1016/j.jiec.2012.09.019,10.1016/j.progpolymsci.2007.01.008,10.1039/c3ta00927k,10.1002/9781119951438.eibc2219,10.1016/j.micromeso.2012.03.012}.
MMMs attempt to combine the excellent flexibility and processability of polymers and the high gas separation performance of other porous materials, such as zeolites \cite{10.1016/S0167-2991(07)80794-4} or MOFs \cite{10.1016/j.ces.2014.10.007},  in order to overcome the Robeson upper bound \cite{10.1016/j.memsci.2008.04.030}.
The addition of fillers to polymer materials can also address other challenges of commercially used polymer materials, like membrane plasticization, low contaminant resistance or biofouling.

MOFs are widely studied as fillers in mixed matrix membranes and compared for their impact on the separation performance of pure polymer materials and such MOF-MMMs have been the subject of recent reviews \cite{10.1039/C2DT31550E,10.1039/c5cs00292c,10.1039/C4CS00437J,10.1021/acs.cgd.7b00595,10.1002/anie.201701109}.
Most reported studies on the preparation and characterization of MOF-based MMMs are still often fundamental and work is still ongoing to develop an understanding of the separation mechanisms, the role of the MOFs and the effect of the polymer/MOF particle interface \cite{10.1039/C7SC04152G}.
This knowledge is needed to develop MOF--MMMs, which are suitable for industrial applications.

\begin{figure}[tbhp]
\begin{center}
\includegraphics[width=1.0\linewidth]{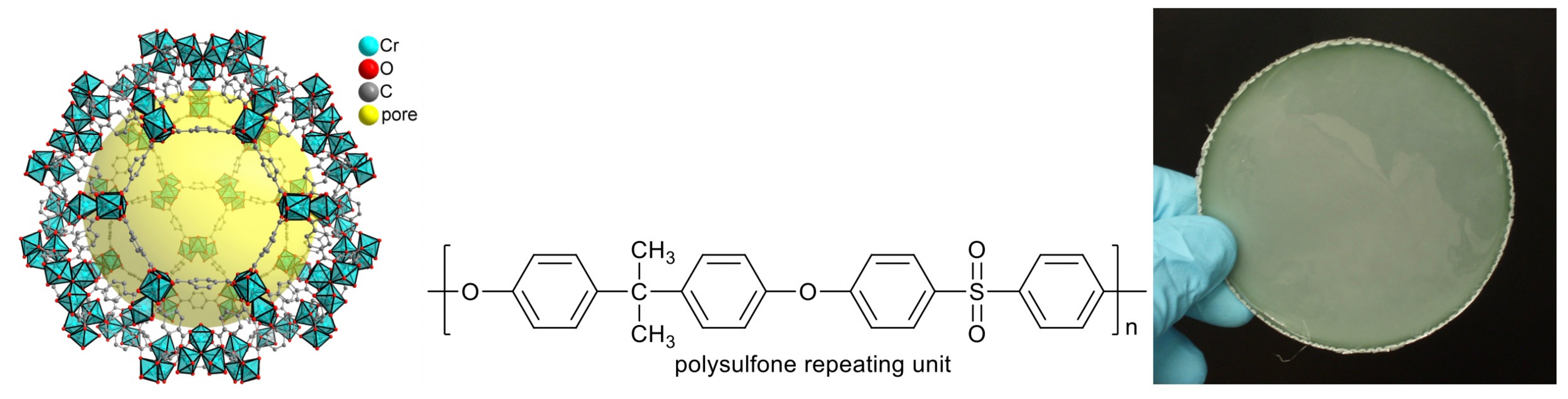}
\caption{Mesoporous cage in MIL-101(Cr) as part of a MTN zeolite network with a pore diameter of ~\SI{34}{\angstrom}, the polysulfone repeating unit and MIL-101-PSF membrane. Adapted from \cite{10.1039/C2CC16628C} with permission from The Royal Society of Chemistry.}
\label{fig:membr3}
\end{center}
\end{figure}

In Germany, the {C. Janiak} group found water-stable MIL-101 microcrystals adhere well to polysulfone (PSF) and yield a very robust mixed-matrix MIL-101-PSF membranes (Figure~\ref{fig:membr3}) \cite{10.1039/C2CC16628C}.
The PSF polymer does not appear impressive in terms of its separation performance.
However, PSF is employed on the scale of several thousand tons per year for membranes for dialysis and water treatment and is one of the most important glassy polymers used in industrial membrane gas separation \cite{10.1021/ie0108088,10.1021/ie8019032}.
PSF is characterized by high temperature stability and good mechanical properties.
Furthermore, it is industrially available with high molecular weights and consistent specifications so that deviations, which are observed from different polymer batches or impurities can therefore be eliminated.

\begin{figure}[tbhp]
\begin{center}
\includegraphics[width=0.5\linewidth]{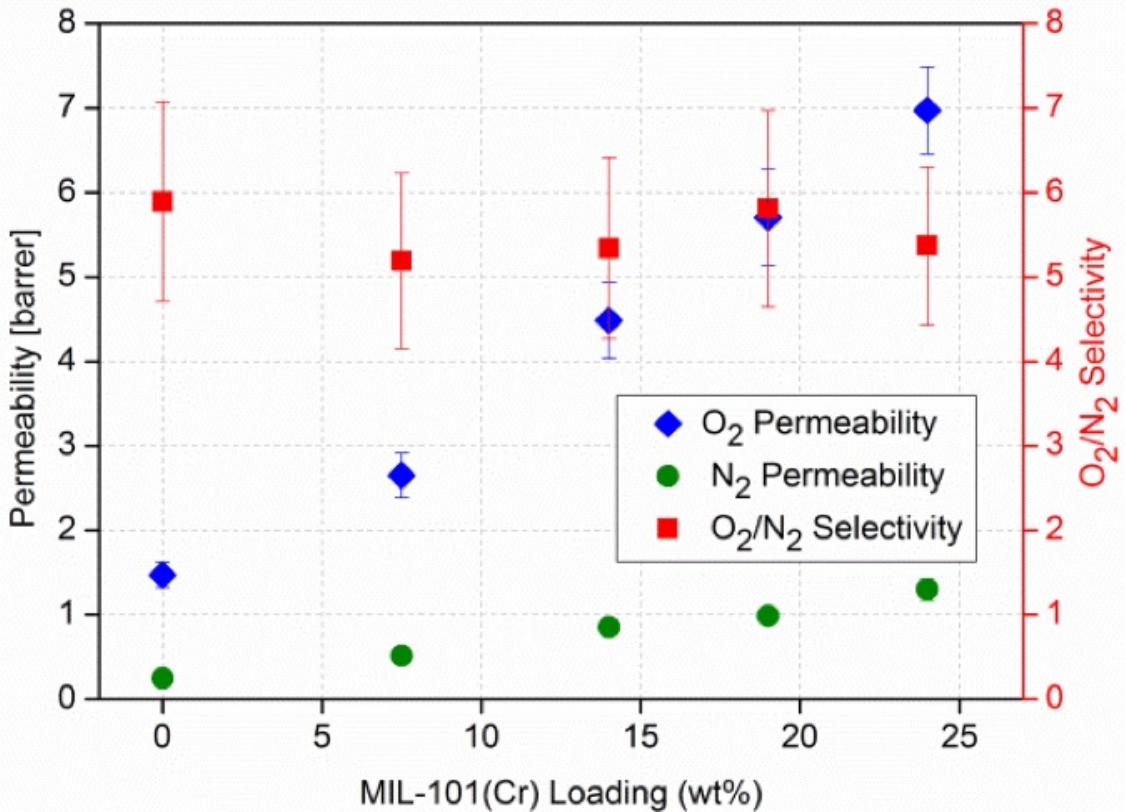}
\caption{\ce{O2}/\ce{N2} permeability and separation performance of pure polysulfone (PSF) and MIL-101-PSF membranes with different {MIL-101} wt\% loadings. Reproduced from \cite{10.3390/membranes3040331}.}
\label{fig:membr5}
\end{center}
\end{figure}

For \ce{O2}/\ce{N2} separation the {MIL-101-PSF} membrane exhibited a remarkable four-fold increase in the permeability of \ce{O2} to required values of above 6~barrer for MIL-101 loadings of 24 wt\%, thereby, retaining the PSF selectivity for \ce{O2} over \ce{N2} (Figure~\ref{fig:membr5}).
Also, the presence of some agglomeration of the MIL-101 particles in the polymer at 24 wt\% loading does not alter these properties.
The essentially unchanged \ce{O2}/\ce{N2} selectivity is consistent with literature results where permeability enhancements for {MOF--MMMs} could be found, but selectivity enhancements were less pronounced.
High loading of up to 24 wt\% of MIL-101 in PSF could be achieved with the MIL-101 particles showing very good adhesion with polysulfone in the MMM and excellent long term stability.
The comparison of these results with a compilation from other MOF-containing MMMs, previously reported, show that the MIL-101-PSF membranes exhibited much higher \ce{O2} permeabilities ($>4$ barrer) than any other MOF-based mixed-matrix membranes \cite{10.1039/C2CC16628C}.

Further single gas (\ce{N2}, \ce{CO2} and \ce{CH4}) permeation tests with {MIL-101-PSF} MMMs showed a significant increase of gas permeabilities of the MMM without any loss in selectivity.
Isochorus, single gas \ce{N2}, \ce{CO2} and \ce{CH4}, permeation experiments of {MIL-101-PSF} membranes show an almost linear increase in permeability for the fastest studied gas \ce{CO2} and at the most slight increases in ideal \ce{CO2}/\ce{N2} and \ce{CO2}/\ce{CH4} selectivities for increasing {MIL-101} content \cite{10.3390/membranes3040331}.

\begin{figure}[tbhp]
\begin{center}
\includegraphics[width=0.9\linewidth]{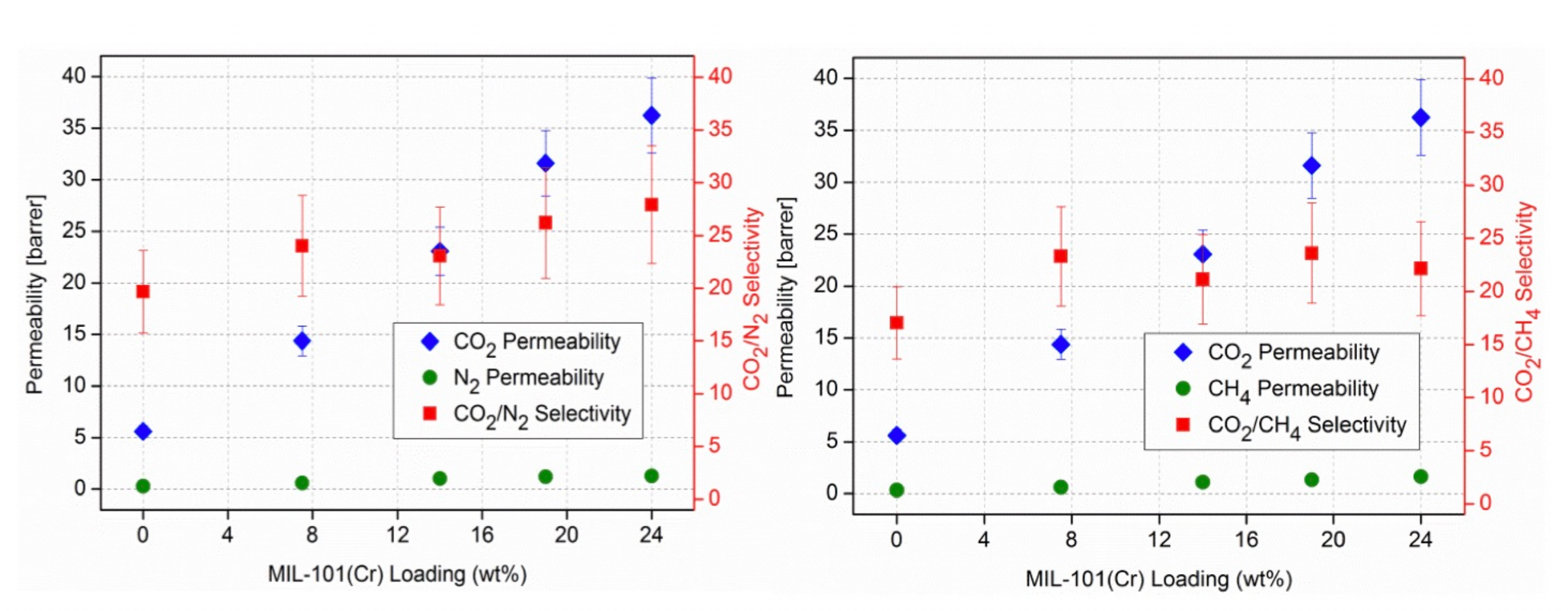}
\caption{\ce{CO2}/\ce{N2} (left) and \ce{CO2}/\ce{CH4} (right) permeability and separation performance of pure PSF and MIL-101/PSF membranes with different MIL wt\% loadings. Reproduced from \cite{10.3390/membranes3040331}.}
\label{fig:membr6}
\end{center}
\end{figure}

Notably, the \ce{CO2}/\ce{N2} separation performance of MIL-101/PSF with 19 wt\% filler was higher than those of most other known {MOF--MMMs} \cite{10.3390/membranes3040331}.
The Maxwell model was able to reproduce approximate effective permeabilities and the ideal selectivity for the \ce{O2}/\ce{N2} and \ce{CO2}/\ce{N2} separation performance of the MIL-101(Cr)/PSF membranes with different filler loadings, under the assumption of a higher gas permeability $P$ in the dispersed MOF phase ($P_d$) than in the continuous polymer phase ($P_c$), that is, $P_d >> P_c$.
Positron annihilation lifetime spectroscopy (PALS) indicated that the increased gas permeability is not due to free volume changes in the PSF polymer but due to the added large free volume inside the {MIL-101} filler particles \cite{10.3390/membranes3040331}.

A number of other MOFs have also been investigated for use in MMMs.
As described previously, ZIF-8 has a pore aperture with a diameter of \SI{3.4}{\angstrom} allowing for preferential adsorption of small molecules \cite{10.1073/pnas.0602439103}.
Combination of ZIF-8 and MIL-101(Cr) in a 1:1 ratio in PSF yielded a MMM with higher selectivity for \ce{CO2} in mixed-gas \ce{CO2}/\ce{CH4} permeation tests for an MMM with combined 16 wt\% MIL-101(Cr)/ZIF-8 (in 1:1 mass ratio, that is 8+8wt\%) compared to analogous 16 wt\% single-filler MMMs (either MIL-101(Cr) or ZIF-8) (Figure 9).
The \ce{CO2}/\ce{CH4} selectivity also increased significantly from 23 for pure PSF to 40 for the 16 wt\% combined MIL-101(Cr)/ZIF-8 membrane \cite{10.1002/ejic.201600190}.
SEM images reveals the breaking of polymer chain packing and linking due to the presence of fillers leading to an increase in polymer matrix free volume which, together with the large porosity of MIL-101(Cr), can explain the permeability improvement in agreement with previous publications related to ZIF-8 containing MMMs \cite{10.1002/cphc.201100583}.

\begin{figure}[tbhp]
\begin{center}
\includegraphics[width=0.5\linewidth]{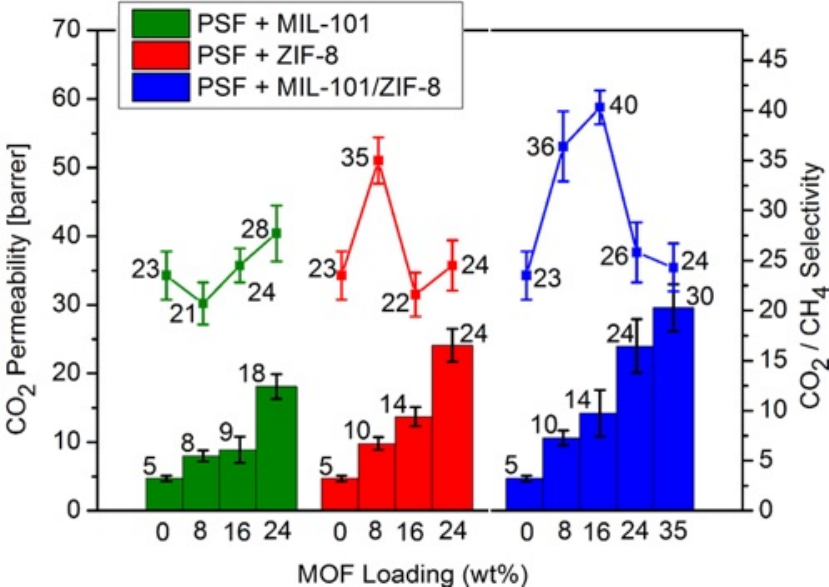}
\caption{\ce{CO2}/\ce{CH4} permeability (given as bottom bars) and mixed-gas separation performance (as top lines) of pure PSF (0 wt\% MOF loading), MIL-101(Cr)@PSF (green), ZIF-8@PSF (red) and combined or mixed-MOF MIL-101(Cr)/ZIF-8@PSF MMMs (blue), for different MOF loadings. The loading of the combined MIL-101(Cr)/ZIF-8@PSF MMMs comes through MIL-101/ZIF-8 in a 1:1 ratio. Good quality membranes with 35 wt\% MOF loading were only possible in the case of the filler combination with 17.5 wt\% MIL-101 (Cr) + 17.5 wt\% ZIF-8. Reproduced from \cite{10.1002/ejic.201600190}, with permission.}
\label{fig:membr7}
\end{center}
\end{figure}

\begin{figure}[tbhp]
\begin{center}
\includegraphics[width=0.7\linewidth]{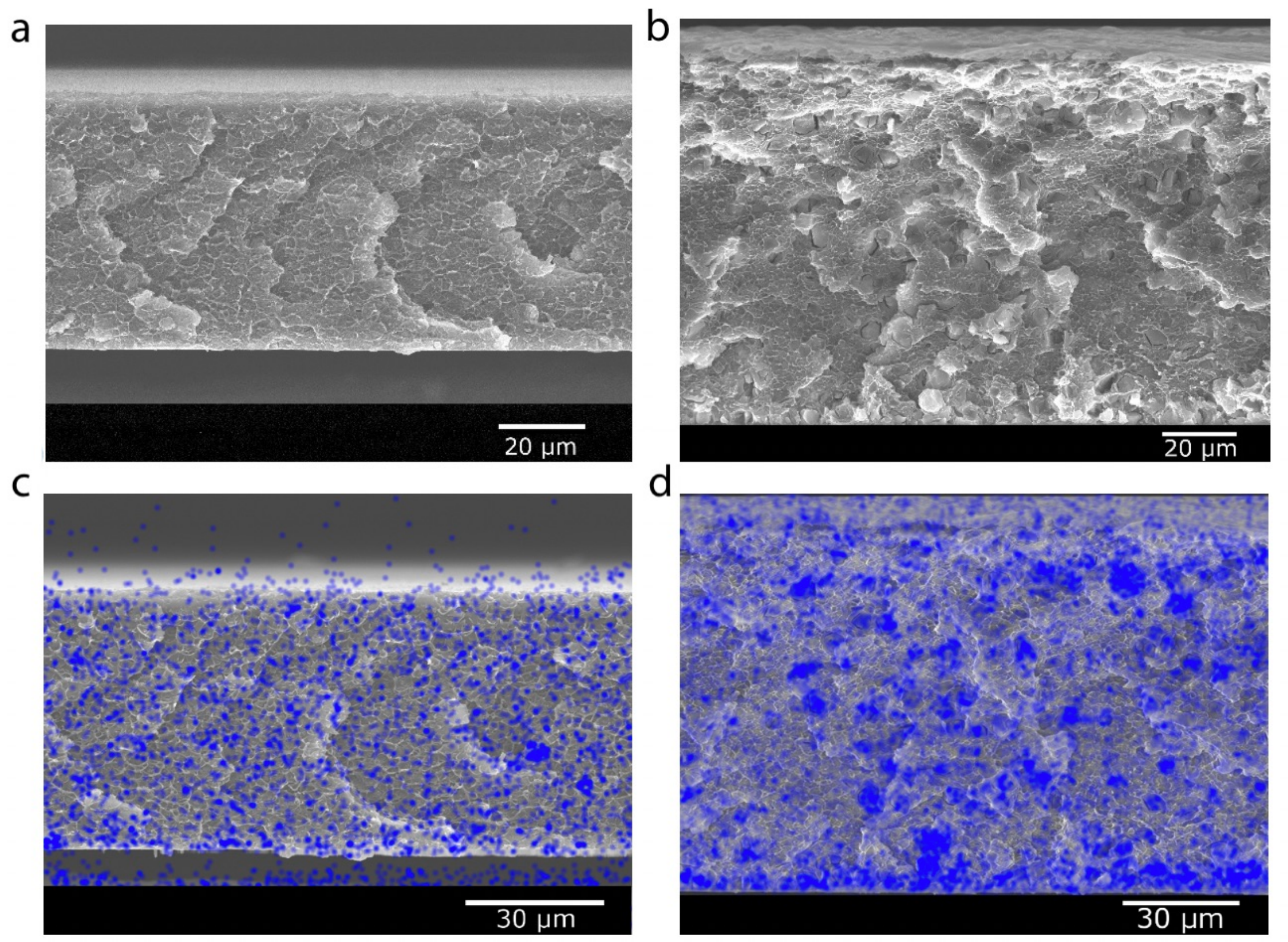}
\caption{SEM images of cross-section of Matrimid with different loadings of \ce{[Co4(\mu_4-O)(Me2pzba)3]} as filler (a) 8 wt\%, (b) 24 wt\%. EDX-mapping of cobalt (blue) in MMM cross-sections (c) 8 wt\%, (d) 24 wt\% \ce{[Co4(\mu_4-O)(Me2pzba)3])}. Some detection of Co is found outside of the membrane cross section due to reflection of the electron beam on the sample holder. The bottom of the images corresponds to the bottom of the membrane when casted. Adapted with permission from \cite{10.1021/acs.cgd.7b00202}. Copyright 2017 American Chemical Society.}
\label{fig:membr8}
\end{center}
\end{figure}

Pure MOF-5 membranes and MOF-5 MMMs with different polymer materials have shown good separation properties for dry \ce{CO2}/\ce{CH4} gas mixtures \cite{10.1016/j.memsci.2008.12.006,10.1016/j.jiec.2013.12.091}.
However, the low water stability of MOF-5 is an insurmountable problem for possible real-world applications.
The decomposed material is no longer porous towards \ce{N2} adsorption, which is seen as a fundamental prerequisite for MOF fillers in MMMs in order to have an effect on the separation process.
An isostructural MOF-5 analogue, 3D-\ce{[Co4(\mu4-O)(Me2pzba)3]} with the bi- or mixed-functional pyrazolate-carboxylate ligand, allowed for the preparation of moisture-tolerant \ce{[Co4(\mu4-O)(Me2pzba)3]}/Matrimid mixed-matrix membranes by the {C. Janiak} group \cite{10.1021/acs.cgd.7b00202}.
Synthesis of 3D-\ce{[Co4(\mu4-O)(Me2pzba)3]} by microwave-assisted heating gave small particle size and less aggregation compared to the conventional solvothermal synthesis.
The control of the particle size and shape of MOF-fillers is crucial for the preparation of MOF MMMs with improved separation properties \cite{10.1002/adfm.201505352}.
The small MOF particles were uniformly embedded in the polymer without aggregation and with good MOF-polymer compatibility as shown through the SEM images in combination with EDX cobalt element mapping (Figure~\ref{fig:membr8}).
The good MOF-polymer compatibility was further affirmed by the improved \ce{CO2}/\ce{CH4} separation performance of the MMMs over neat Matrimid membranes.
This improvement corresponds to more than 46 \% in \ce{CO2}/\ce{CH4} selectivity for 24 wt\% filler loading and 3~bar transmembrane pressure, relative to the pure Matrimid membrane, together with an enhanced permeability of 49 \% for \ce{CO2}.

More investigations and considerable research effort is required for better understanding of the permeation mechanisms and selectivity properties, the roles of filler loading, geometry, pore size and the nature of the molecular interactions on preferential adsorption when preparing single-MOF-MMMs and mixed-MOF-MMMs.
It is evident that the combinations of different types of fillers may lead to substantial improvements in gas separation using MMMs.
The efforts of German researchers and other talented researchers from the rest of the world have produced a number of membrane materials that provide exceptional separation performance.

\subsection{Luminescent sensors}

Tunable emission property of suitable metal--organic frameworks has made them an attractive candidate for selective sensing of hazardous chemicals, such as solvents, ions and biomolecules.
Contribution for this application comprises of the development of suitable MOFs and the processes for real time application. \cite{10.1039/C6CS00930A}
As discussed in several literature reports, the emission property of a MOF arises from 5 major pathways: a) ligand based emission, b) metal based emission, c) metal to ligand charge transfer (MLCT), d) ligand to metal charge transfer (LMCT) and e) metal to metal charge transfer (MMCT) \cite{10.1039/c4cc00848k}.
Of these methods, the ligand and metal based emissive MOFs are most straightforward and can be effectively achieved by choosing suitable components.
Thus, the change in the emission property directly reflects their change from inside the framework, in the presence of an analyte.
A comprehensive review on the application of MOFs for sensing and a global overview has been previously addressed \cite{10.1016/j.micromeso.2015.03.036}.

Using the emissive property of a Zn based MOF ({DUT-25}), the S. Kaskel group have demonstrated a successful sensor to detect different solvents \cite{10.1002/chem.201202352}.
The MOF has \textit{nbo-b} topology, cross linked with second linker, and contains a \ce{[Zn4O(CO2)6]} SBUs.
DUT-25 shows intense emission and the emission intensity of the MOF is observed to decrease in the presence polar solvents such as MeOH and acetonitrile.
Importantly, a sequential bathochromic shift of the emission maxima corresponds to the presence of solvents in increasing polarity.
This distinct change in emission maxima and intensity adds to the appropriateness of DUT-25 for optical sensing, as the material is transparent in nature.
This transparency allowed for visual monitoring as key sensor feature and subsequently a responsive molecule (Nile Blue, a solvatochromic dye) was loaded into the framework.
The dye loaded MOF showed a prominent visual change under influence of different solvents with prompt action.
The interaction of the pore encapsulated dye with less polar solvents such as n-heptane, diethyl ether, 1,4-dioxane, and acetone produces a light blue color.
This blue color changed to turquoise to sense the presence of acetonitrile and light blue for MeOH.
Moreover, identical color change has been shown on exposure of the respective vapor, demonstrating the usefulness of the ordered pores of the structure.
Apart from the pure solvatochromic property, the encapsulated dye has been deprotonated on interaction with NMP, producing a clear and distinct color change to purple.
On protonation from protic solvent, such as MeOH, it shows a slow and hypsochromic transformation to a green color.

Similar emission enhancement to detect different solvents was reported by {K. Müller-Buschbaum} et al. \cite{10.1039/c4dt03578j}.
This report demonstrated the applicability of a Ce based MOF for rapid detection of water and oxygen through “turn-off” of emission. While exposure to \ce{CH3CN} causes a ``turn-on'' emission demonstrating dual performance for multiple analyte detection.

\begin{figure}[tbhp]
\begin{center}
\includegraphics[width=0.6\linewidth]{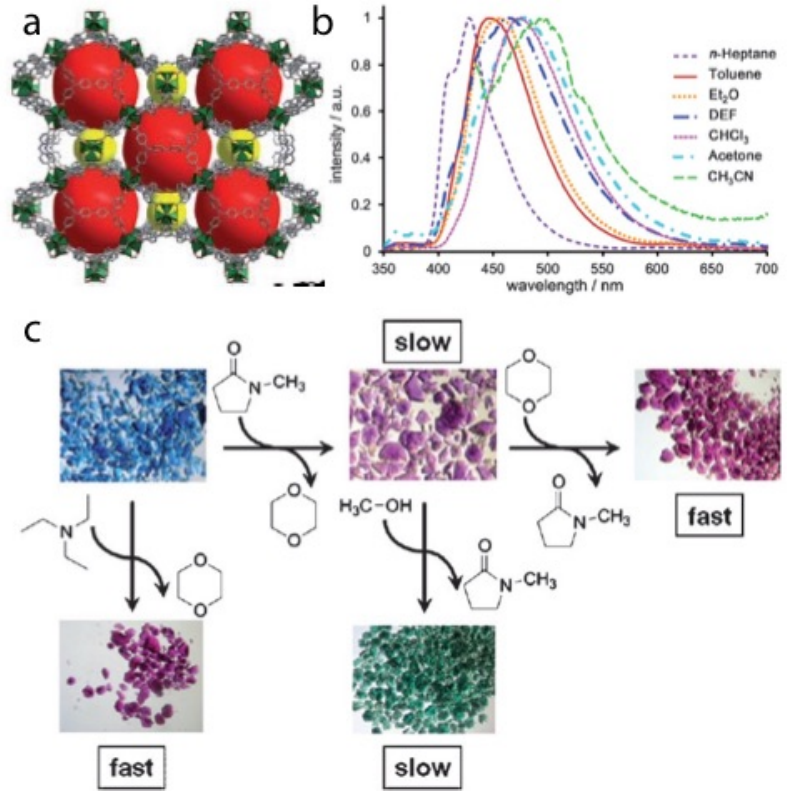}
\caption{(a) Structural representation of DUT-25 MOF and (b) its emission spectra under influence of different solvents. c) Solvatochromic color change of Nile-Blue loaded DUT-25. Adapted from \cite{10.1002/chem.201202352}, with permission.}
\label{fig:lum1}
\end{center}
\end{figure}

On a further illustration, chemical stability of Zr based MOFs has been applied to develop a new solvatochromic MOF (DUT-122) \cite{10.1002/ejic.201600261} from 9-fluorenone-2,7-dicarboxylic acid, based on the well-known solvatochromic behavior of fluorene derivatives.
Formation of the porous framework leads to blue shifting of the emission maxima of the linker from inside the framework by \SI{21}{\nano\metre}.
Expected solvatochromic behavior was promptly observed for the MOF with a variation of emission maxima from \SI{477}{\nano\metre} for THF to \SI{521}{\nano\metre} for MeOH.
Here the incapsulation of the solvent into the MOF pores and the effective interaction plays a crucial role in the resulting color of the MOF.
This shows the effectiveness of solvent interaction with pore walls leading to a shift of the emission maxima upon a change of solvent polarity.
Non-radiative transitions occur when OH groups from the analyte induce vibrational coupling, accounting for a red shift in such cases.
The inherent stability of the Zr cluster makes {DUT-122} suitable for detecting the presence of water.
DUT-122 shows a sharp change in color (darkening of the yellow color) as well as emission property (red shift of emission maxima to \SI{536}{\nano\metre}) in water and the kinetic analysis of the interaction revealed a fast response profile.

The challenge for detection of trace amount of water in organic solvent has been recently addressed by {K. Müller-Buschbaum} and coworkers.
They have reported the development of a composite material for facile detection of water from organic solvents, using a luminescent MOF as the active sensor.
The chosen lanthanide MOF \cite{10.1021/acsami.5b11965} shows intense fluorescence because of the lanthanide SBU, sensitized by the bipyridine containing linkers.
The luminescent MOF was then applied to superparamagnetic microparticles of \ce{Fe3O4}/\ce{SiO2}, forming a core/shell structure having the MOF layer as the active shell.
This composite provides the capability to interact with applied magnetic field, a prerequisite for easy and efficient separation from any medium.
Notably, the oxophilic nature of the lanthanides makes the MOF susceptible towards hydrolysis and the hydrolyzed product was found to lose the luminescence property.
Thus, the loss in emission property has a direct correlation with the subsequent water concentration responsible for the hydrolysis.
A detectable lowering of emission intensity was observed by the presence of a mere 0.1\% water in organic solvents like toluene and hexane.
The emission continued to quench efficiently in proportion with the water concentration, demonstrating possible quantitative estimation.
Theoretical detection limits of 0.01\% (\SI{9}{\micro\gram}) and 0.03\% (\SI{20}{\micro\gram}) water in toluene and hexane, respectively, was achieved.

In further work, a red emitting Eu based MOF (Eu-bipy or Eu-BDC) was added in addition to a green emitting Tb-bipy MOF, forming a tri-component microparticle \cite{10.1039/c7tc03312e}.
The complex microparticles contain a shell comprising of two differently emissive MOFs, and a superparamagnetic core.
Yellow emission of these composites is a sum of the red and green emission from individual MOFs and is suitable as a ratiometric sensor.
Loss of emission from hydrolysis of the MOFs, the key working principle for the sensors, shows different emission at different concentration based on the difference in hydrolysis kinetics of the individual luminescent MOFs.
A gradual shift of the emission from yellow towards red region of the spectra thus indicates the water content in hexane from anhydrous to contaminated.
Furthermore, a detection limit of 0.03\% (\SI{20}{\milli\gram}) water in anhydrous hexane was observed, comparable to the well-known Karl Fischer titration.

\begin{figure}[tbhp]
\begin{center}
\includegraphics[width=0.7\linewidth]{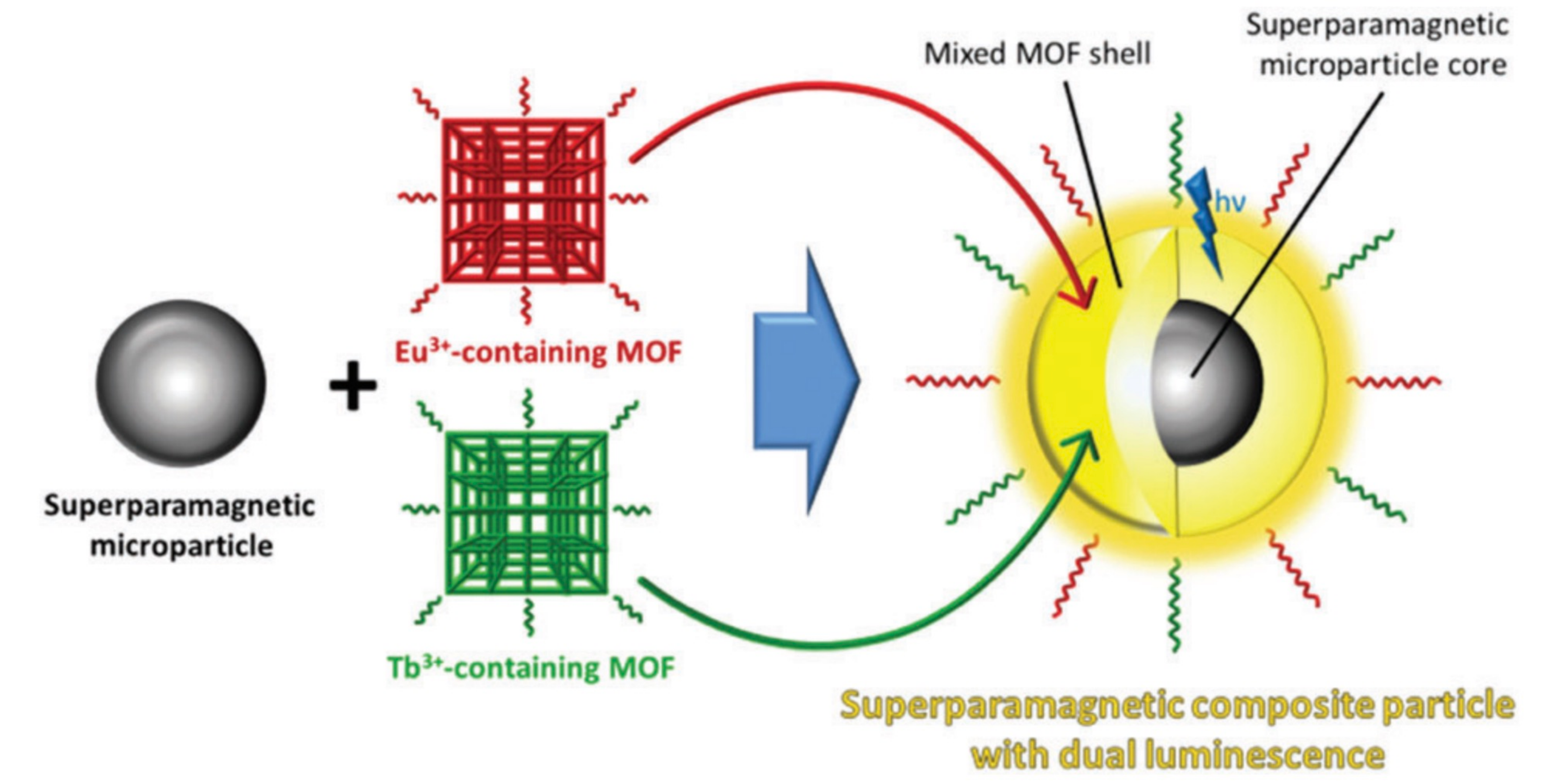}
\caption{Scheme for generation of microparticle with dual emission, featuring a luminescent and superparamagnetic core/shell. Reproduced from \cite{10.1039/c7tc03312e} with permission of The Royal Society of Chemistry.}
\label{fig:lum2}
\end{center}
\end{figure}

Detection of fluoride in water \cite{10.1038/srep02562} has been demonstrated by the group of {T. Bein} using a hybrid composite of the metal-organic framework \ce{NH2}-MIL-101(Al) and fluorescein.
The reporter moiety containing fluorescein as fluorophore remains covalently attached to the linker and doesn’t show any emission when inserted in the MOF.
However, on treatment with water, the MOF structure decomposes, leading to the release of the reporter molecules with a possibility for complex formation with the incoming analyte.
On selective complexation with fluoride ions in water, it shows a ``turn-on'' of the emission.
The ability to selectively detect fluoride from a mixture of other anions and with a sensitivity of 15~ppb makes this sensor remarkable.

Fabrication of thin films, a compulsory prerequisite for device based application has been attempted by the {R. A. Fischer} group \cite{10.1002/adfm.201500760}.
They demonstrated the development of quartz crystal microbalance (QCM) substrates for the application of sensing many volatile organic compounds, such as benzene, toluene, ethylbenzene, isomers of xylene and hexane and alcohols.
Growth of ZIF thin films on the QCM substrate allowed for analytes to pass through the ZIF pores at different rates, based on the pore size and surface functionality.
Selectivity of ZIFs towards different analytes on such devices has potential to be beneficial in industrial purposes including biofuel recovery and isomer separation at near future.

\subsection{Catalysis}

Owing to the unique property of chemical tunability from inside a defined framework MOFs are considered as promising candidates for catalysis.
Being heterogenous phase, MOFs also bear advantages, in terms of recyclability and ease of catalyst removal, over conventional homogenous catalysts.
These highly attractive features of MOFs led to many German chemists to investigate their capability in catalyzing numerous important reactions.
For the ease of discussion we classify the MOF based catalysts into 2 major categories, based on the chemical form of the active catalyst: MOFs having metal center as catalytic active site and composite materials using the MOF as support and/or active form of catalyst.

The metal ions present in the structure often act as structural nodes, providing structural integrity and rigidity to the framework.
However, based on the coordination environment, the metal ion containing SBUs have high potential to serve as catalytically active centers.
A chromium(III) cluster, the building unit for MIL-101, has proven highly effective for the cyanosilylation reaction of aldehydes \cite{10.1039/b718371b}.
The role of unsaturated sites for catalysis has been studied by {N. Stock} and coworkers \cite{10.1039/c2jm15592c}, using the Co-based microporous framework (Co-BTT), catalyzing ring opening reaction of styrene oxide along with oxidation of cycloalkanes and benzyl compounds.
Similar strategies incorporating a Rh paddlewheel during MOF construction has proven valid in achieving good performance for hydrogenation reactions; achieving high conversion of styrene into ethylbenzene without any side products \cite{10.1039/c3ta12795h}.

Taking advantage from the well known redox active property of cerium oxide, the {N. Stock} group also has shown the catalysis property of the UiO-66 architecture, when constructed using Ce node.
The thermal and chemical stability of the MOF allowed for catalytic  aerobic oxidation of benzyl alcohol through a redox pathway \cite{10.1039/C5CC02606G}.
They also demonstrated the effect of \ce{-NH2} groups on the ligand using an Al-based MOF towards Knoevenagel condensation reaction, an important reaction for industrial processes \cite{10.1039/c6ce02664h}.
Likewise decoration of the organic linker with \ce{-SO3H} functionalities led to an increase in the acidic character.
The ordered presence of functional groups in the MOF framework made it capable to dehydrating ethanol vapor towards formation of ethylene with an enhancement from 7\% conversion for non-functionalized linker containing MOF up to 91\% for the sulfonate functionalized MOF \cite{10.1002/chem.201501502}.
Solvent-free hydroxymethylation has been shown to be catalyzed using a Bi based MOF \cite{10.1002/anie.201204963}.
The acidity from the MOF structure is found to prevent any consecutive condensations and polymerization, showing the advantage of the designed structure of MOFs.

Another important strategy to improve the metal center based catalytic activity of MOFs has been developed by {R. A. Fischer} and coworkers through selective defect formation \cite{10.1002/chem.201602641}.
Formation of defects inside the framework causes more sites to be exposed and available as an active site.
This improves the already active Ru based MOFs for catalyzing the ethylene dimerization reaction and Paal–Knorr pyrrole synthesis, a reaction of acute industrial importance.

Liquid phase oxidation of cyclohexene using tert-butyl hydroperoxide as oxidant has been thoroughly studied by {D. Volkmer} and coworkers.
Co(II) complexes are known to be active for such catalysis under homogenous conditions thus incorporation of catalytically active Co(II) sites, inside a MOF, demonstrates a good turnover number (TON) with recyclability as achieved using \ce{[Co(II)(BPB)]*3DMF} \cite{10.1002/zaac.200800158}, MFU-1 \cite{10.1002/anie.200901241} and MFU-2 \cite{10.1002/chem.201003173} MOFs as the catalysts.
Additional investigations of MFU-1 have demonstrated the ability to act in a biomimetic cascade reaction using molecular oxygen from air to carry out oxidation reaction of several organic compounds using N-hydroxyphthalimide (NHPI) as a cocatalyst loaded into the framework.
Similar catalytic activity was observed when vanadium based MOFs were employed for the catalysis \cite{10.1002/ejic.201101099}, with no leaching of metal ion and efficient recyclability.
Aerobic oxidation of tetralin has been achieved using a Cu-based MOF (CFA-5), as another mimic to biological copper-containing enzymes \cite{10.1039/c4dt01880j}.

The porous characteristic of MOFs provides the opportunity to include active species inside the framework.
This strategy has been proven useful by many scientists to employ MOF structures for application as a support to catalytically active moieties.
The initial report for MOF based catalysis from Germany by {S. Kaskel} and workers shows the applicability of MIL-101 as a support for Pd \cite{10.1039/b718371b}.
The Pd supported on MIL-101 showed a high efficiency support for catalytic hydrogenation of styrene.
{R. A. Fischer} et al. have also presented a catalyst using a MOF as support for metal nanoparticles.
When Pd nanoparticles were encapsulated into the frameworks of UiO-66 and UiO-67 through chemical vapor infiltration of (allyl)Pd(Cp), a composite material having Pd nanoparticles loaded on MOF was obtained \cite{10.1002/ejic.201500299}.
This composite material has been tested to show efficient catalytic property for reactions, such as selective hydrogenation of ketones to secondary alcohols.
Composite materials have been further investigated to encapsulate bimetallic nanoparticles inside the chemically robust frameworks of Zr based MOFs \cite{10.1002/chem.201603984}.
The resulting insertion of PdPt or RuPt nanoparticles inside the UiO-66 framework causes the generation of catalytically active composite materials with uniform distribution of the nanoparticles.
The PdPt nanoparticles were found to perform better for catalysis in nitrobenzene hydrogenation, when immobilized inside the MOF framework.
The RuPt nanoparticles inside the MOF composite are found to be active for preferential oxidation of CO in excess hydrogen.

Recently, the paddlewheel SBU of HKUST-1 was modified with catalytically active sites, such as Pd, and the activity improved \cite{10.1039/c6dt02893d}.
The observed catalytic activity was further improved when Pd NP are loaded into the Pd substituted MOF.
Thus aqueous-phase hydrogenation of 4-nitrophenol to 4-aminophenol using \ce{NaBH4} as a reducing agent proceeds at an extremely fast rate when the Pd content inside the framework is significantly higher.
This highlights a novel method to overload Pd for providing an improved catalytic activity.

\subsection{MOFs for cyclic water adsorption for heat transformation}
\begin{figure}[tbhp]
\begin{center}
\includegraphics[width=0.5\linewidth]{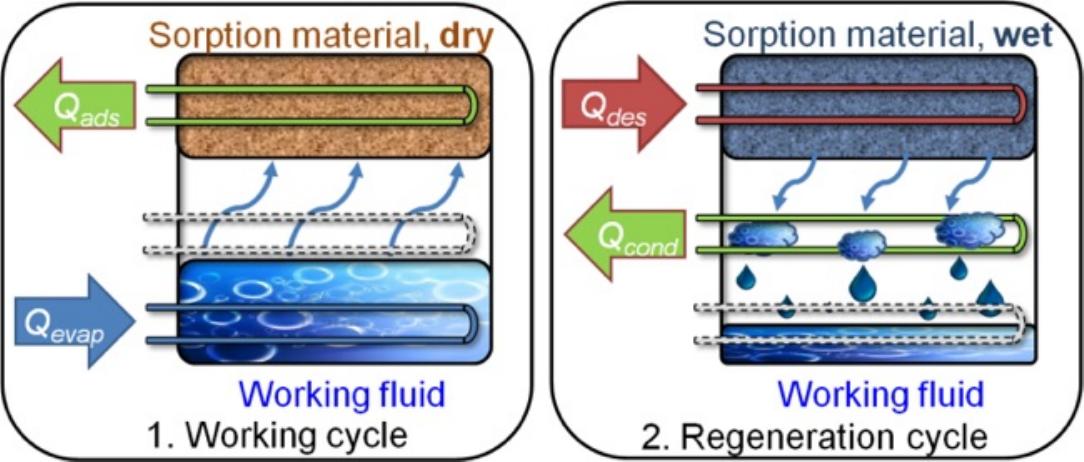}
\caption{Principle of adsorption chilling or adsorption heat pump. Working cycle: A working fluid (typically \ce{H2O}) is evaporated at low pressure by application of evaporation heat $Q_\mathrm{evap}$ (useful cold), and adsorbed at a microporous material, releasing adsorption heat $Q_\mathrm{ads}$. Regeneration cycle: When the adsorbent is saturated, driving heat $Q_\mathrm{des}$ is applied for desorption of the working fluid. The vapor then condenses in a cooler, and condensation heat $Q_\mathrm{cond}$ is released. \cite{10.1039/C3NJ01556D} - Published by The Royal Society of Chemistry (RSC) on behalf of the Centre National de la Recherche Scientifique (CNRS) and the RSC.}
\label{fig:water1}
\end{center}
\end{figure}

Microporous materials with high water stability, high water uptake capacity at low but not too low relative humidity and with tunable hydrophilicity, such as MOFs, are currently in the focus for reversible cycling water adsorption in order to achieve low temperature heat transformation applications \cite{10.1039/C3NJ01556D,10.1002/ejic.201101056,10.2533/chimia.2013.419}.
Heat transformations in thermally driven adsorption chillers (TDCs) or adsorption heat pumps (AHPs) are an alternative to traditional air conditioners or heat pumps operating on electricity or fossil fuels. Thermally driven adsorption chillers and heat pumps feature a two-step process, which is depicted in Figure~\ref{fig:water1}.

By using solar or waste heat as the driving energy, TDCs or AHPs can significantly help to minimize primary energy consumption and greenhouse gas emissions generated by industrial or domestic heating and cooling processes, especially when run on electricity.
TDCs and AHPs are based on the evaporation and consecutive adsorption of coolant liquids, preferably water, under specific conditions.
TDCs and AHPs are established technologies and are commercially available with small scale capacities of 5--\SI{25}{\kilo\watt} produced by InvenSor (Germany), Fahrenheit (ex SorTech, Germany), Mitsubishi Plastics (Japan), SolabCool (Netherlands) and Jiangsu Huineng (China) and large scale capacities of 50--\SI{1000}{\kilo\watt} manufactured by Power Partners (USA), GBU mbH (Germany), HIJC (ex-Nishiyodo, USA) and Mayekawa (Japan) \cite{10.1016/j.energy.2016.03.036}.
The ranges of application of TDCs and AHPs, as well as their efficiencies, power densities and total costs, are substantially influenced by the microporosity and hydrophilicity of the employed adsorption materials.
At present TDCs and AHPs use either silica gel or zeolites. Here, we briefly summarize current investigations by Germany-based researchers for development and possibilities of MOFs compared to classical materials.

\begin{figure}[tbhp]
\begin{center}
\includegraphics[width=0.5\linewidth]{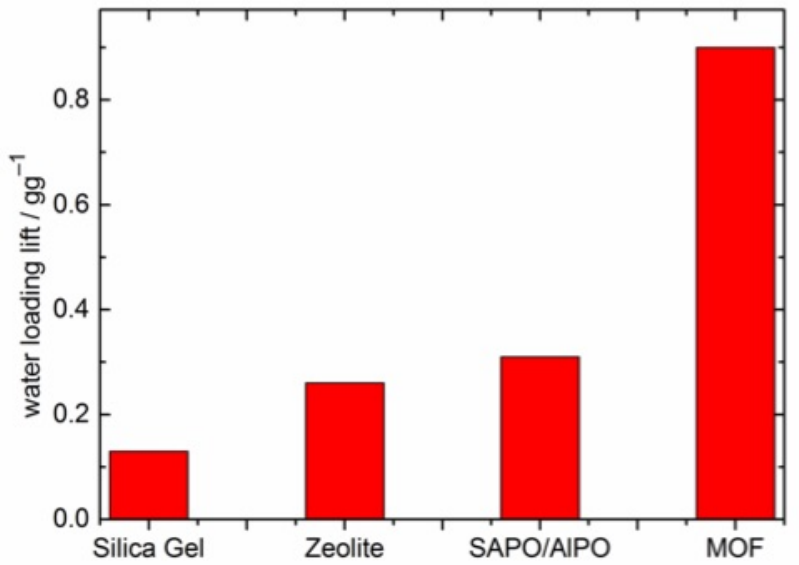}
\caption{Comparison of typical water uptake capacities as loading lifts (in g water/g dry sorption material) within an adsorption/desorption cycle for traditional porous materials and MOFs (SAPO/AlPO = silica-aluminophosphate/aluminophosphate). Reproduced from \cite{10.1002/ejic.201101056}, with permission.}
\label{fig:water2}
\end{center}
\end{figure}

Several classes of materials are potentially promising for adsorption heat transformation applications \cite{10.1016/j.applthermaleng.2011.09.003}.
These materials include metal aluminophosphates (AlPOs, SAPOs, MeAPOs), ordered porous solids, porous carbons and various composites (SWSs, AlPO-Al foil).
With their large water uptakes, MOFs surpass those materials (Figure~\ref{fig:water2}), at the same time, the variability of the MOF building blocks (metal node and organic linker) allows for tuning the microporosity and hydrophobicity/hydrophilicity, depending on the specific application.

Water stability is, however, a problem for MOFs.
Many are not stable at all and decompose upon prolonged contact with water even at room temperature \cite{10.1021/ja9061344,10.1016/j.micromeso.2015.11.055}.
At present MIL-101, CAU-10-H, Al-fum and UiO-66 appear to be the most water stable MOFs \cite{10.1016/j.micromeso.2015.11.055}.
That is why a considerable amount of testing of MOFs for water adsorption for heat transformation is for inital stability investigation.
It should be noted, water is the obvious fluid of choice for thermally driven chillers and adsorption heat pumps, however, alcohols, such as ethanol, could be an alternative in the context of the hydrolytic instability of MOFs \cite{10.1039/C3NJ01556D,10.1016/j.energy.2014.11.022}.

\begin{figure}[tbhp]
\begin{center}
\includegraphics[width=0.5\linewidth]{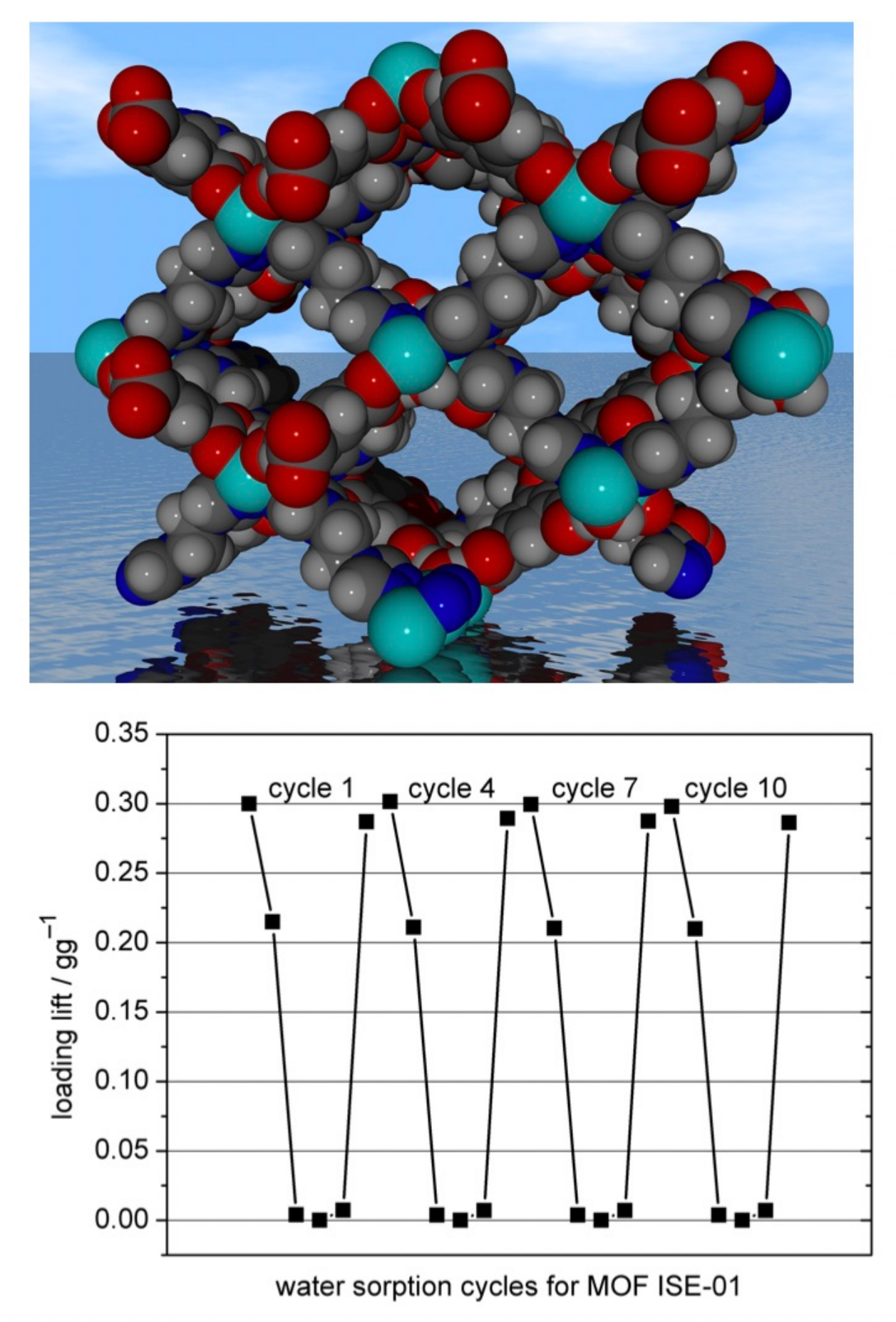}
\caption{Top: Space-filling presentation of the mixed-ligand MOF ISE-1 with 52\% water-filled volume. The crystal water in the channels is not shown. \cite{10.1039/b715812b} - Reproduced by permission of The Royal Society of Chemistry. Bottom: Water loading lift over 10 different cycles with four of them shown for stability test. Reprinted (adapted) with permission from \cite{10.1021/ja808444z}. Copyright 2009 American Chemical Society.}
\label{fig:water3}
\end{center}
\end{figure}

The {C. Janiak} group began testing MOFs for heat-transformation applications in 2009.
This group initially reported water cycling of the mixed-ligand MOF \ce{[Ni3(\mu_3-btc)2(\mu_4-btre)2(\mu-H2O)2]} (btre = 1,2-bis(1,2,4-triazol-4-yl)ethane), also named ISE-1.
This MOF was synthesized from water with an initial water content of ca. 30 wt\% \cite{10.1039/b715812b}. This water content can be reversibly desorbed and adsorbed over several cycles (Figure~\ref{fig:water3}) \cite{10.1021/ja808444z}.

The adsorption of water vapor in the highly porous MIL-101(Cr) of over \SI{1}{\gram\per\gram} (between \SI{140}{\celsius} and \SI{40}{\celsius} under a water vapor pressure of \SI{5.6}{\kilo\pascal}) together with the stability over several cycles makes MIL-101(Cr) the record holder for water uptake \cite{10.1002/ejic.201001156}.
MIL-101(Cr) is synthesized in quite harsh acidic aqueous reaction conditions (\SI{220}{\celsius} in water) \cite{10.1126/science.1116275}.
Thus, fundamental water stability of this material can be expected and was verified over 40 water adsorption cycles.
Unfortunately, MIL-101(Cr) is too hydrophobic as the main uptake of the otherwise advantageous s-shaped isotherm occurs only at $p/p_{0} \approx 0.2$ \cite{10.1002/ejic.201001156}.

There are many MOFs that demonstrate improved water adsorption compared to ISE-1.
For example, the water uptake of \SI{0.75}{\gram\per\gram} for MIL-100(Fe) and MIL-100(Cr), at small relative pressures of $p$/$p_{0}$ $< 0.4$, with a steep s-shaped isotherm and comparatively small hysteresis, including very good cycle stability, make these MOFs ideal candidates for cycling water adsorption \cite{10.1039/C2JM15615F,10.1016/j.micromeso.2008.11.020,10.1246/cl.2010.360}.
The isostructural MIL-100(Al) features a similar isotherm, albeit with a smaller loading lift of slightly less than \SI{0.5}{\gram\per\gram} \cite{10.1039/C2JM15615F}.
Notably, MIL-101(Cr) can be functionalized with amino, nitro or other groups on the benzene-1,4-dicarboxylate ligand through time-controlled postsynthetic modification of the parent material \cite{10.1039/C0CC04526H}.
Hydrophilic nitro (\ce{-NO2}) or amino (\ce{-NH2}) functionalities were introduced into MIL-101(Cr) in order to achieve increased water loading at lower $p$/$p_{0}$ values.
Fully and partially (p) aminated MIL-101(Cr)-(p)\ce{NH2} showed an unchanged water uptake capacity with respect to parent MIL-101(Cr) (about \SI{1}{\gram\per\gram}) and very high water stability over 40 adsorption-desorption cycles \cite{10.1021/cm304055k}.

\begin{figure}[tbhp]
\begin{center}
\includegraphics[width=0.5\linewidth]{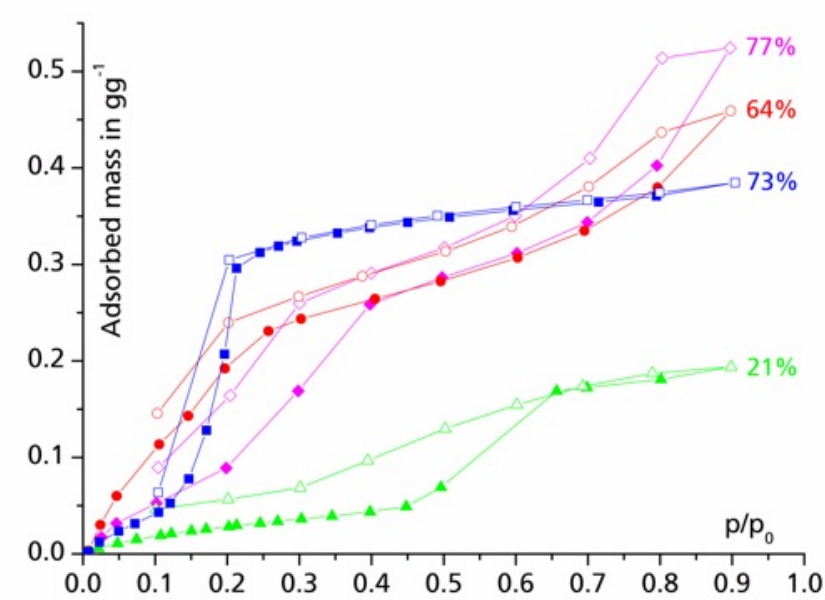}
\caption{Water adsorption/desorption isotherms of UiO-66, \ce{H2N}-UiO-66, UiO-67 and \ce{H2N}-MIL-125, acquired at T = \SI{25}{\celsius}. Adsorption: filled symbols, desorption: empty symbols. \cite{10.1039/C3DT51471D} - Reproduced by permission of The Royal Society of Chemistry.}
\label{fig:water4}
\end{center}
\end{figure}

Further, the feasibility of the metal-organic frameworks UiO-66, UiO-67, \ce{H2N}-UiO-66 and \ce{H2N}-MIL-125(Ti) as adsorption materials in heat transformations were investigated, since at the start of the project zirconium MOFs of the UiO-66 family were regarded as rather water stable.
The amino-modified compounds \ce{H2N}-UiO-66 and \ce{H2N}-MIL-125 feature indeed a high heat of adsorption (89.5 and \SI{56.0}{\kilo\joule\per\mole}, respectively) and a very promising \ce{H2O} adsorption isotherm due to their enhanced hydrophilicity (Figure~\ref{fig:water4}). For \ce{H2N}-MIL-125 the very steep rise of the \ce{H2O} adsorption isotherm in the $0.1 < p/p_{0} < 0.2$ region is especially beneficial for the intended heat pump application \cite{10.1039/C3DT51471D}.
Yet, \ce{H2N}-UiO-66 appears of limited stability during a multi-cycle hydrothermal stress test while the water uptake of \ce{H2N}-MIL-125 remains unchanged during short, non-equilibrium cycles.
\ce{H2N}-MIL-125 appears to show hydrothermal stability comparable to MIL-100(Fe) or MIL-101(Cr) \cite{10.1039/C3DT51471D}.

The distinctive water adsorption properties of microporous aluminum fumarate (Al-fumarate, Basolite A520) with an s-shaped isotherm, a narrow hysteresis and a loading \textgreater \SI{0.3}{\gram\per\gram} at relative pressures as low as $ p/p_{0} = 0.3$ under realistic working conditions represented a significant advancement of MOF-based adsorption heat transformation processes \cite{10.1039/C4RA03794D}.
Furthermore, the {C. Janiak} group demonstrated aluminum fumarate shows favorable and unprecedented cyclic hydrothermal stability \cite{10.1039/C4RA03794D}.
For the application of heat transformation, unhindered heat and mass transfer are crucial for fast adsorption/desorption cycles and high power density.
A \SI{300}{\micro\metre} thick, polycrystalline, thermally well coupled and highly accessible coating of microporous aluminum fumarate was deposited onto an aluminum metal substrate via the thermal gradient approach.
This was found to be stable for the first 4500 adsorption/desorption cycles with water vapor, as judged by unchanged crystallinity by PXRD \cite{10.1039/C4RA03794D}.
The maximum water exchange of \SI{0.35}{\gram\per\gram} of Al-fumarate is lower than for other MOFs, however, it can be fully utilized under realistic, isobaric working conditions due to the hydrophilicity of Al-fumarate.
The desired, steep s-shape of the isotherm in a relative pressure band of $0.2 < p/p_{0} < 0.35$  had not been observed for other water-stable MOFs before.
In this context, Al-fumarate compares well with hydrophilic, zeolitic adsorbents such as AlPO-18 or AlPO-4 \cite{10.1016/j.applthermaleng.2010.03.028}.
Importantly, Al-fumarate can be produced relatively cost-efficiently via a simple precipitation reaction in a continuous reactor or even by extrusion.
This is a strong advantage and in contrast to, for example, the template-based AlPO-18 synthesis, or various other hydro/solvothermal synthesis routes required for other MOFs.
Al-fumarate also does not contain heavy metals, fluorine or critical organic compounds.

\begin{figure}[tbhp]
\begin{center}
\includegraphics[width=0.5\linewidth]{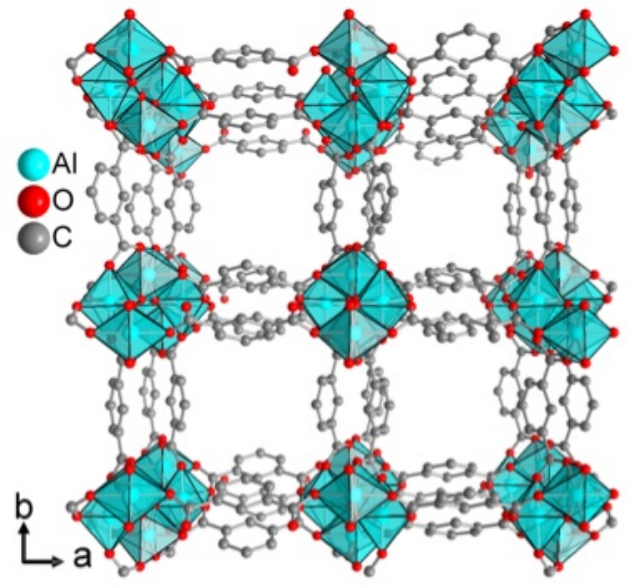}
\caption{The three-dimensional centrosymmetric tetragonal framework structure of CAU-10-H with its square shaped one-dimensional channels. \cite{10.1039/C6TA01757F} - Published by The Royal Society of Chemistry.}
\label{fig:water5}
\end{center}
\end{figure}

Another aluminium based MOF, aluminium isophthalate MOF {CAU-10-H}, reported by {N. Stock} and coworkers \cite{10.1021/cm3025445} exhibits promising water adsorption characteristics for the application in heat-exchange processes with water as working fluid (Figure~\ref{fig:water5})\cite{10.1021/cm3025445,10.1039/c4dt02264e,10.1039/C6TA01757F}.
CAU-10-H coating with Silikophen was evaluated for its long-term stability under closed-cycle conditions for 10,000 water adsorption/desorption cycles, which are approaching typical expected lifetimes for adsorptive heat transformation applications \cite{10.1039/C6TA01757F}.
No degradation of the adsorption capacity could be observed which makes CAU-10-H the most stable MOF under these humid cycling conditions reported.
While the water uptake of \SI{0.26}{\gram\per\gram} for the coated sample is lower compared to other stable MOFs, the successful coating procedure and high stability up to 10,000 cycles under working conditions make CAU-10-H the current top performing MOF for heat pump applications \cite{10.1039/C6TA01757F}.
Since CAU-10-H is the second MOF demonstrating such long-term stability and since the first MOF with such property (aluminum fumarate) is also based on \ce{Al^{3+}}, we expect that Al-based MOFs are the most promising MOF adsorbents for application in water based heat pumps \cite{10.1039/C6TA01757F}.

It is evident that shaping MOFs, which are normally obtained as powders or microcrystals, is one indispensable factor for potential applications of MOFs \cite{10.1039/C3NJ01556D}.
In order to make use of MOFs for cyclic water adsorption applications, shaping into monoliths, granules or coatings could be envisioned.
Thereby, the essential porosity of the MOF should not be lost and the amount of MOF in the shaped formed should be as high as possible.
MIL 101(Cr) was successfully embedded into a macroporous and monolithic oil-water (o/w) high internal phase emulsion (HIPE) foam, based on crosslinked poly(2 hydroxyethyl methacrylate) (HEMA) \cite{10.1016/j.micromeso.2014.11.025} or cross-linked poly(N-isopropyl acrylamide) (NIPAM) \cite{10.1016/j.micromeso.2015.09.008}.
These hierarchical and mechanically stable monolithic composite materials with up to 59 wt\% of MIL-101(Cr) for the HEMA-HIPE and a realistic 71 wt\% for the NIPAM-HIPE show higher methanol and water vapor uptake capacities compared to the pure HIPE.
Pre-polymerization of the HIPE emulsions was shown to be an indispensable factor for synthesizing highly porous composites where the micro- and mesopores of MIL 101(Cr) remain partially unblocked.
The maximum vapor exchange of the MIL-101@HEMA-HIPE composite and the NIPAM-HIPE composite are certainly lower than in pure MIL-101(Cr) but most can be utilized under realistic working conditions up to $p/p_{0} = 0.4$ \cite{10.1016/j.micromeso.2014.11.025,10.1016/j.micromeso.2015.09.008}.

Monolithic MOF composites were also synthesized using micro-to-mesoporous MIL-100(Fe,Cr) and MIL 101(Cr) with a pre-polymerized mesoporous resorcinol-formaldehyde based xerogel as binding agent \cite{10.1016/j.micromeso.2015.05.017}.
The monolithic bodies could be loaded with up to 77 wt\% of powdery MOF material with retention of the MOF surface area and porosities (from \ce{N2} adsorption) by pre-polymerization of the xerogel solution.
These MIL-101(Cr)@xerogel-\ce{H2O} composites matched the wt\%-correlated BET values and water uptakes within experimental error.
As an indication of the hierarchical nature the 77 wt\% MIL-101(Cr)@xerogel-\ce{H2O} composite achieved \SI{0.79}{\gram\per\gram} water uptake at $p/p_{0} = 0.5$ while for bulk MIL-101(Cr) only \SI{0.57}{\gram\per\gram} water uptake could be achieved \cite{10.1016/j.micromeso.2015.05.017}.

\begin{figure}[tbhp]
\begin{center}
\includegraphics[width=0.5\linewidth]{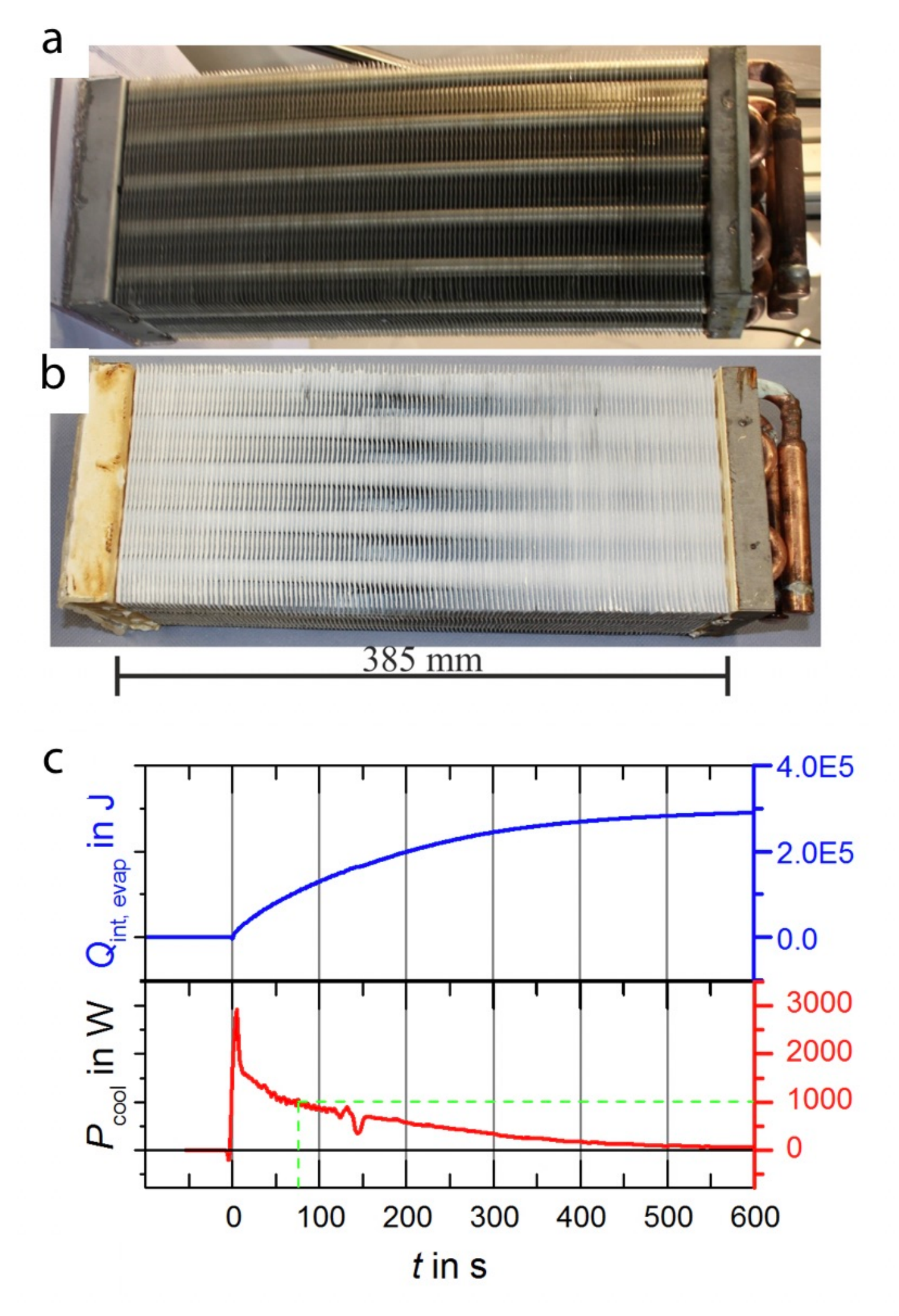}
\caption{Heat-exchanger (a) before and (b) after coating with \SI{493}{\gram} of the aluminum fumarate MOF and drying. (c) Gross cooling power P$_{\mathrm{cool}}$ of the aluminum fumarate-coated heat exchange for operating conditions \SI{90}{\celsius} – \SI{30}{\celsius} – \SI{18}{\celsius} (red line, lower part), calculated from the integral heat of evaporation $Q_{\mathrm{int,evap}}$ (blue line, upper part). The green line indicates P$_{\mathrm{cool}}$ = \SI{1000}{\watt} for a half cycle time of t = 74 s. Adapted with permission from \cite{10.1021/acs.iecr.7b00106}. Copyright 2017 American Chemical Society.}
\label{fig:water6}
\end{center}
\end{figure}

Recently, a step towards application was made by the group of {S. K. Henninger} at the Fraunhofer Institute for Solar Energy Systems by fabricating a functional, full-scale heat exchanger coated with \SI{493}{\gram} of the microporous aluminium fumarate MOF (Basolite A520) using a polysiloxane-based binding agent (Figure~\ref{fig:water6}) \cite{10.1021/acs.iecr.7b00106}.
The function of the heat exchanger was evaluated to a gross cooling power of \SI{2900}{\watt} (at the beginning of the adsorption cycle) or, respectively an average cooling power of \SI{690}{\watt} (up to a limit of 90\% equilibrium loading in 7 minutes) under the working conditions of a realistic adsorption chiller of \SI{90}{\celsius} – \SI{30}{\celsius} – \SI{18}{\celsius} (temperature level of heat source, heat rejection/condenser and evaporator).
With the inherent multi-cycle stability of microporous aluminum fumarate and the excellent long-term stability of polysiloxane coatings, reported in the literature, these results clearly suggest that the technology has the potential for industrial application and can significantly advance adsorption-based chilling \cite{10.1021/acs.iecr.7b00106}.

\subsection{Biomedical applications}

Another advanced application of porous MOFs includes biomedical applications which has been showcased by {S. Wuttke} and coworkers.
The porous nature with tunable pore environment has made MOFs attractive candidate for host-guest chemistry \cite{10.1021/nn405469g,10.1016/j.biomaterials.2017.01.025}.
The responsiveness of bare or modified frameworks helps to produce the desired control of guest encapsulation and their release.
This triggered release process has proven useful in release of important biologically active cargo, towards successful application of MOFs as drug delivery vehicles.

Nanoparticles of mesoporous MOFs, MIL-100(Fe) and MIL-101(Cr), posses significant porosity (3205 and \SI{2004}{\square\metre\per\gram}, respectively) and thus were chosen for developing a delivery vehicle system \cite{10.1039/c5cc06767g}.
On treatment with dispersion of lipids (1,2-dioleoyl-sn-glycero-3-phosphocholine, DOPC), a lipid bilayer was observed to develop around the MOF nanoparticles.
Thus, a system for dye loaded MOF nanoparticles with lipid bilayer coating was easily achieved for loading into target cancer cells.
The active dyes from such nanoparticles were then shown to release from their core on triggering with suitable stimuli like a nonionic surfactant (triton X-100).
Application of such trigger causes an immediate release of the loaded cargo, ready for action inside the cancer cell.
This approach serves as a backbone for later studies to include other dyes like fluorescein \cite{10.3390/ma10020216}.

Over the time, systematic development has been made with these model systems, including investigations of biocompatibility.
Studies for such lipid bilayer coated MOF nanoparticles have been carried out to investigate any potential health risk, using human endothelial cells and mouse lung cells as model system \cite{10.1002/adhm.201600818}.

\subsection{MOF Thin Films}

\begin{figure}[tbhp]
\begin{center}
\includegraphics[width=0.7\linewidth]{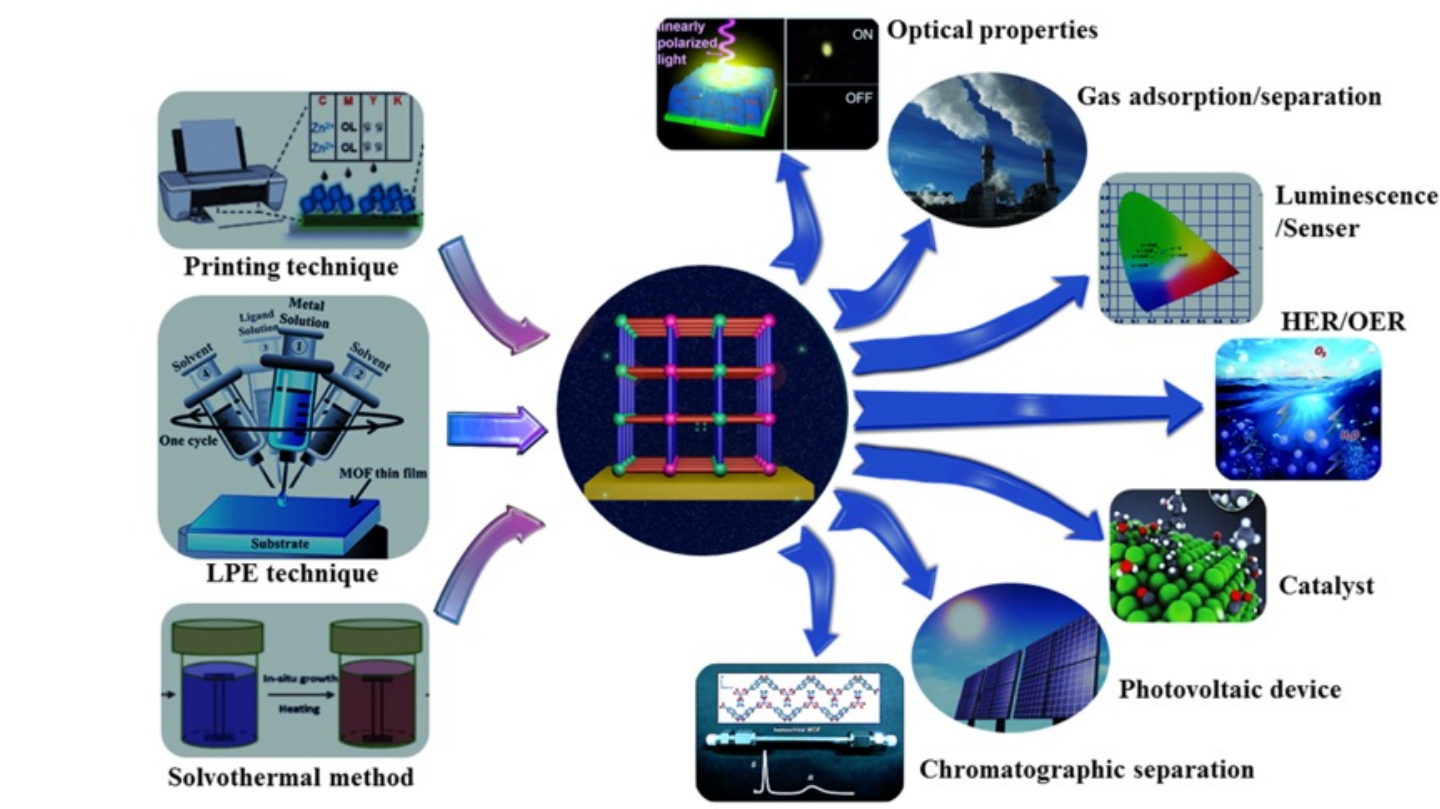}
\caption{Scheme of typical deposition techniques and applications that were developed in Germany in last decade.}
\label{fig:thinfilm1}
\end{center}
\end{figure}

As discussed, MOFs have invoked continuous interest for academic and industrial applications in the last decade due to their intrinsic porous characteristics, unique structural diversities and tunable functionalities \cite{10.1039/B804680H}.
There is great demand for fabricating powdered porous materials as films because of the obvious advantages of surfaces and interfaces for use as smart films, chemical sensors, electronic devices (Figure~\ref{fig:thinfilm1}) \cite{10.1016/S1387-1811(00)00345-0,10.1016/S0167-2991(97)80684-2,10.1021/cm960148a}.
In this prospect, MOF thin films were developed with the pioneering work of {R. A. Fischer}, {C. Wöll} and their coworkers in 2005.
The conventional solvothermal method was used to deposit patterned crystalline MOF-5 films on COOH/CF-terminated self-assembled monolayers (SAMs) on Au(111) surfaces \cite{10.1021/ja053523l}.

In 2007, MOF systems were extended to HKUST-1 on SAMs terminated with a different functional group under solvothermal crystallization conditions by {T. Bein} and coworkers \cite{10.1021/ja0701208}.
They also studied the water adsorption behavior of HKUST-1 films at different temperatures monitored by a quartz crystal microbalance (QCM) instrument \cite{10.1016/j.micromeso.2008.01.024}, which offers an alternative strategy for researchers to study gas adsorption behavior.
Also, as described previously, using the conventional solvothermal method {J. Caro} and coworkers further developed a seeded secondary growth method to obtain dense MOF membranes, such as ZIF-7, ZIF-8 and ZIF-90 membranes, with high multi-gas selectivity \cite{10.1002/anie.200905645}.
This work therefore provides an alternative way to fabricate solid support-MOF membrane, which is likely superior to gas separation via molecular sieves.

Another achievement in the field of MOF thin films was the development of a stepwise layer-by-layer (LBL) or liquid-phase epitaxy (LPE) growth technique for a variety of applications.
{O. Shekhah} et al. described the application of LPE to grow a prototypical MOF, HKUST-1 on SAMs modified Au substrates \cite{10.1021/ja076210u}.
The LPE technique enables one to obtain MOF thin films (or SURMOFs) with high porosity, good crystallinity, and preferred orientation \cite{10.1021/cr200167v}.
The deposition process can be monitored by surface plasmon resonance (SPR) or QCM frequency as a function of the time, which suggests a controllable growth progress.
In an optimized case, the surface roughness can be reduced to the nm scale, which is beneficial for the application of MOF thin films in nanotechnology \cite{10.1002/cphc.201000488}.
Furthermore, the LPE technique can control the interpenetration behavior of MOFs mounted on surfaces.
{C. Wöll} and coworkers fabricated non-interpenetrated MOF-508 SURMOFs on SAM modified Au surfaces via the LPE method \cite{10.1038/nmat2445}.
In this way, the non-interpenetrated MOF-508 can be obtained by using the pyridine-terminated organic surface as a nucleation template to suppress the second and interpenetrating lattice in the SURMOFs.
Kr Brunauer-Emmett-Teller (BET) surface area was used to study the non-interpenetration behavior of SURMOFs and the BET surface area is calculated based on the MOF films thickness, obtained from atomic force microscopy.
The non-interpenetrated MOF-508a shows a value of \SI[separate-uncertainty = true]{930\pm15}{\square\metre\per\cubic\centi\metre}, which is smaller than twice the Kr BET value for the interpenetrated MOF-508a, \SI[separate-uncertainty = true]{627\pm15}{\square\metre\per\cubic\centi\metre}.
This implies that MOF-508a with a small Kr BET adsorption value behaves like an interpenetrated structure, whereas the MOF-508a with a large Kr BET adsorption value behaves like a non-interpenetrated structure.
This discovery shows an alternative way to improve the tunability of the porosity and surface areas as well as the flexibility of MOF films, facilitating the application of SURMOFs in gas adsorption, drug storage and release and guest molecules encapsulation \cite{10.1038/nmat2445}.

The advantages of the LPE technique enable fabricated MOF films to be employed for the application of selective adsorption, gas/vapor sensors, catalysis, luminescence sensors and photovoltaic device \cite{10.1021/cr200167v,10.1002/9783527635856.ch13,10.1039/C7CS00315C}.
For example, from 2012--2017, {R. A. Fischer} and coworkers integrated SURMOFs on SAM functionalized gold substrates for enantioselective gas/volatile organic compound adsorption \cite{10.1002/anie.201104240,10.1002/anie.201207908,10.1002/adfm.201500760}.
Here the uptake of guest molecules can be directly monitored by QCM.
The chosen MOFs for SURMOFs were developed from single MOFs to multi-variant heteroepitaxially grown MOFs (MOF-on-MOF concept) \cite{10.1007/s11426-011-4406-8,10.1039/C3TA13812G,10.1002/chem.201002381}.
The heteroepitaxial growth of MOF-on-MOF can greatly improve the adsorption kinetics of the SURMOF composites \cite{10.1039/C3TA13812G,10.1002/chem.201002381,10.1039/C0DT01818J}.
The integration of different MOFs onto one SURMOF system can inherit their own characteristics, imparting synergetic functions for the SURMOF.
Thus, the MOF-on-MOF concept can expand the potential application of SURMOFs in various research areas.

{C. Wöll} and coworkers investigated the biocompatibility of SURMOFs by encapsulating bioactive molecules into the pores of SURMOFs using the LPE technique.
SURMOF-2 (\ce{[Cu(bdc)2]_n}) was found to be remarkably stable in water and artificial seawater, however, SURMOF-2 is found to degrade very fast in typical cell culture media.
The results suggest that SURMOFs can act as a release matrix for pharmaceuticals and other drugs.
This study opens up the possibility of applying SURMOFs in the biochemical field \cite{10.1021/la300457z,10.1186/1559-4106-8-29}.

Moreover, {C. Wickleder}, {C. Wöll} and coworkers fabricated HKUST-1 SURMOFs loaded with europium $\beta$-diketonate complexes via LPE.
Luminescence spectra of these SURMOFs before and after the loading Eu-based compounds suggest that MOF thin films are promising as materials for photonic antennae \cite{10.1002/cphc.201200262}.
Similarly, SURMOFs can also encapsulate other fluorescent molecules, which also show interesting optical properties \cite{10.1039/C7CC00961E}.
{D. Schlettwein}, {C. Wöll} and coworkers immobilized ferrocene (Fc) as a redox mediator into the lattices of SURMOFs \cite{10.1039/C1CC16580A}.
Because Fc is easily oxidized to \ce{Fc^+} when applying a voltage, Fc molecules in proximity of an electrode will be quickly oxidized.
The charges transfer through the film by a hopping transport mechanism between adjacent loaded Fc molecules in the SURMOF lattices.
Thus, the Fc molecules in the SURMOFs will be oxidized.
This study suggests that SURMOFs loaded with redox mediators are promising in the field of catalysis and electronic devices \cite{10.1039/C1CC16580A}.

Furthermore, {C. Wöll} and coworkers also reported the post-synthetic modification of LPE deposited SURMOFs for standard click chemistry with high conversion yields \cite{10.1021/la403854w}.
For this, \ce{[Zn2(N3-bdc)2(dabco)]} was first deposited on SAM modified Au substrates via LPE method and a strain-promoted metal-free click reaction was used to functionalize the SURMOF with different reactants containing strained triple bonds.
Considering the advantages of strain-promoted azide-alkyne cycloaddition (SPAAC), the alkyne products were obtained with a yield of up to 100\%.

In 2015, {C. Wöll} and coworkers reported that SURMOFs fabricated on magnetic nanoparticles via the LPE method showed promising applications in catalysis, chromatographic separation and drug delivery systems \cite{10.1021/acsnano.5b00483}.
Additionally, they also developed SURMOFs with promising dielectric, optical and sensing properties from 2013--2016 \cite{10.1039/C7CS00315C,10.1063/1.4819836}.
In 2013, the optical constant (dielectric constant) of HKUST-1 SURMOFs was measured before and after removing the water/EtOH trapped in the structure \cite{10.1063/1.4819836}.
It was observed that the optical constant of HKUST-1 SURMOFs decreases to 1.39 after the removal of the solvent in the voids of the SURMOFs.
However, the optical constant of HKUST-1 SURMOFs increases again to 1.55-1.6 after re-exposure to water/EtOH atmosphere.
The dependence of the optical properties on water/EtOH adsorption reveals the potential application of SURMOFs for optical sensing devices.
In 2015, {J. Liu} et al. fabricated crystalline porphyrin MOFs on SAM modified fluorine-doped tin oxide {FTO} via the spray LPE method \cite{10.1002/anie.201501862}.
FTO supported SURMOFs were assembled as a photovoltaic device with conductive FTO as the bottom electrode and iodine/triiodide as the top electrode. As studied by current-voltage (I-V) curves, the assembled porphyrin Zn-SURMOF based photovoltaic device shows a photocurrent efficiency of 0.45\% with an open-circuit voltage of \SI{0.7}{\volt}, a short circuit current density of \SI{0.71}{\milli\ampere\per\square\centi\metre} and a fill factor of 0.65.
These values demonstrate the efficiency of porphyrin Zn-SURMOF based photovoltaic devices is more than a factor of two compared to photovoltaic devices without porphyrin Zn-SURMOFs (open circuit voltage of \SI{0.57}{\volt}, short-circuit current density of \SI{0.45}{\milli\ampere\per\square\centi\metre}, fill factor of 0.55 and photocurrent efficiency of 0.2\%).
Porphyrin-based SURMOFs show promising application for photovoltaic devices, which play a crucial role in energy storage and conversion.

To summarize, Germany-based researchers have pioneered the development of liquid phase epitaxy technique for the fabrication of MOF thin films ranging from understanding the growth mechanism to various applications in the last decade.
The SURMOFs fabricated by liquid phase epitaxy approach have demonstrated promising applications in various area, especially in nanotechnology.

Additionally, {T. Bein} and coworkers demonstrated approaches for the growth of MOF thin films via crystallization at room temperature and thin gel-layer synthesis method \cite{10.1002/anie.201001684}.
Highly orientated MOF thin films, such as MIL-88B, MIL-53, HKUST-1 and UiO-68-\ce{NH2}, were obtained on various SAM modified solid surfaces by crystallization at room temperature \cite{10.1039/C2DT12265K}.
Fluorescent dye molecules were covalently modified inside the pores of MOF thin films by post-synthesis methods, which demonstrate size-selective fluorescence quenching behavior.
MOF thin films of MIL-88B also show high water/gas adsorption ability \cite{10.1039/B916953A}.
Furthermore, {A. Terfort} and coworkers, based at Goethe University Frankfurt, developed a simple precipitation route, an evaporation method to deposit MOF thin films at room temperature, allowing for the precise localization of MOF particles on a solid surface \cite{10.1002/adfm.201002529}.
The {A. Terfort} group also developed evaporation method and inkjet-printing approaches for integrating patterned MOFs onto solid surfaces \cite{10.1002/adma.201301626}.
These contributions have potential for different micro-printing and nanotechnological areas.

Apart from the above mentioned methods, a variety of advanced deposition techniques, such as Langmuir-Blodgett layer-by-layer deposition \cite{10.1038/nmat2769}, electrochemical deposition \cite{10.1021/cm900069f}, microwave-induced thermal deposition \cite{10.1039/B800061A}, and chemical vapor deposition \cite{10.1038/nmat4509}, have also been developed to fabricate MOF thin films with well-defined morphology and structures.

MOF thin films fabricated by different techniques will perform different functions due to the differences in the resulting particles sizes, surface roughness, film thickness and film stabilities obtained by the different deposition approaches.
This exciting research ranging from various deposition techniques to applications of MOF thin films has been carried out all over the world.
Despite these encouraging reports, the study of MOF thin films is still at an emerging stage of vigorous development and considerable effort is required before MOF thin films can be industrialized and used commercially.

\section{Conclusion}
MOFs are clearly a promising class of materials displaying many advanced functions and Germany-based researchers have invested significant effort into understanding these materials.
In this review, we have particularly outlined the reports of German research groups for the synthesis of new MOFs, approaches for producing MOFs on an industrial scale, the development of advanced measurement methods for characterization and finally examples of potential functions and properties.

\rev{We believe the collaborative nature of MOF research in Germany will continue to flourish and enable the discovery of new materials and significant scientific advancement.
For example, MOF thin film preparation and characterization of conductive or semiconducting MOFs is being pioneered by researchers across Germany and this work is poised to revolutionize electronic and photonic devices.}

While this work has highlighted the work of German investigators we note these studies have benefited tremendously from fundamental and groundbreaking studies from researchers all around the world.
Moreover, much of this work is the product of excellent international collaboration with other European and international researchers which we hope will continue into the future.

\section{Acknowledgements}

This work was supported by the Deutsche Forschungsgemeinschaft [grant number FOR2433]; European Research Council (ERC) under the European Union’s Horizon 2020 research and innovation programme [grant number 742743]; Bundesministerium für Wirtschaft und Energie (BMWi) [grant number 0327851B]; Bundesministerium für Bildung und Forschung in the project Optimat [grant number 03SF0492C].

\section{References}

\bibliography{article}

\end{document}